\documentclass[11 pt]{article}
\usepackage[margin=1.06in]{geometry}
\usepackage{graphicx}
\usepackage{amsmath}
\usepackage{amsthm}

\usepackage[usenames,dvipsnames]{color}
\definecolor{LightGray}{gray}{0.85}
\definecolor{VeryLightGray}{gray}{0.95}
\usepackage{multirow}
\usepackage{rotating}
\usepackage{colortbl}

\usepackage[numbers,square,comma,sort&compress]{natbib}
\usepackage{ulem}
\usepackage[colorlinks]{hyperref}

\usepackage{lineno}

\theoremstyle{plain}

\providecommand{\Ro}{\mathcal{R}_0}

\newcommand{\beginsupplement}{%
        \setcounter{table}{0}
        \renewcommand{\thetable}{S\arabic{table}}%
        \setcounter{figure}{0}
        \renewcommand{\thefigure}{S\arabic{figure}}%
     }

\newcommand{\captionfonts}{\small}
\makeatletter  
\long\def\@makecaption#1#2{%
  \vskip\abovecaptionskip
  \sbox\@tempboxa{{\captionfonts #1: #2}}%
  \ifdim \wd\@tempboxa >\hsize
    {\captionfonts #1: #2\par}
  \else
    \hbox to\hsize{\hfil\box\@tempboxa\hfil}%
  \fi
  \vskip\belowcaptionskip}
\makeatother   

\newcommand{\figpath}{Figures}

\newcommand{\comment}[1]{}

\title{\sffamily Forecasting and Uncertainty in Modeling the \\2014-2015 Ebola Epidemic in West Africa}
\author{Marisa C. Eisenberg$^{\dagger \ddagger}$, Joseph N.S. Eisenberg$^\dagger$, Jeremy P. D'Silva$^{*\dagger}$, Eden V. Wells$^\dagger$, \\Sarah Cherng$^\dagger$, Yu-Han Kao$^\dagger$, and Rafael Meza$^\dagger$\\ 
\small $^\dagger$ Department of Epidemiology, School of Public Health, University of Michigan, Ann Arbor\\
\small $^\ddagger$ Department of Mathematics, University of Michigan, Ann Arbor\\
\small $^*$ Father Gabriel Richard High School, Ann Arbor
}

\date{}

\begin{document}
\maketitle

\begin{abstract} \sffamily
The 2014-2015 Ebola epidemic in West Africa is the largest ever recorded, with over 27,000 cases and 11,000 deaths as of June 2015. 
The public health response was challenged by difficulties with disease surveillance (particularly in more remote regions), which impacted subsequent analysis and decision-making regarding optimal interventions. 
We developed a stage-structured model of Ebola virus disease (EVD). A key feature of the model is that it includes a generalized correction term accounting for factors such as the fraction of cases reported and fraction of the population at risk (e.g. due to contact patterns, interventions, spatiotemporal spread, pre-existing immunity, asymptomatic cases, etc.). We generated a range of short term forecasts for Guinea, Liberia, and Sierra Leone, which we then validated using subsequent data. We also used the model to examine the uncertainty in the relative contributions to transmission by the different stages of infection (early, late, and funeral). 
We found that a wide range of forecasted trajectories fit approximately equally well to the early data. By including the correction factor term, however, the best-fit models correctly forecasted EVD activity for all three countries, both individually and for all countries combined. In particular, the models correctly forecasted the slow-down and stabilization in Liberia, as well as the continued exponential growth in Sierra Leone through November 2014. 
Parameter unidentifiability issues hindered estimation of the relative contributions of each stage of transmission from incidence and deaths data alone, which poses a challenge in determining optimal intervention strategies, and underscores the need for additional data collection. 
Even with these limited data, however, it is still possible to accurately capture and predict the epidemic dynamics by using a simplified correction term that approximately accounts for the complex underlying factors driving disease spread. 
\end{abstract}


\section{Introduction}
Ebola virus disease (EVD) was first identified in 1976, in two simultaneous outbreaks in Sudan and the Democratic Republic of the Congo \cite{WHO_EbolaFactsheet}. The 2014-2015 Ebola epidemic is unprecedented in both its size and complexity, exceeding the numbers of cases and deaths for all previous Ebola outbreaks combined \cite{WHO_NEJM}. The outbreak began in Guinea in December 2013 \cite{baize2014emergence}, expanding to yield widespread and intense transmission in Guinea, Liberia, and Sierra Leone, as well as cases in seven additional countries (Italy, Mali, Nigeria, Senegal, Spain, UK, USA) \cite{WHO_EbolaSitReps}. 
The devastation of the 2014-2015 Ebola epidemic in West Africa presented the public health community with many surprises and many challenges. Unparalleled explosion of cases within urban regions and in a new part of Africa was unexpected; the Ebola virus' ecologic niche was the exploitation of a combination of urbanization, regional instability, extreme poverty and insufficient infrastructure to contain spread \cite{Alexander2014}. The challenges associated with containment have been varied and included accurate reporting of cases and deaths; accurate tracking of behavioral changes that put people at increased or decreased risk, and accurate assessment of conferred immunity within the population. These and other factors associated with surveillance, behavior patterns affecting risk, and biological patterns affecting risk vary over time and impact how well incidence patterns are estimated. 

\begin{figure}
\centering
\includegraphics[width=0.9\textwidth]{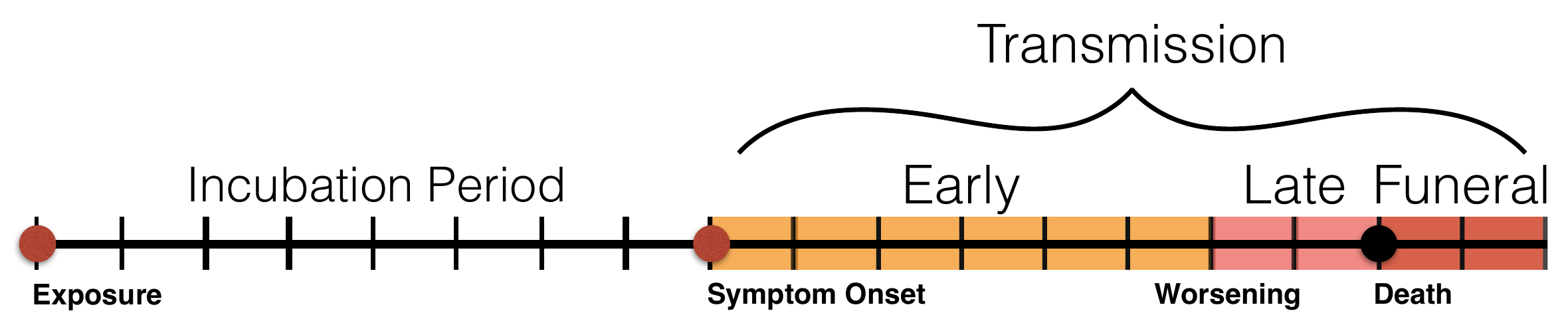}
\caption{Progression of EVD from exposure, through onset of symptoms, worsening, and eventually death.}
\label{fig:progression}
\end{figure}

The difficulties of surveillance have resulted in frequent reclassification and changes in case reporting as the outbreak evolved \cite{WHO_EbolaSitReps}. Reporting fraction estimates have been difficult, although some have been presented \cite{Meltzer2014, scarpino2014epidemiological}. Often regions were lacking in laboratory capacity to confirm cases, which when combined with widespread underreporting \cite{Meltzer2014,WHO_EbolaSitRep29Oct14} results in the potential for wide variation in reporting fraction with either over- or under-estimates of cases. Rural areas in particular saw significant difficulties in both data collection and healthcare response \cite{dallatomasinas2015ebola}. The levels of asymptomatic infection and preexisting immunity are also largely unknown for the current outbreak, and have been noted as factors which can lead to overestimation of incidence when forecasting \cite{Bellan2014}. Additionally, behavior change has been noted as a major factor affecting the course of the epidemic, and ultimately helping to curtail it \cite{funk2014ebola,ReutersEbolaBehaviorChange}, however there is little quantitative data on behavioral changes over time. Clearly, good surveillance requires appropriate mechanisms to collect data on these issues of reporting, behavior, and population immunity.  Although we aspire to collect better and more complete data, in the midst of an outbreak data collection is often not and arguably should not be considered a priority over providing health care.  

There have been a wide range of simulation models developed for EVD in recent months to assist in decision-making and public health response efforts \cite{Meltzer2014, Bellan2014, nishiura2014early, chowell2014transmission, gomes2014assessing, rivers2014modeling, browne2014model, fisman2014early, chowell2014western, pandey2014strategies, kiskowskithree, yamin2014effect}, many of which are based on early models by Chowell, Legrand, and others \cite{chowell2004basic, legrand2007understanding, lekone2006statistical}. Several models have been used for forecasting and for understanding disease dynamics \cite{Meltzer2014, Bellan2014, gomes2014assessing, nishiura2014early}, disease progression and immunity \cite{yamin2014effect, Bellan2014}, and evaluating alternative interventions \cite{lewnard2014dynamics, rivers2014modeling}.
In this manuscript, we add to this growing literature by presenting a stage-structured model of EVD transmission, where we incorporate a correction term that broadly accounts for the fraction of cases and deaths reported, as well as the symptomatic fraction, pre-existing immunity levels, and other components determining the overall fraction of the population at risk. This correction factor provides a simple parametric adjustment by which we can account for  limitations common in data sparse outbreak settings, allowing us to accurately capture and predict the epidemic dynamics.

\section{Data}
Cumulative incidence of suspected cases and deaths from Guinea, Liberia, Sierra Leone, and all countries combined as reported by the World Health Organization (WHO) \cite{WHO_EbolaSitReps} from May 2014 to February 2015 was used in this analysis. In estimating the model parameters, we used start dates roughly corresponding to when each location began consistent exponential growth, to avoid issues due to initial stochasticity and hurdles in setting-up reporting systems (e.g. some early data shows decreases in cumulative cases, likely due to changes in surveillance systems, case definitions, etc.), as shown in the Figures. For the end dates for used in parameter estimation, we note that intervention efforts escalated significantly after October 1, 2014, including the beginning of the first-ever UN emergency health mission, the UN Mission for Ebola Emergency Response (UN-MEER) on October 1, 2014 \cite{UNMEER}. Additionally, the data beginning in October had increased uncertainty and changes in reporting, with incomplete data and significant decreases in reported cumulative cases (likely due to changes in reporting) \cite{WHO_EbolaSitReps}. A WHO Roadmap update from October stated that ``the capacity to capture a true picture of the situation in Liberia remains hamstrung by underreporting of cases'' \cite{WHO_EbolaSitRep29Oct14}. Thus, we primarily fitted data up through October 1, 2014, and used the subsequent data for validation, as described below.

\section{Methods}

\begin{figure}
\centering
\includegraphics[width=0.55\textwidth]{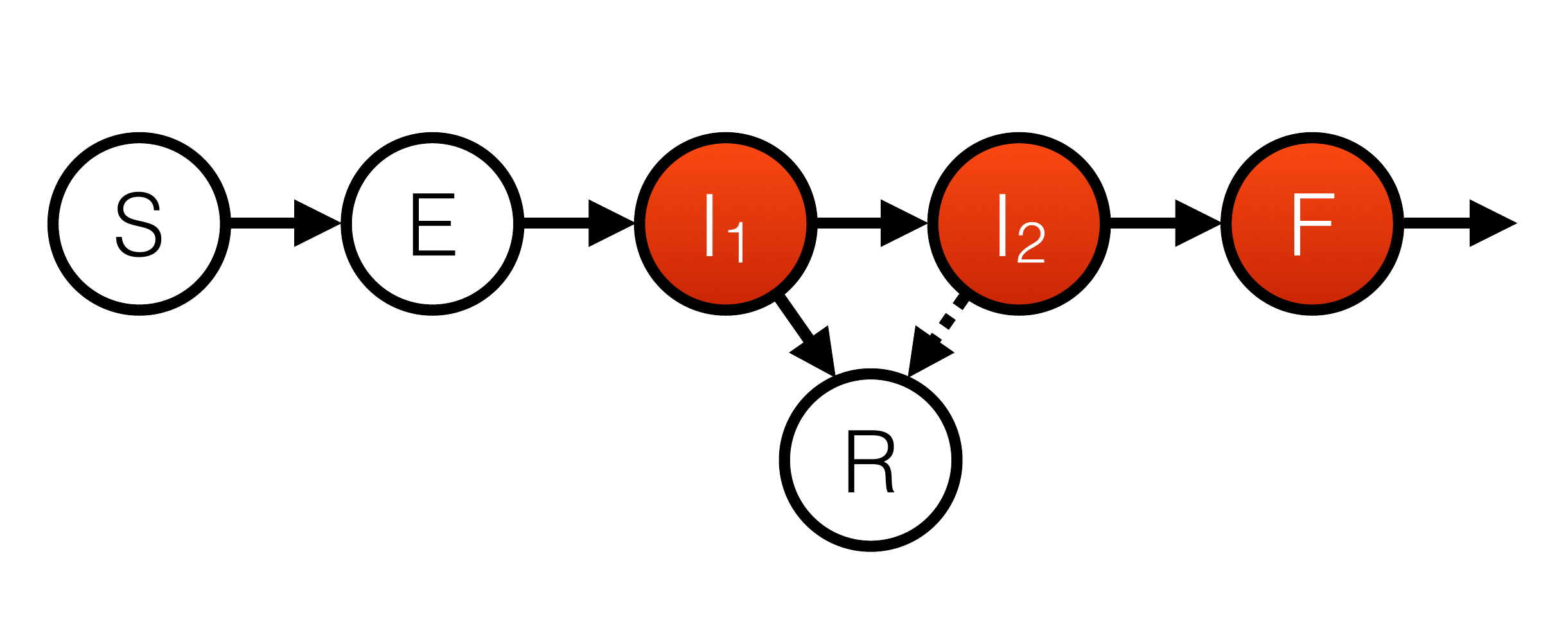}
\caption{Stage-structured compartmental model of EVD, which splits the population into susceptible ($S$), exposed ($E$), first-stage infected ($I_1$), late-stage infected ($I_2$), recovered $R$, and funeral transmissible ($F$). Red compartments are transmissible, and recovery rates are greater from $I_1$ than from $I_2$ \cite{Ndambi1999}.}
\label{fig:SimpleModel}
\end{figure}

\subsection{Model Structure} 
EVD is transmitted by contact with body fluids \cite{CDC_Ebola, WHO_EbolaFactsheet}, with the overall course of EVD infection shown in Figure \ref{fig:progression}.  Healthcare workers are at particular risk for transmission due to their frequent contact with patients and body fluids. Burial ceremonies during which mourners contact the body of the deceased have also been identified as playing an important role in community-based transmission \cite{CDC_Ebola, WHO_EbolaFactsheet, Francesconi2003}. EVD has an average incubation period of 8-10 days (range of $\sim$2-21 days), with individuals believed to become infectious after the onset of symptoms \cite{WHO_EbolaFactsheet}.
EVD progresses in two broad stages---an initial infectious stage ($I_1$) in which symptoms tend to be milder (such as fever, headache, sore throat, muscle aches), often progressing to diarrhea, vomiting, and a second, more intense stage ($I_2$) during which these as well as more advanced symptoms (such as hemorrhaging and multi-organ failure) manifest \cite{WHO_EbolaFactsheet, Ndambi1999}. Typical time to progression between infection stages is approximately 5-7 days \cite{CDC_EbolaClinicanInfo2014}. 
As second stage symptoms tend to have more release of body fluids, transmissibility increases as the disease progresses \cite{Francesconi2003}.
Most recoveries occur from the first stage, while the second stage is often fatal \cite{Ndambi1999, beeching2014ebola, isaacson1978clinical, CDC_EbolaClinicanInfo2014}. Among fatal cases, death occurs 6-16 days after symptom onset (averaging 7.5 days for the 2014 outbreak) \cite{CDC_EbolaClinicanInfo2014}. Thus, we used a two-staged infection process to reflect the increasing symptoms and transmissibility as the infection progresses \cite{Ndambi1999, WHO_EbolaFactsheet}, with death occurring after symptom progression to the second stage ($I_2$). The model structure is shown in Figure \ref{fig:SimpleModel} and the model equations are:
\begin{equation}
\begin{aligned}
\frac{dS}{dt} &= - (\beta_I I_1 + \beta_2 I_2 + \beta_F F)S \\
\frac{dE}{dt} &= (\beta_I I_1 + \beta_2 I_2 + \beta_F F)S - \alpha E \\
\frac{dI_1}{dt} &= \alpha E - \gamma_{1} I_1\\
\frac{dI_2}{dt} &= \delta_1 \gamma_{1} I_1 - \gamma_2 I_2 \\
\frac{dF}{dt} &= \delta_2 \gamma_2 I_2 - \gamma_F F\\
\frac{dR}{dt} &= (1-\delta_1) \gamma_1 I_1 + (1-\delta_2) \gamma_2 I_2 - \gamma_R R\\
\end{aligned}
\label{eq:simplemodel}
\end{equation}
where $S$ is the fraction of the population which is susceptible, $E$ the fraction exposed, $I_1$ the fraction in the first stage of infection, $I_2$ the second stage of infection, $R$ the fraction of the population which is recently recovered (i.e. who would still require a bed if hospitalized), and $F$, the fraction of the population that has died and is in the process of being buried. 
Because we considered a relatively short time frame ($<1$ year), for simplicity we ignored population background births and deaths. 

The parameter $\delta_1$ represents the fraction of infected who progress to $I_2$ and $\delta_2$ the fraction who subsequently progress to death, with $\delta = \delta_1 \cdot \delta_2$ thus defined as the overall fraction of cases resulting in death (case fatality risk (CFR)). As the overall CFR is a more commonly measured quantity than the progression rate from $I_1$ to $I_2$, we primarily worked using $\delta$ rather than $\delta_1$ (setting $\delta_1 = \delta/\delta_2$). 

The model structure is somewhat similar to the SEIHFR model of Legrand et al. \cite{legrand2007understanding}, in which infected individuals in the community ($I$) can be admitted to the hospital ($H$), however the progression in this model is through the natural history of the disease rather than from household to hospital. Thus, there are different mechanistic assumptions underlying the two models, so that even though the resulting compartmental diagrams are similar, there are important differences in the flows between compartments. For example, SEIHFR models typically assume lower transmission rate parameters in the hospital ($H$) stage \cite{rivers2014modeling}, which contrasts with the higher transmission rate parameters in the second stage of infection used in our model; additionally, mortality is significantly higher for infected individuals in the community than in the hospital, whereas there is higher recovery in the first stage and death only after the second stage of infection in our model. 
\\

\noindent \emph{Basic Reproduction Number}. Using the second-generation matrix approach \cite{diekmann1990definition, van2002reproduction}, the basic reproduction number for the model is given by: 
\begin{equation}\Ro = \frac{\beta_1}{\gamma_1}+\frac{\beta_2 \delta_1}{\gamma_2}+\frac{\beta_F \delta_1 \delta_2}{\gamma_F}. \label{eq:R0}\end{equation}
where $\Ro$ for the system breaks into three portions based on each transmission stage, weighted by the fraction of individuals who reach that stage and the amount of time spent in each stage (as is typical for stage structured models \cite{van2002reproduction}). 
\\

\noindent \emph{Measurement Equations}. To connect the model to the data on cumulative cases and deaths provided by the WHO, we append the following measurement equations to Eq. \eqref{eq:simplemodel}:
\begin{equation}
\begin{aligned}
y_C &= k N \int_0^t \alpha E d\tau, \\
y_D &= k N \int_0^t  \delta_2 \gamma_2 I_2 d\tau, 
\end{aligned}
\label{eq:meas}
\end{equation}
where $y_C$ represents cumulative cases, and $y_D$ cumulative deaths. The parameter $k$ represents a combination of several factors, including: the fraction of the population at risk (whether due to social contact structure, spatial heterogeneity, immunity from asymptomatic infection, or other factors), a rough correction for errors in our initial condition assumptions, and the fraction of cases and deaths reported (reporting fraction). Although in reality the reporting fraction is likely to be different for cases and deaths, it is not known how or in which direction, so for simplicity we set the two to be equal.

\subsection{Parameter Estimation Methods} Table \ref{tab:params} gives the parameter definitions, values, ranges, and sources. For the total population size, $N$, we used a total population of 11,745,189 for Guinea, 4,294,077 for Liberia, and 6,092,075 for Sierra Leone, as determined by the World Bank \cite{WorldBankWAfrica2014}. The model initial conditions were approximated from the data as described in the Supplementary Information.

To fit the model, we used least squares fitted to cumulative cases and deaths as reported by the WHO via the measurement equations \eqref{eq:meas}, using Nelder-Mead optimization in MATLAB \cite{MATLAB2014b}. We would expect that a wide range of model parameters may fit the data equally well---particularly as the data up through October 1 was still in the exponential growth phase, during which it can be fitted with only two parameters (i.e. as one would fit a line on a log scale). Fitting to cumulative incidence measurements has been shown to potentially result in artificially small confidence bounds on parameter estimates and forecasts \cite{king2014avoidable}, although the  potential unidentifiability due to exponential growth-phase data means there may be a large or even infinite range of parameter values that fit the data well, for both cumulative or daily incidence. We used cumulative incidence in our analyses because at the time of the outbreak, cumulative incidence was reported rather than daily case counts \cite{WHO_EbolaSitReps}, and the cumulative incidence provided often decreased significantly from one time point to another (due to reclassification and changes in reporting). Differencing to yield non-cumulative incident cases would thus generate some negative case counts. 


To address both the issue of artificially smaller confidence intervals due to the use of cumulative incidence data, and the fact that the data are largely exponential (yielding inherent identifiability issues in fitting), we used a Latin Hypercube (LH) sampling approach. We sampled all parameters across realistic ranges given in Table \ref{tab:params}, and then for each sample, we fit only two parameters, the transmission parameter $\beta_1$ and the overall mortality fraction $\delta$. 
We note that all of the sampled regions could generate qualitatively good fits that were consistent with the observed data once $\beta_1$ and $\delta$ were fitted. 
Thus, the method allows us to examine the full range of parameters and model trajectories consistent with both the data and the known realistic ranges of the parameters. 
We then used these bounds rather than traditional confidence intervals for our forecasting and estimates. Essentially, this method uses the biological ranges for the parameters to generate our ranges of estimates for the parameters and forecasts, rather than likelihood-based criteria, which may suffer from issues of unidentifiability and lack of independence in the data. 
Since realistic values for $k$ are largely unknown (as $k$ accounts for a wide range of factors), the range for $k$ was set very broadly, with an upper bound of 1 (representing perfect reporting and a completely at-risk population) and an approximate lower bound determined by taking the lowest value which still yielded qualitatively good fits. We LH-sampled 1000 parameter sets for each of four cases: all countries combined, Guinea, Liberia, and Sierra Leone.
We also tested the extrema of each range in Table \ref{tab:params}, to illustrate the full range of potential behaviors generated by the model in the realistic ranges.

\begin{table*}
\centering
\def\arraystretch{1.5}
\begin{tabular}{| c | m{2.1in} | c | c | c |}

\hline
\rowcolor{LightGray}  &\bf Definition                    	&\bf Units 			    & \bf Range &\bf Source\\
\hline

$\beta_1$   & Transmission parameter for first stage of illness  & \small people$^{-1}$ days$^{-1}$  & Estimated  & Estimated  \\

$\frac{\beta_2}{\beta_1}$ & Ratio of infectiveness in first vs. second stage of illness   &   unitless   & 1.5 - 5	&   \cite{yamin2014effect, dowell1999transmission, Francesconi2003} \\

$\frac{\beta_F}{\beta_1}$ & Ratio of infectiveness in first stage vs. funeral transmission   &   unitless   & 1.5 - 5	&   \cite{yamin2014effect, rivers2014modeling, dowell1999transmission, Francesconi2003} \\

$\alpha^{-1}$ & Average incubation period  						& days 			& 8-10 & \cite{WHO_EbolaFactsheet, CDC_Ebola} \\

$\gamma_1^{-1}$ & Average length of first stage of illness & days & 5 - 7 & \cite{WHO_EbolaFactsheet, Ndambi1999} \\

$\gamma_2^{-1}$ & Average length of second stage of illness & days & 1 - 2 & \cite{WHO_EbolaFactsheet, Ndambi1999} \\

$\gamma_F^{-1}$ & Average time from death until burial & days & 1 - 3 & \cite{WHO_EbolaFactsheet, rivers2014modeling} \\

$\gamma_R^{-1}$ & Average duration of ETU bed occupancy after recovery & days& 5 - 15 & \cite{Meltzer2014},$^*$ \\

$\delta$ & Overall mortality fraction & unitless & 
Estimated & Estimated \\

$\delta_2$ & Mortality fraction among those in second stage of illness & unitless & 0.9 - 1 & \cite{Ndambi1999} \\

$k$ & Fraction of the population at risk, symptomatic fraction, and reporting fraction & unitless & 0.001 - 1$^\dagger$ & See methods \\

\hline
\end{tabular}

\caption{Model parameters, definitions, units, sources, and ranges used in the simulations. $^*$Personal communication with clinicians on the ground in Liberia. $^\dagger$This range was used for all simulations except when fitting Guinea in Figure \ref{fig:simplefit}, where a wider range from 0.00025 - 1 also yielded equally good fits to the data and was used. ETU = Ebola treatment unit. }
\label{tab:params}
\end{table*}

\begin{table*}
\centering
\def\arraystretch{1.4}

\begin{tabular}{| c |  cc  cc  c |}

\hline

&                       &&\bf \centering Overall && \bf Range of \\
&\bf Parameter &&\bf Best Estimate && \bf  Estimates (Median) \\
\cline{2-2} \cline{4-4} \cline{6-6}

\multirow{3}{*}{\begin{sideways}\footnotesize ~All Countries \end{sideways}} 

 & $\beta_1$ && 0.11  && 0.079 - 0.18 (0.12) \\
 & $\delta$    &&  0.56  && 0.55 - 0.6 (0.57)\\
&\cellcolor{VeryLightGray} $\Ro$ &\cellcolor{VeryLightGray}&\cellcolor{VeryLightGray}  1.60 &\cellcolor{VeryLightGray} & \cellcolor{VeryLightGray} 1.43 - 1.64 (1.53) \\
\hline

\hline
\multirow{3}{*}{\begin{sideways}Guinea~ \end{sideways}} 
 & $\beta_1$ && 0.16 && 0.078 - 0.18 (0.11)\\ 
 & $\delta$ && 0.65 && 0.64 - 0.72 (0.68)\\ 
&\cellcolor{VeryLightGray} $\Ro$ &\cellcolor{VeryLightGray}& \cellcolor{VeryLightGray} 1.79 &\cellcolor{VeryLightGray} & \cellcolor{VeryLightGray} 1.47 - 1.79 (1.61) \\ 
\hline

\hline
\multirow{3}{*}{\begin{sideways}Liberia~ \end{sideways}}
 & $\beta_1$ && 0.12 && 0.0725 - 0.17 (0.10)\\
 & $\delta$ &&0.63 && 0.63 - 0.72 (0.67)\\
&\cellcolor{VeryLightGray} $\Ro$ &\cellcolor{VeryLightGray}& \cellcolor{VeryLightGray} 1.81 &\cellcolor{VeryLightGray} & \cellcolor{VeryLightGray} 1.34 - 2.75 (1.47)\\
\hline

\hline
\multirow{3}{*}{\begin{sideways}\footnotesize Sierra Leone~ \end{sideways}}
 & $\beta_1$ && 0.12 && 0.085 - 0.17 (0.12) \\
 & $\delta$ && 0.38 && 0.36 - 0.38 (0.37) \\
&\cellcolor{VeryLightGray} $\Ro$ &\cellcolor{VeryLightGray}& \cellcolor{VeryLightGray} 1.32 &\cellcolor{VeryLightGray} & \cellcolor{VeryLightGray}  1.19 - 1.37 (1.25)\\
\hline

\end{tabular}
\caption{Estimated values for $\beta_1$, $\delta$, and $\Ro$, using the stage structured model. All remaining parameters were LH sampled from the ranges in Table \ref{tab:params}. The overall best fit values of $\beta_1$ and $\delta$ across all samples are given in the middle, and the range and median of estimates are given in the right column. }
\label{tab:simplefit}
\end{table*}

\section{Model Simulations}
\subsection{Fitting and Forecasting} Table \ref{tab:simplefit} shows the estimated values for $\beta_1$, $\delta$, and $\Ro$ for each data set. 
The remaining LH sampled (non-fitted) parameters that resulted in the overall best fit in Table \ref{tab:simplefit} are given in the Supplementary Information. Since $\gamma_R$ does not affect the model goodness-of-fit (since recovered individuals are assumed not to transmit the virus), we took $\gamma_R$ to be the midpoint of the range in Table \ref{tab:params} for the best-fit simulations.

\begin{figure}
\centering
\rule{2cm}{0.4pt} . Guinea . \rule{2cm}{0.4pt}\\
\includegraphics[width=0.32\textwidth]{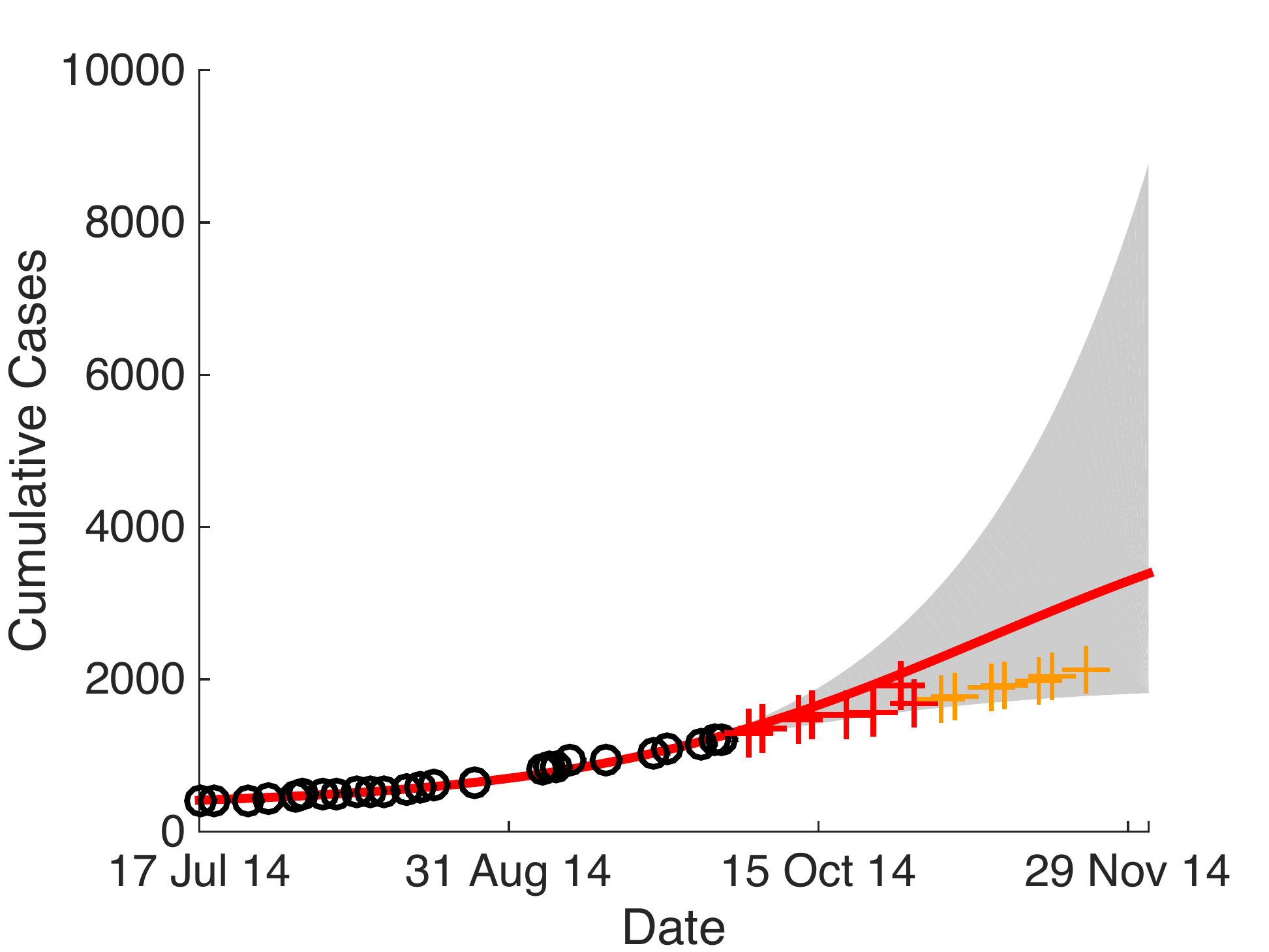}
\includegraphics[width=0.32\textwidth]{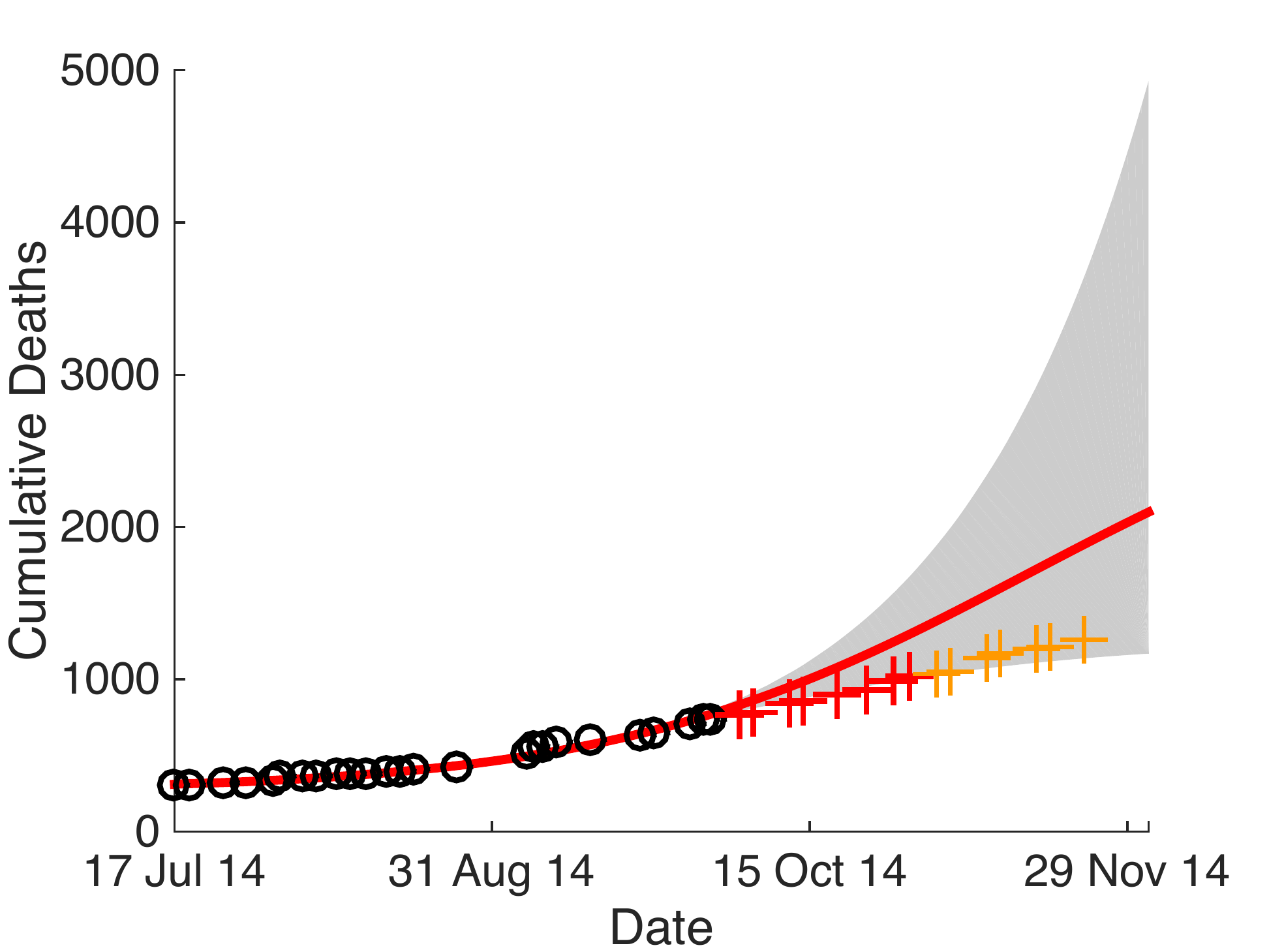}
\includegraphics[width=0.32\textwidth]{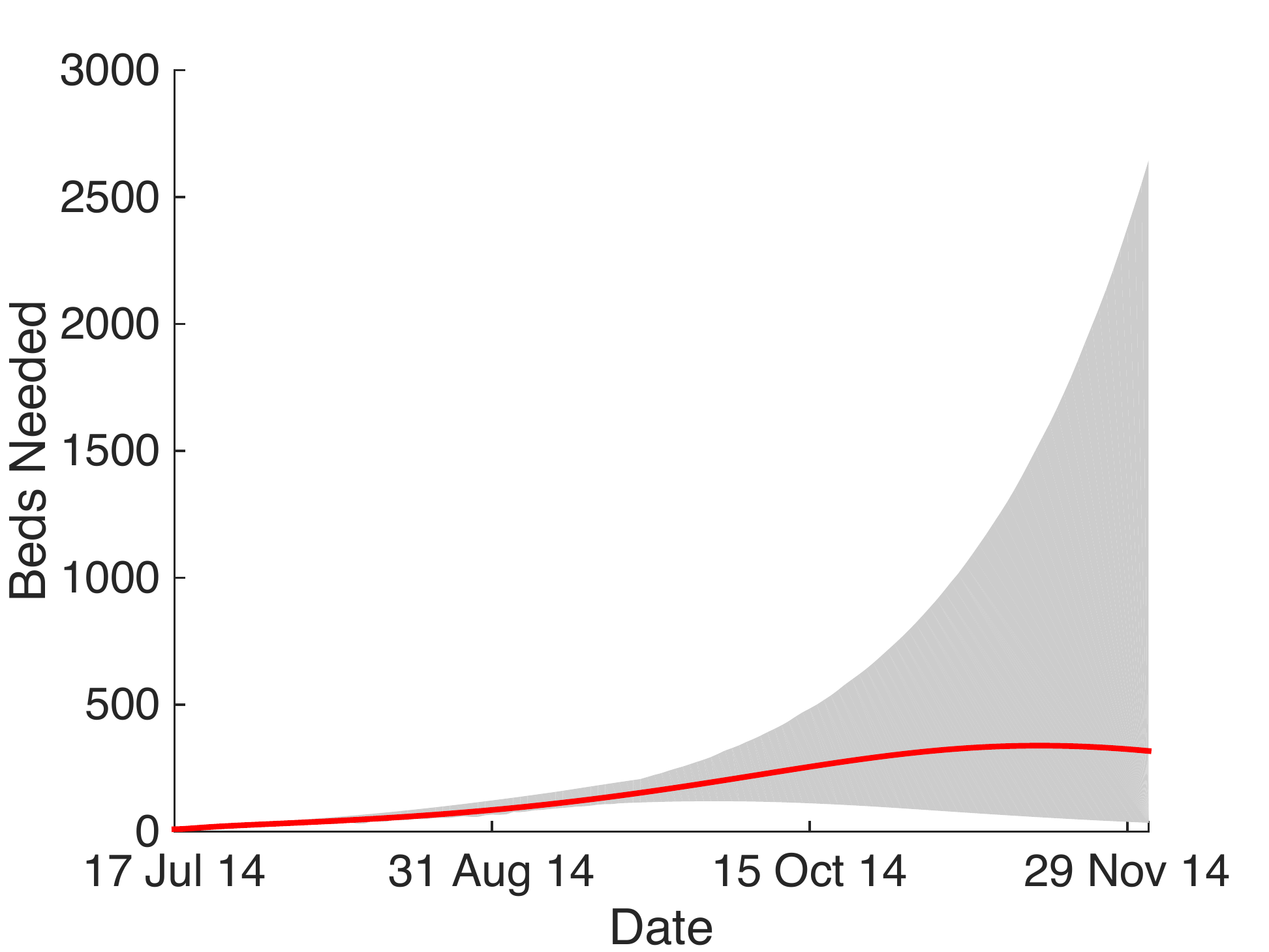}
\\
\vspace{0.25cm}
\rule{2cm}{0.4pt} . Liberia . \rule{2cm}{0.4pt}\\
\includegraphics[width=0.32\textwidth]{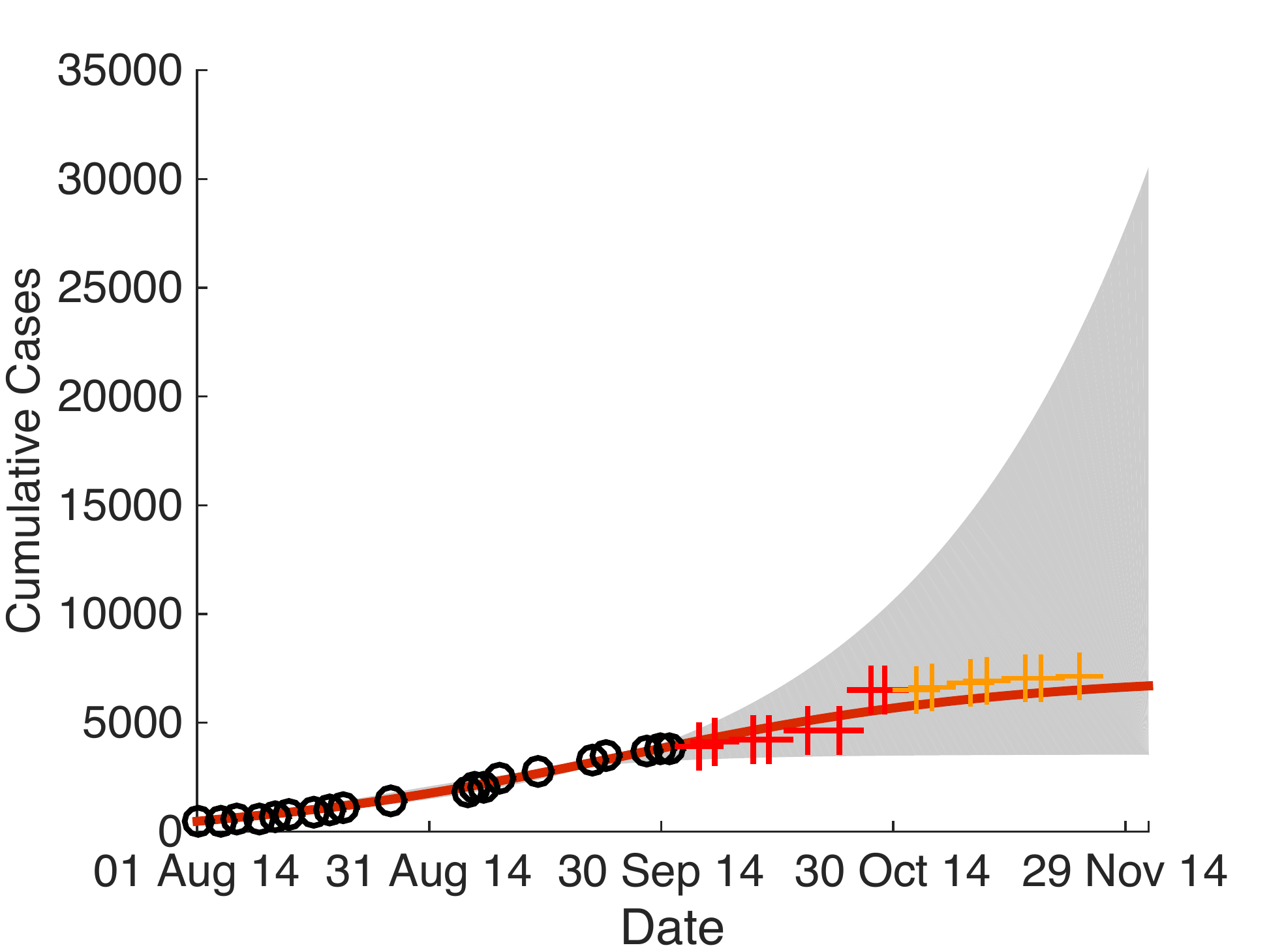}
\includegraphics[width=0.32\textwidth]{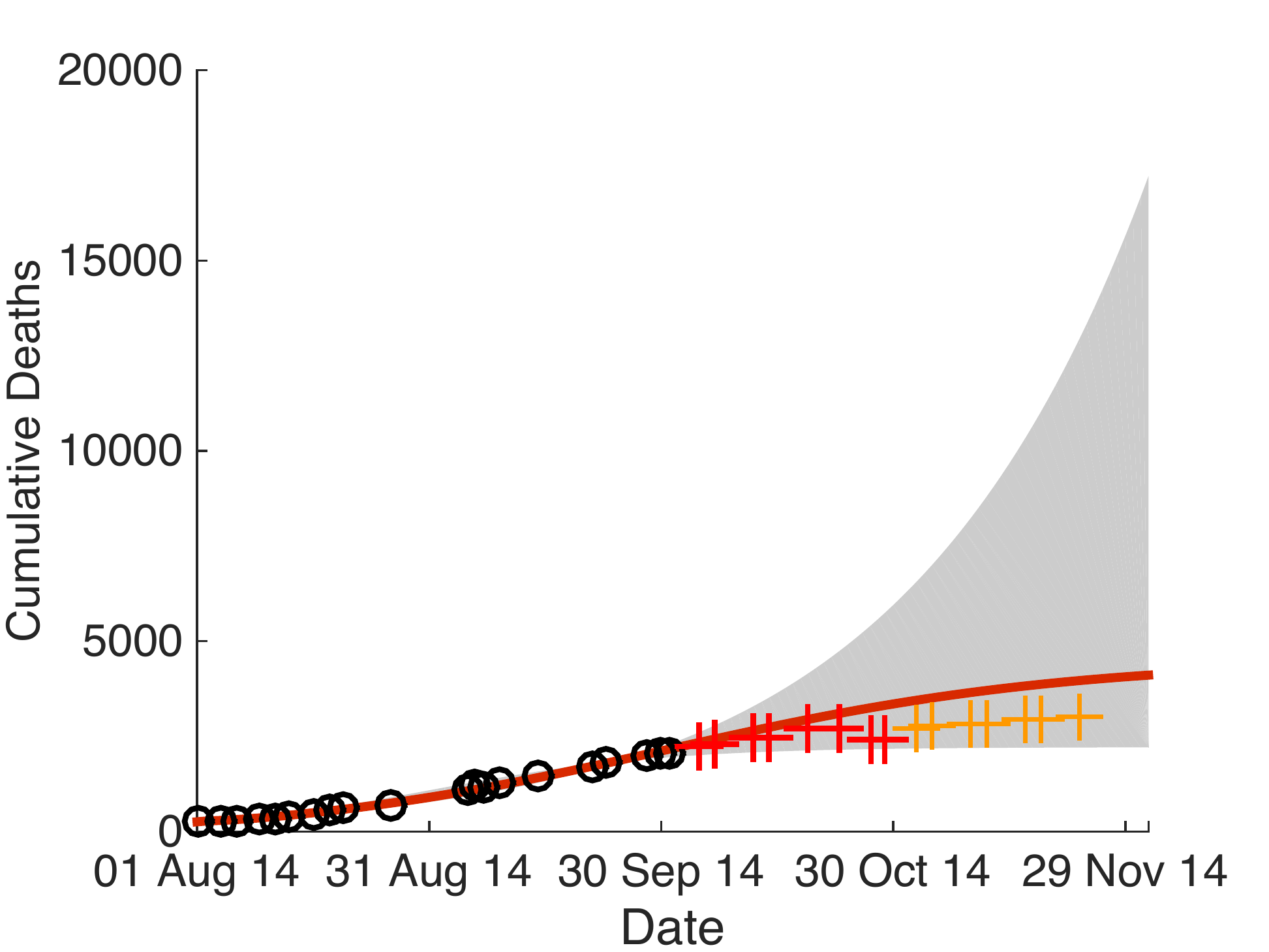}
\includegraphics[width=0.32\textwidth]{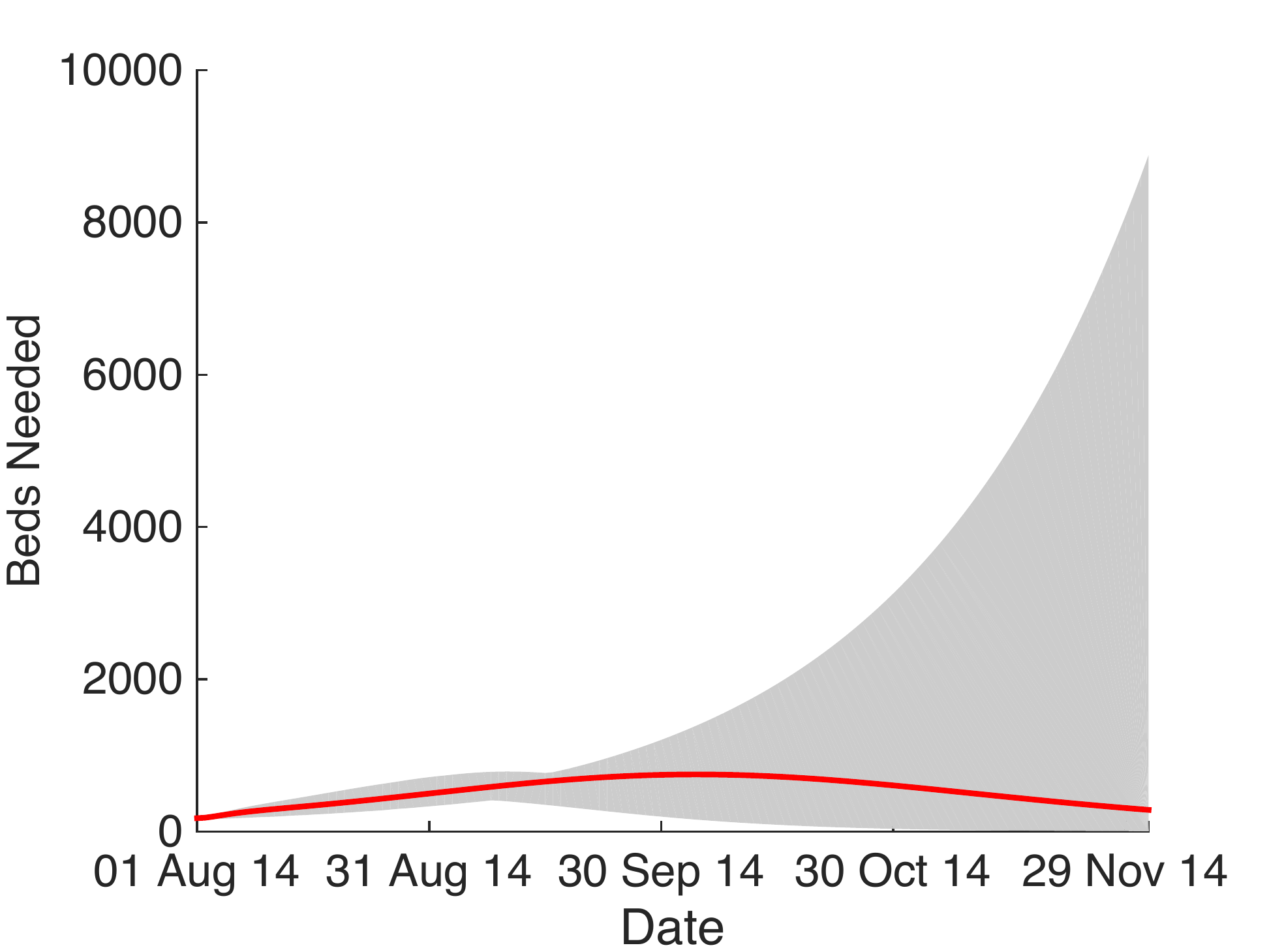}
\\
\vspace{0.25cm}
\rule{2cm}{0.4pt} . Sierra Leone . \rule{2cm}{0.4pt}\\
\includegraphics[width=0.32\textwidth]{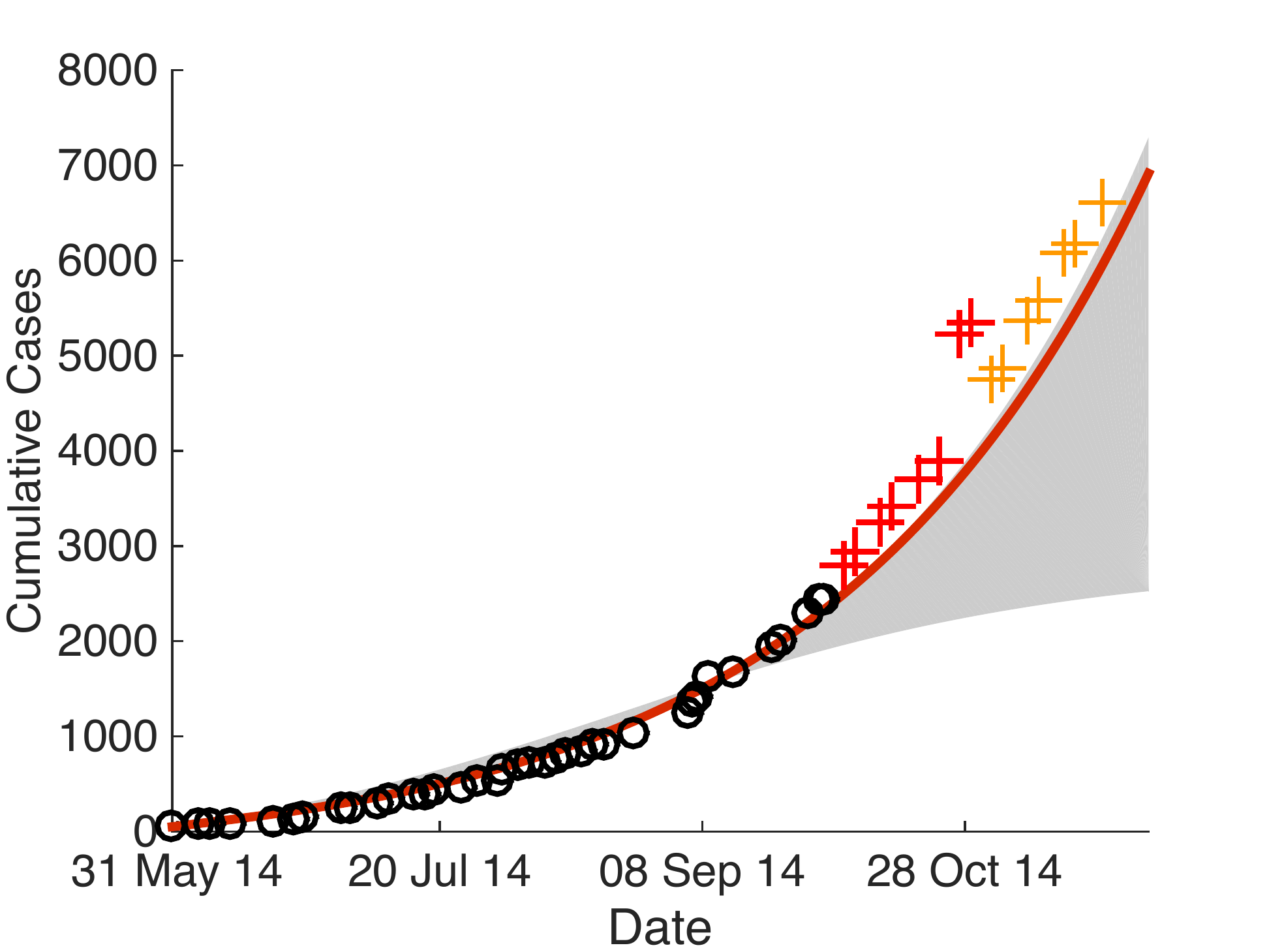}
\includegraphics[width=0.32\textwidth]{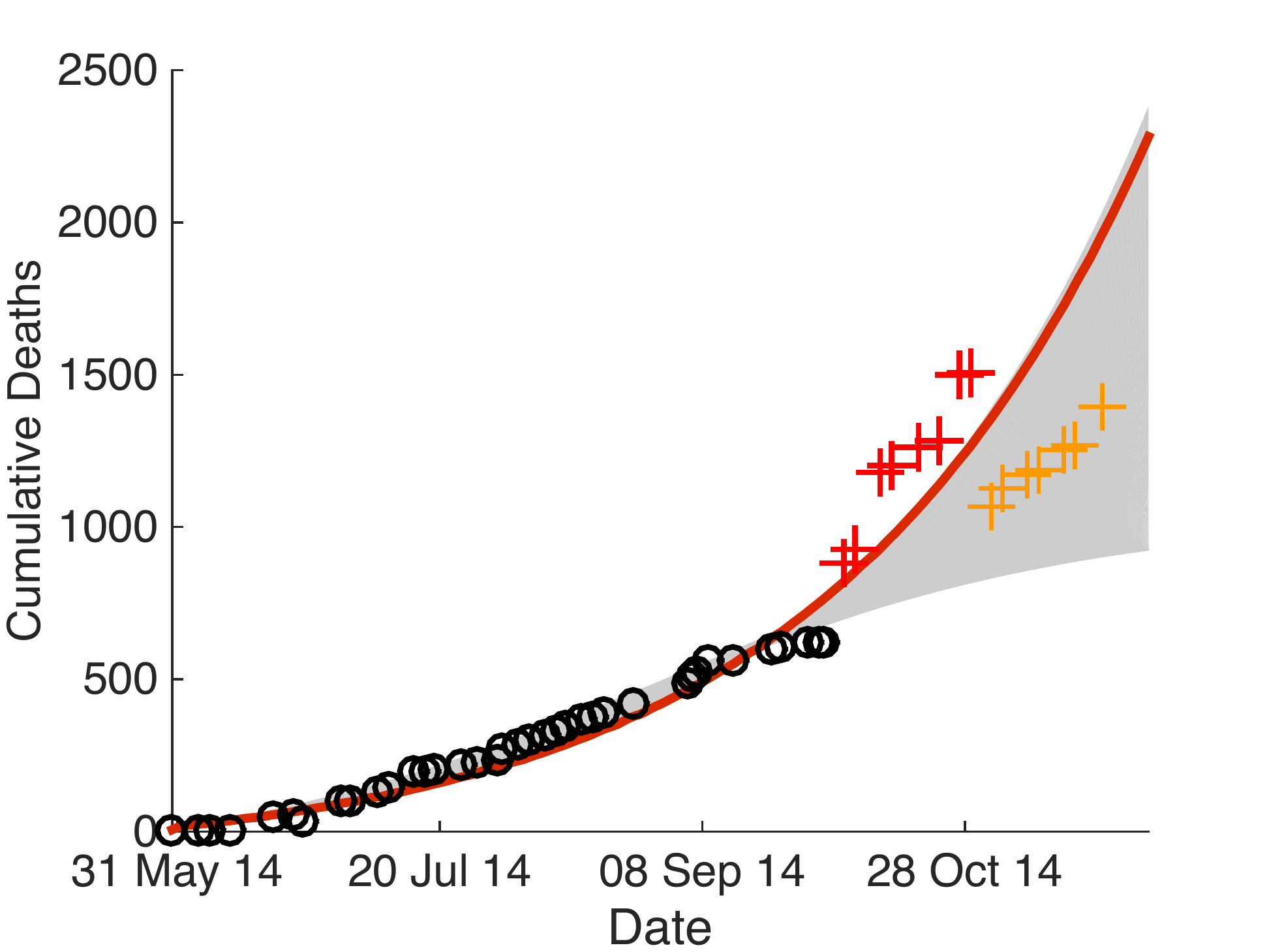}
\includegraphics[width=0.32\textwidth]{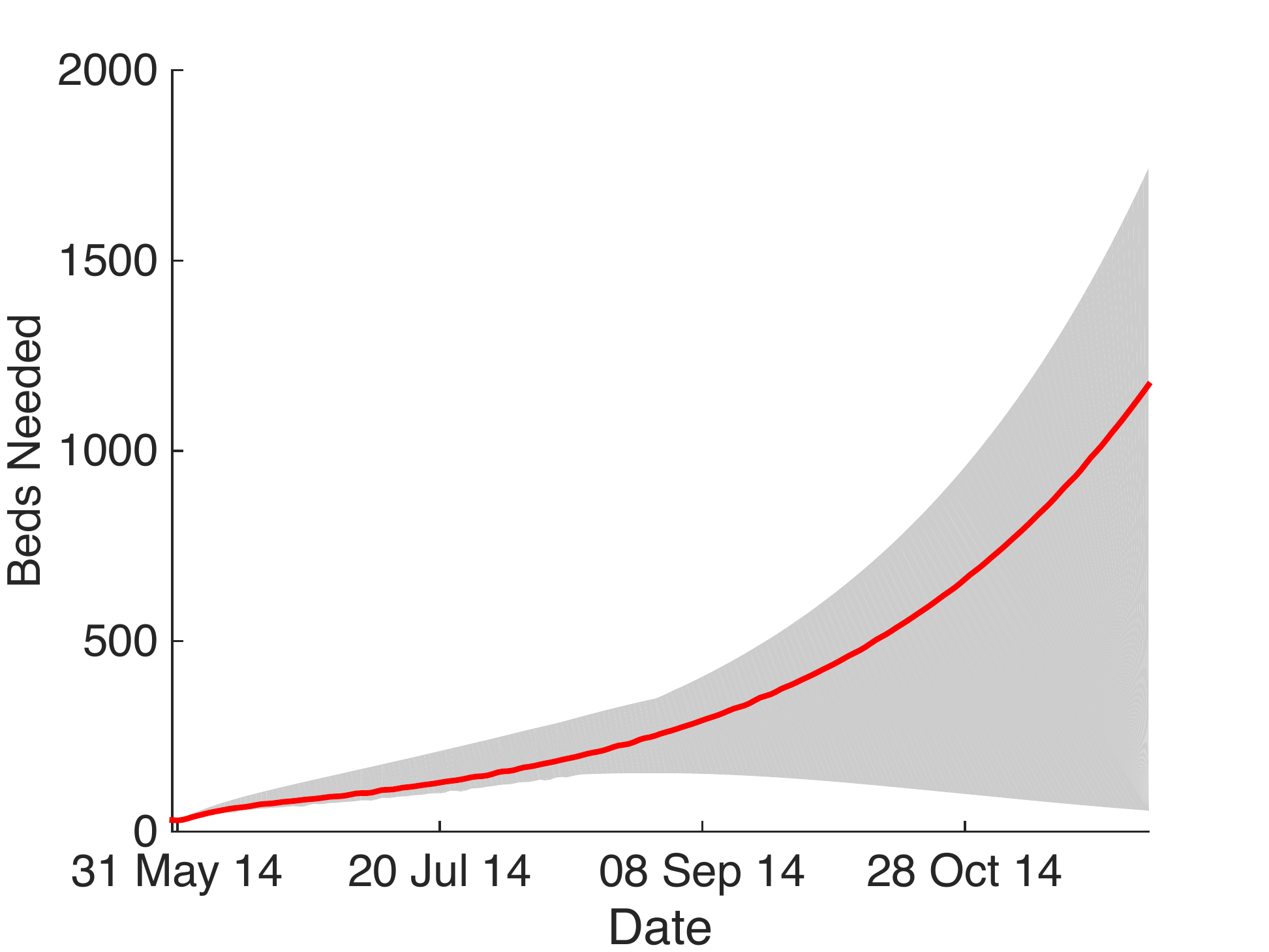}
\\
\vspace{0.25cm}
\rule{2cm}{0.4pt} . All Countries Combined . \rule{2cm}{0.4pt}\\
\includegraphics[width=0.32\textwidth]{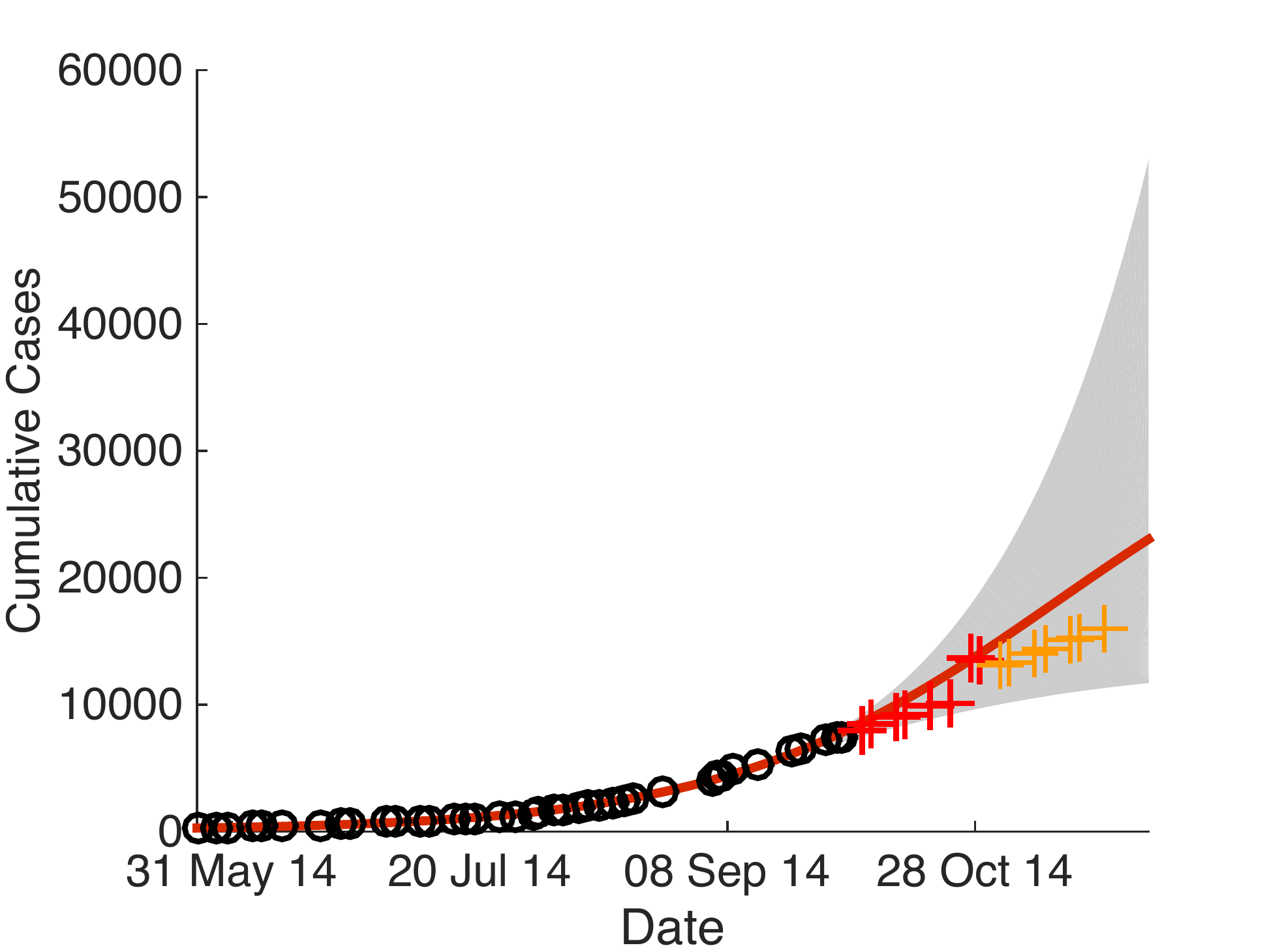}
\includegraphics[width=0.32\textwidth]{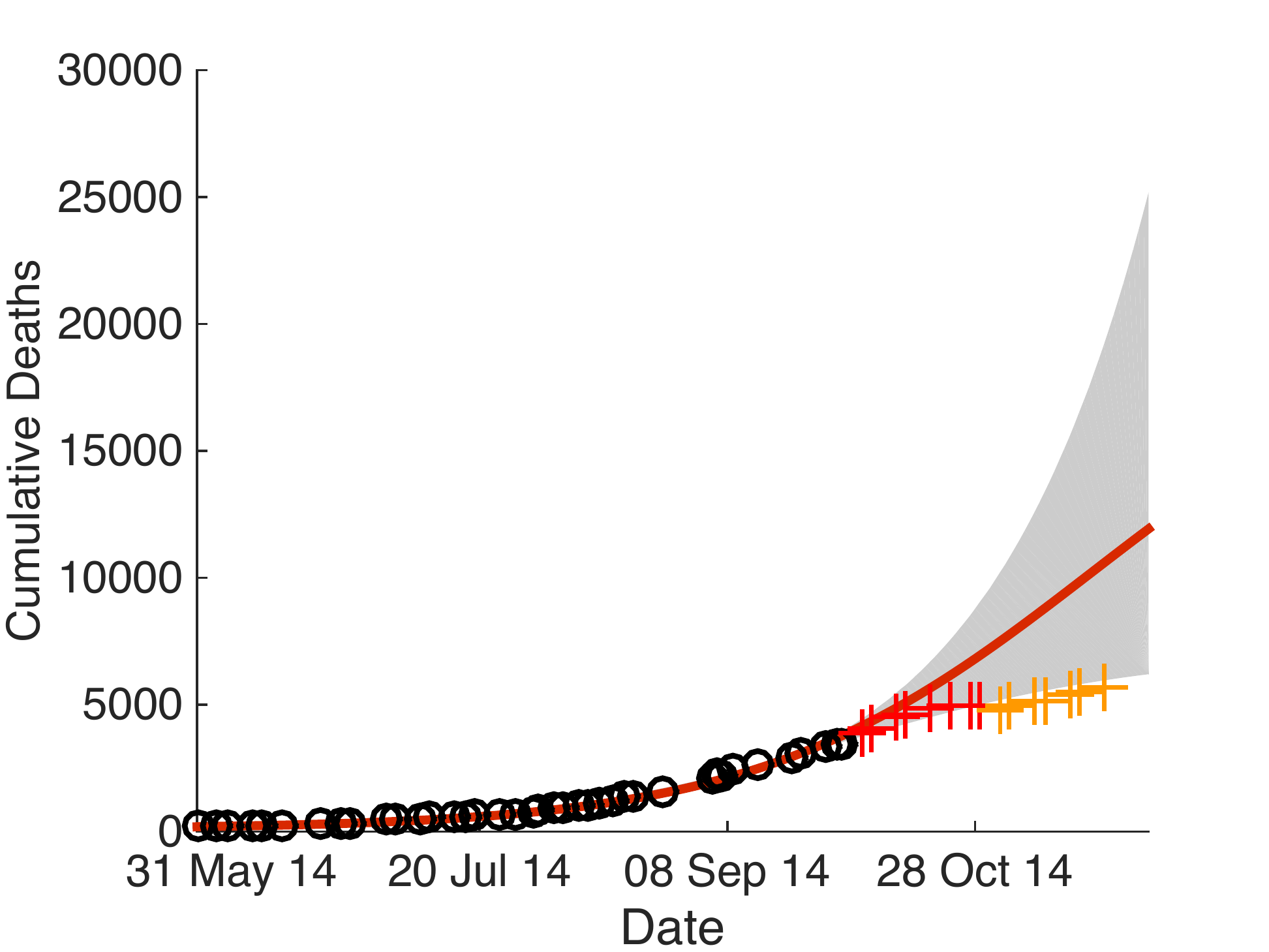}
\includegraphics[width=0.32\textwidth]{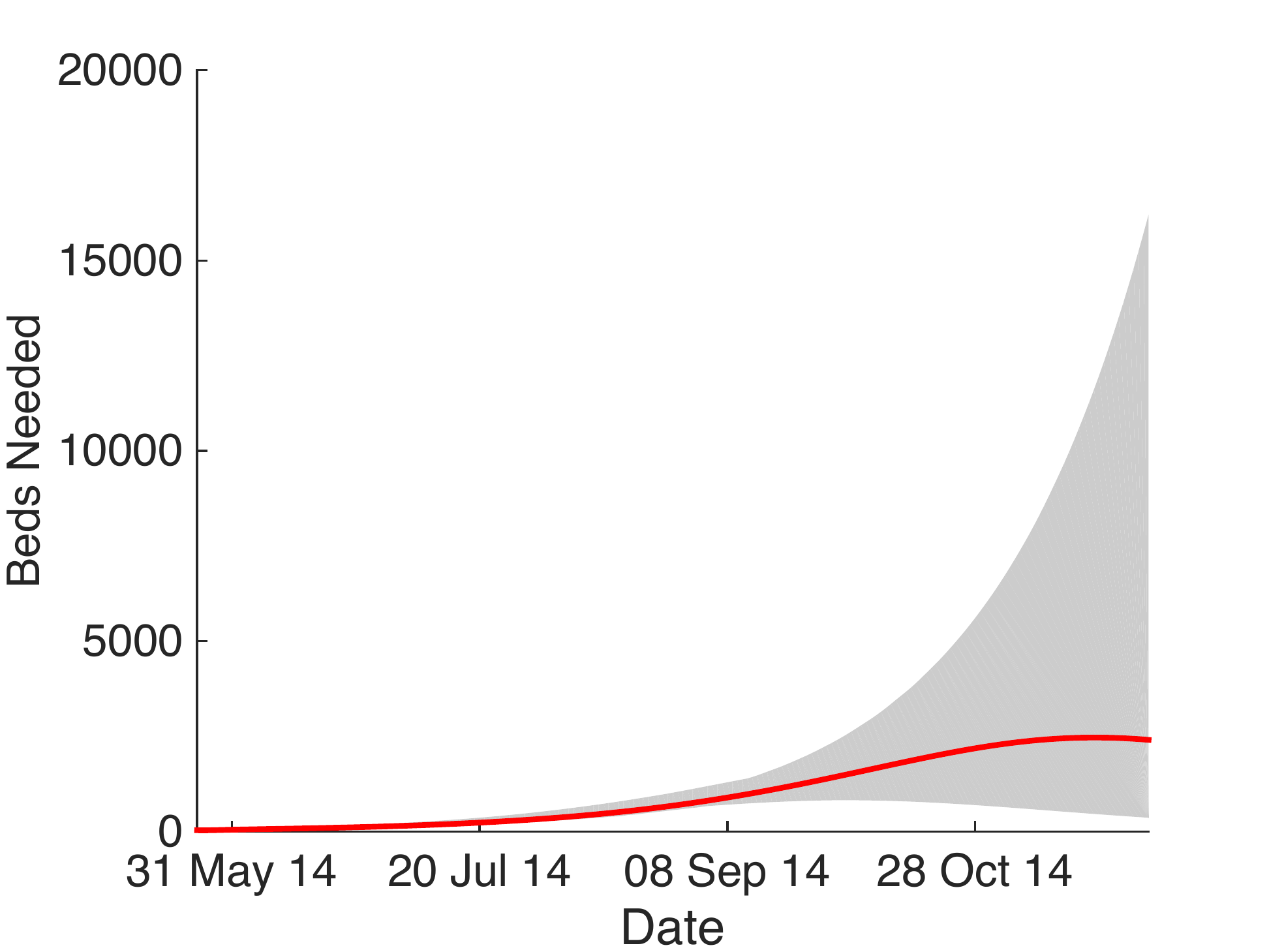}
\caption{Model fit to total cumulative cases and deaths up through October 1, 2014. Grey region shows the full range of best-fit trajectories for the parameter ranges in Table \ref{tab:params}. The model dynamics using overall best fit parameters across all LH-sampled values is shown in red. Data used for fitting is shown as circles, subsequent data not used for fitting shown as red +'s for October and orange +'s for November.}
\label{fig:simplefit}
\end{figure}

Figure \ref{fig:simplefit} shows the fits to data with projections of the numbers of cases, deaths, and beds needed through December 1, 2014. The number of beds needed was calculated as the total infected and recently recovered, $kN(I_1 + I_2 + R)$ (though we note that $k$ includes the reporting rate, so that the true number of beds needed may be larger). The model predictions showed a wide range of potential trajectories (grey shaded regions), with the overall best fit trajectory shown as a solid line.
For comparison purposes, we also show the reported data for cases and deaths from October (red +'s), and from November (orange +'s), although we note that our model projections did not explicitly account for changes in reporting rate or effects of interventions, both of which are known to have been significant during October and November \cite{WHO_EbolaSitReps, UNMEER}. In particular, there are several sudden increases and decreases in the cumulative case data that are due to changes in reporting \cite{WHO_EbolaSitReps}, which the model does not capture as we use a constant reporting rate (in particular, the model cannot capture decreases in cumulative cases). Nonetheless, the forecast from the LH sampled best-fit trajectory (red line) was generally close to the reported data, and captured the correct qualitative behavior in all cases (i.e. continued exponential growth vs. slowing down of the epidemic). The forecasts matched the first month of data (red +'s) quite closely, with wider deviations in comparing with the second month of data (orange +'s). In particular, the model was able to correctly forecast the downturn in Liberia but continued exponential growth in Sierra Leone.
\\

\noindent\textit{Parameter Uncertainty and Transmission by Stage.} We note that the full ranges of LH-sampled parameter values in Table \ref{tab:params} yielded similarly good fits, showing the expected practical unidentifiability of the parameters when fitting to largely exponential growth phase epidemic data. Supplementary Figure \ref{fig:GofHist} shows histograms of the residual sum of squares values for all fits. The trajectories with the lowest squared residuals did tend to cluster near the best fit line in Figure \ref{fig:simplefit}, as shown in Supplementary Figure \ref{fig:BestTraj} (which shows the best 10\% of all LH-sampled fits). Among these best 10\% of fits, most of the LH-sampled parameters show broad ranges covering the full bounds given in Table \ref{tab:params}, illustrating that the parameter uncertainty/unidentifiability persists when only the best goodness of fit scores are considered (Supp. Fig. \ref{fig:BestHist}). 

\begin{figure}
\centering
\includegraphics[width=0.42\textwidth]{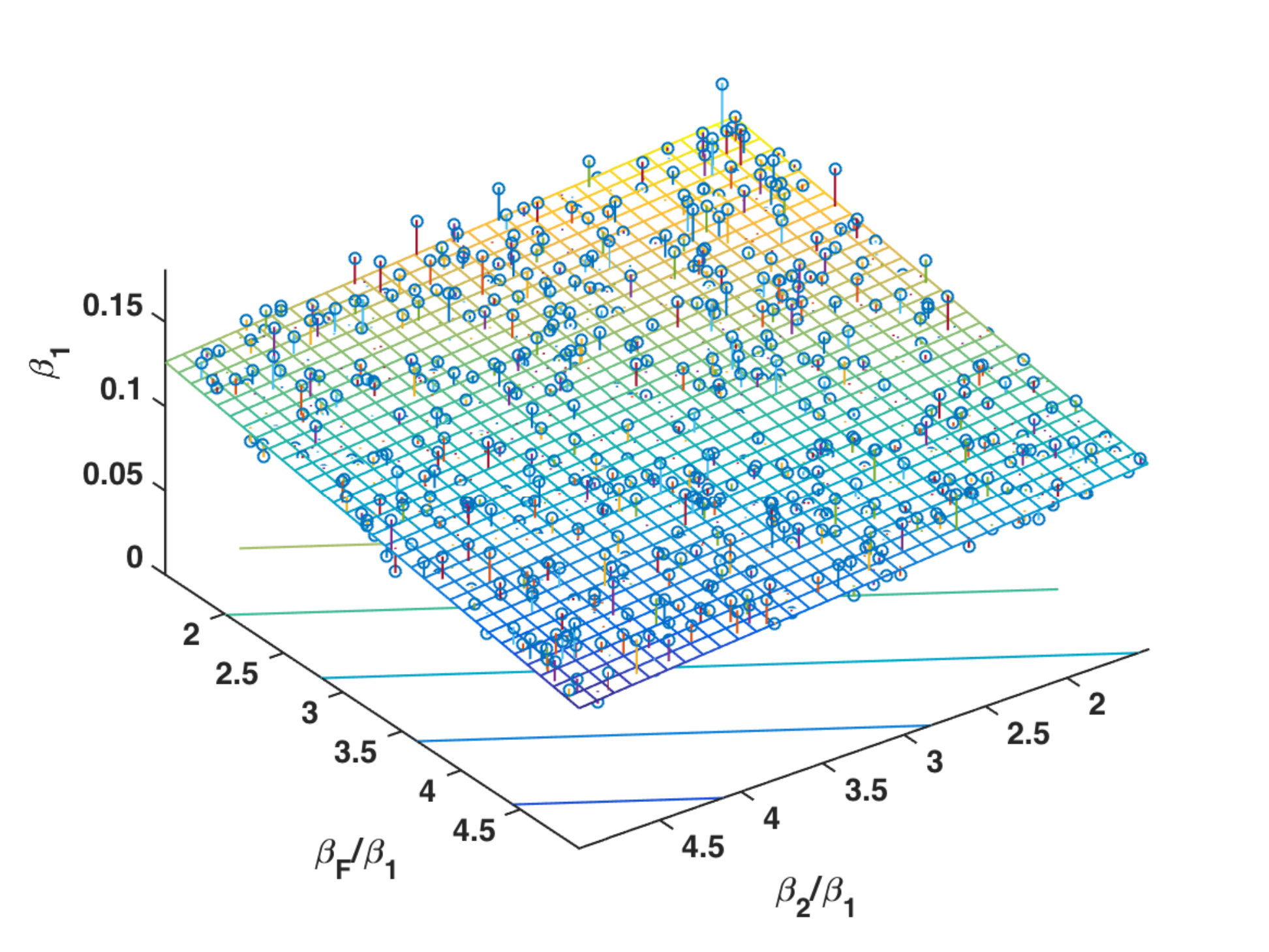}
\includegraphics[width=0.42\textwidth]{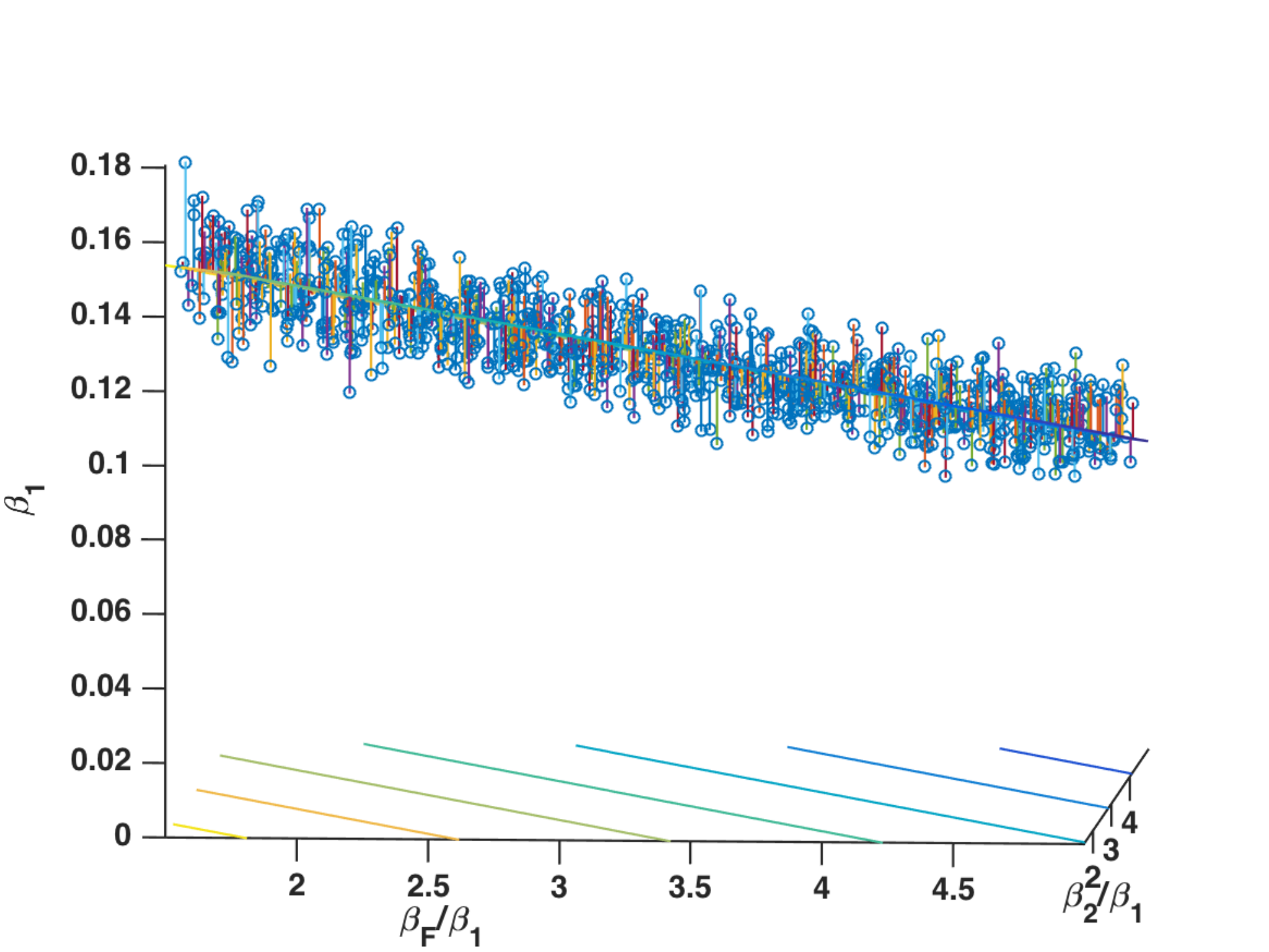}\\
\includegraphics[width=0.4\textwidth]{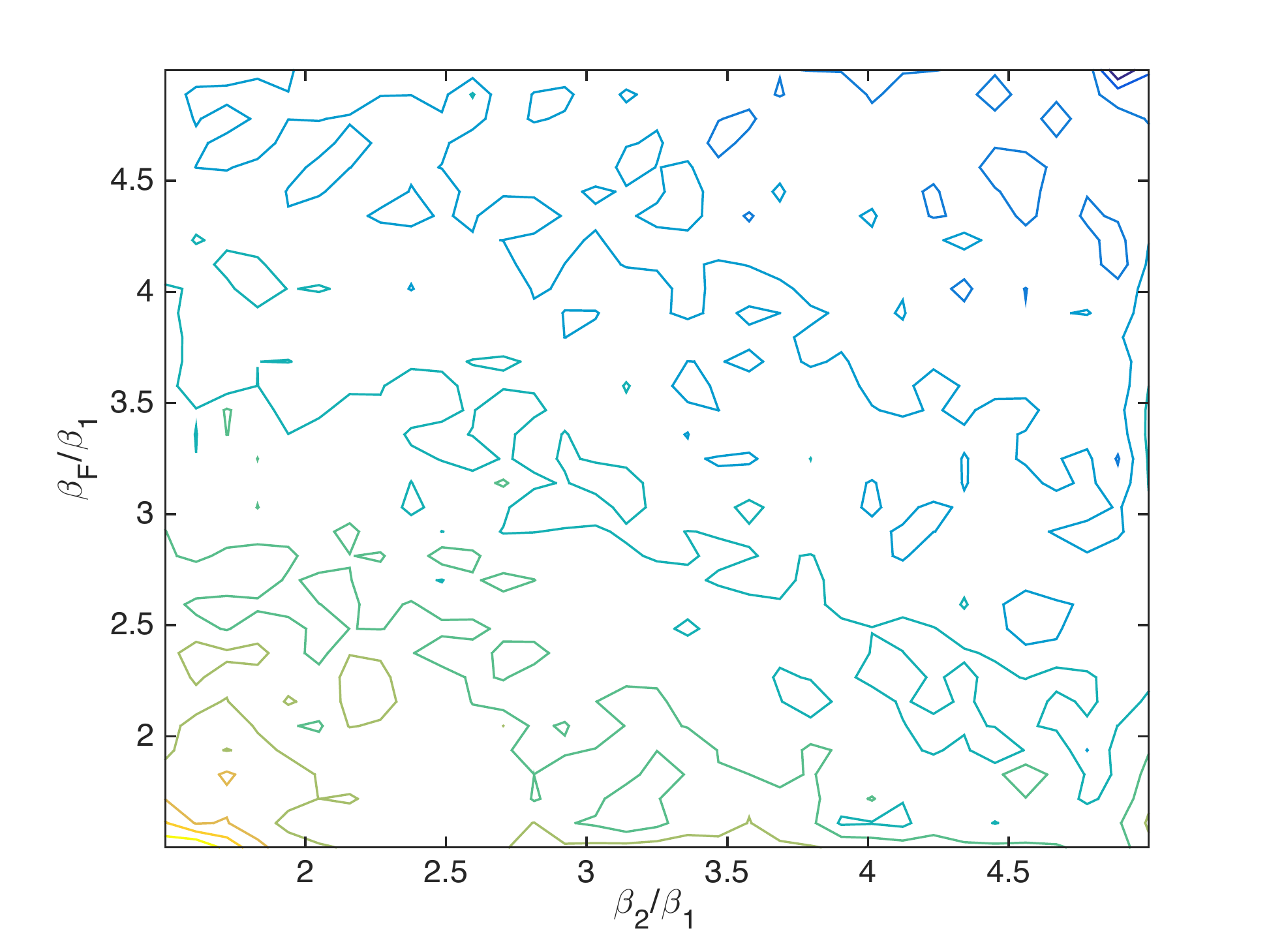}
\includegraphics[width=0.4\textwidth]{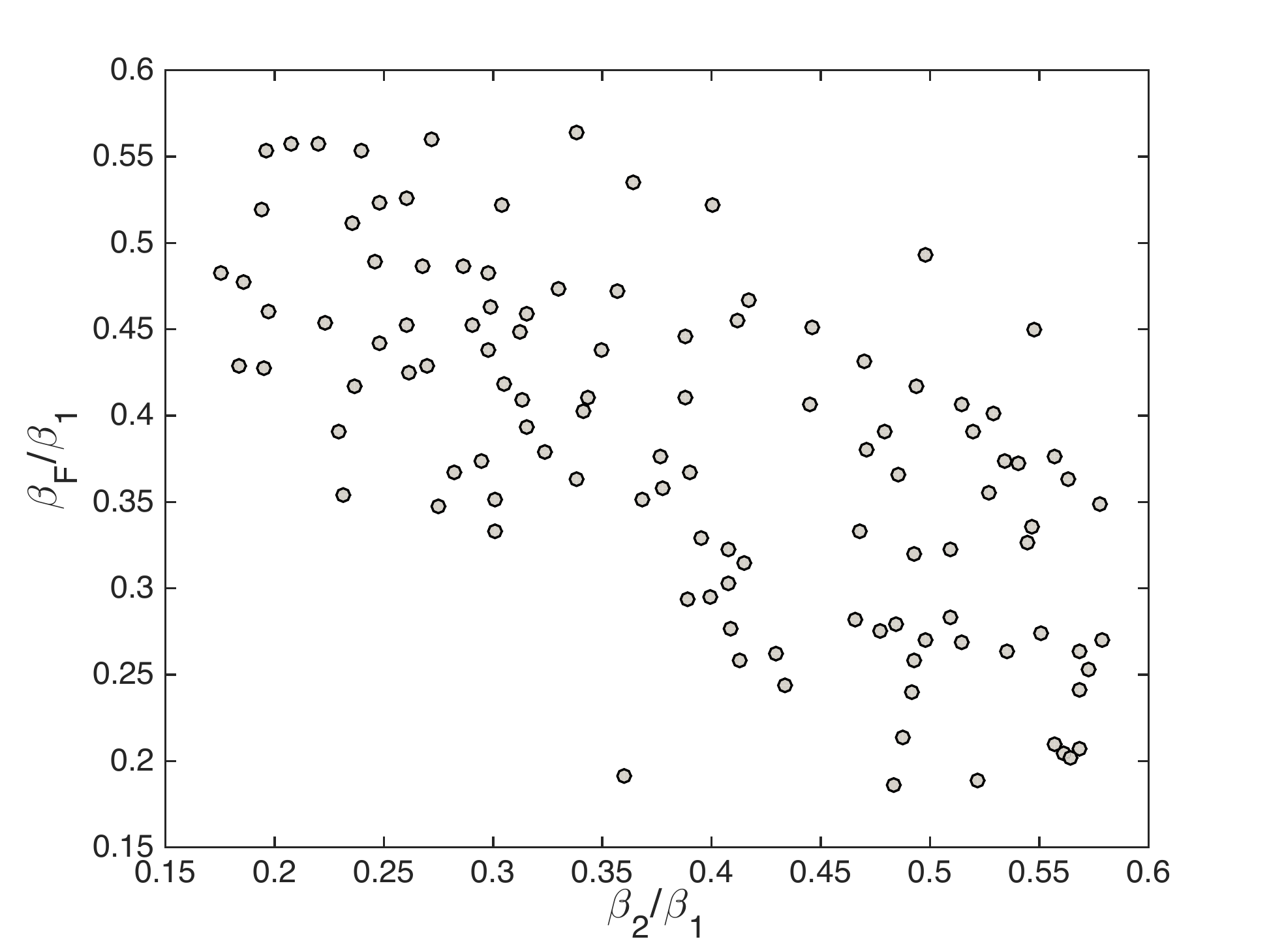}
\caption{Top row: front (left) and side (right) views of the estimates of $\beta_1$, $\beta_2/\beta_1$, and $\beta_F/\beta_1$ for all countries combined (Fig. \ref{fig:simplefit}), with the best-fit plane shown as a mesh. The three parameters have an approximately linear relationship with one another, indicating that the parameters may be unidentifiable individually, but identifiable in combination. Bottom row: left panel shows contour lines of $\beta_2/\beta_1$ and $\beta_F/\beta_1$ for fixed values of $\beta_1$, and right panel shows a scatterplot of $\beta_2/\beta_1$ and $\beta_F/\beta_1$ for a roughly fixed value of $\beta_1$ between 0.115 and 0.12. Both panels highlight the approximately linear compensation between $\beta_2$ and $\beta_F$, such that similarly good fits can be obtained for a wide range of values, so long as roughly the same total transmission level is maintained.
}
\label{fig:Beta12FCombo}
\end{figure}

In particular, a wide range of values for the transmission parameters by stage were able to generate very similar fits and forecasts. This suggests these parameters may form a practically identifiable combination \cite{eisenberg2014determining, landaw1984algorithm}, wherein decreases in transmission levels in one stage can be compensated for by increasing transmission in another stage, so that the individual transmission parameters cannot be estimated (are unidentifiable), but the overall transmission level is estimable. This compensation manifests in parameter space as an approximate plane of best fit parameter estimates, shown in Figure \ref{fig:Beta12FCombo}. 

The parameter uncertainty also results in large uncertainty in the contributions of each stage to $\Ro$. The model $\Ro$ breaks up naturally into terms for each of the three stages (with each term in Eq. \eqref{eq:R0} corresponding to $I_1$, $I_2$, and $F$), representing the relative contributions of each transmission stage to $\Ro$. Each term can be interpreted as the average number of secondary cases generated by an infectious individual while in a particular stage of transmission. For both the full set of LH-sample fits and the best $10\%$ of fits, a similarly wide range of contributions by stage for $\Ro$ were found, as shown for all countries combined in Supplementary Figure \ref{fig:BestR0s} and Supplementary Table \ref{tab:stageR0}. Further discussion of the uncertainty in parameters can be found in the Supplementary Information.
\\

\noindent\textit{Reporting and Population at Risk.} The parameter $k$ was responsible for the much of the breadth in the forecasted trajectories for cumulative cases and deaths in Figure \ref{fig:simplefit}. In general, the forecasts from a single set of parameters could be adjusted to cover the full shaded range of forecasted cases and deaths just by changing $k$ and re-fitting $\beta_1$ and $\delta$, with lower values of $k$ corresponding to earlier deviation from exponential growth and epidemic slowdown (due to exhaustion of susceptibles as the effective population at risk is smaller), and the top edge of the grey shaded regions representing the model fit without including $k$ (i.e. with $k = 1$). The overall best fit trajectories shown in red in Figure \ref{fig:simplefit} coresponded to LHS sampled values for $k$ that were quite low for the all countries combined, Guinea, and Liberia cases ($k=0.0026,0.0006$, and $0.0024$ respectively), but not for Sierra Leone ($k = 0.68$), correctly forecasting epidemic slowdown in the first three cases, with continued exponential growth in Sierra Leone. 
As $k$ includes both the reporting rate and fraction of the population at risk, we hypothesize that $k$ provides a simplified way to account for interventions, existing immunity, and changes in behavior, which can effectively act to shrink the population at risk.

\subsection{Expanded Fitting and Forecasting Simulations} To evaluate the importance of including $k$ when forecasting, we generated multiple fits and forecasts with $k$ either included or not included in the model. For these fits, for simplicity we fixed all parameters to the midpoints of the ranges in Table \ref{tab:params}, and then fit $\beta_1$, $\delta$, and either fit $k$ or fixed $k = 1$ (equivalent to not including $k$ in the model). We fit the models to data sets up through the first of the month for July, August, September, October, November, and December, and then forecast the subsequent two months (so the forecasts were until Sept. 1, 2014 - Feb. 1, 2015).

\begin{figure}
\centering
\rule{2cm}{0.4pt} . Including $k$ . \rule{2cm}{0.4pt}\\
\includegraphics[width=0.4\textwidth]{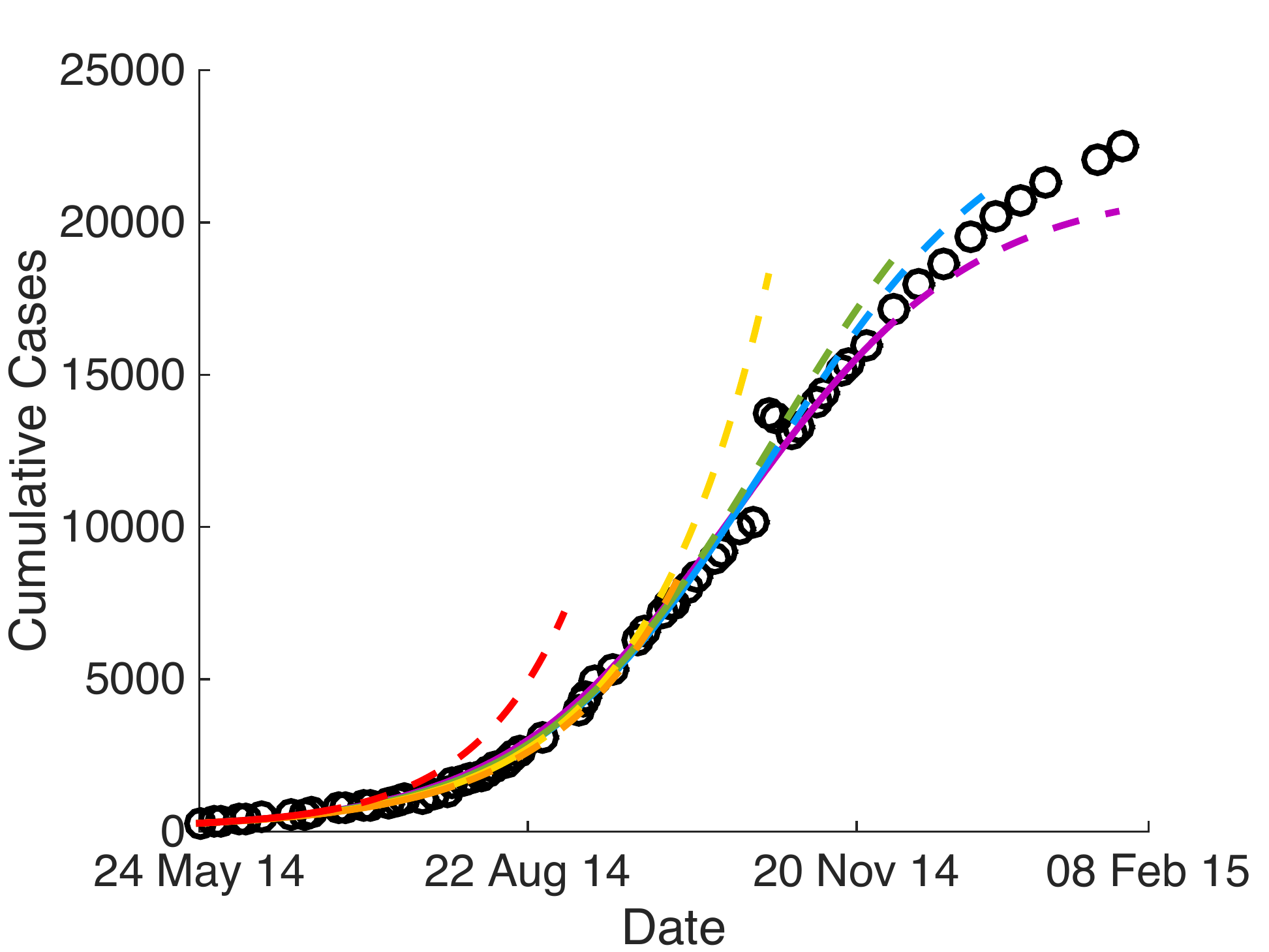}
\includegraphics[width=0.4\textwidth]{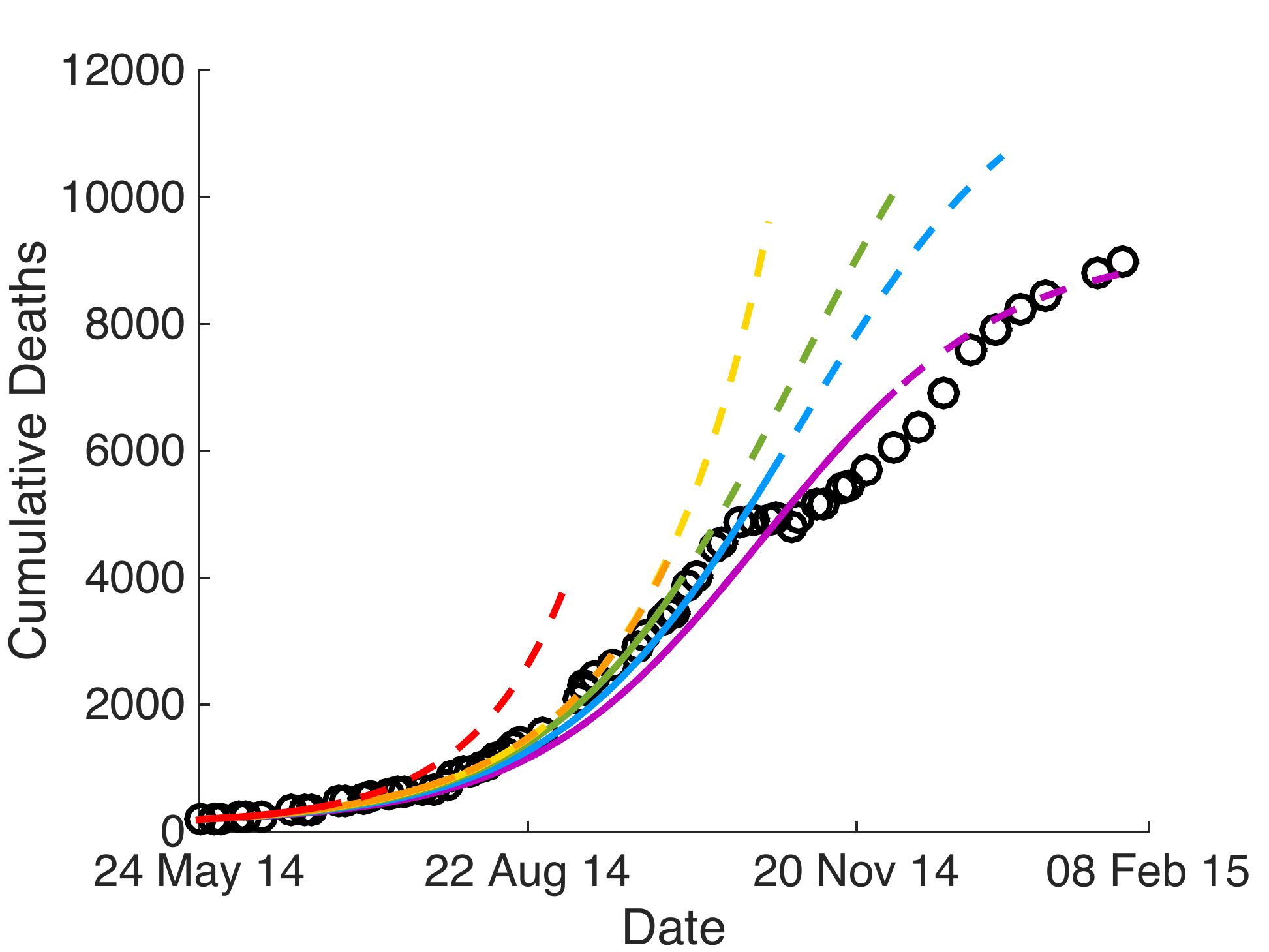}\\
\vspace{0.5cm}
\rule{2cm}{0.4pt} . Without $k$ . \rule{2cm}{0.4pt}\\
\includegraphics[width=0.4\textwidth]{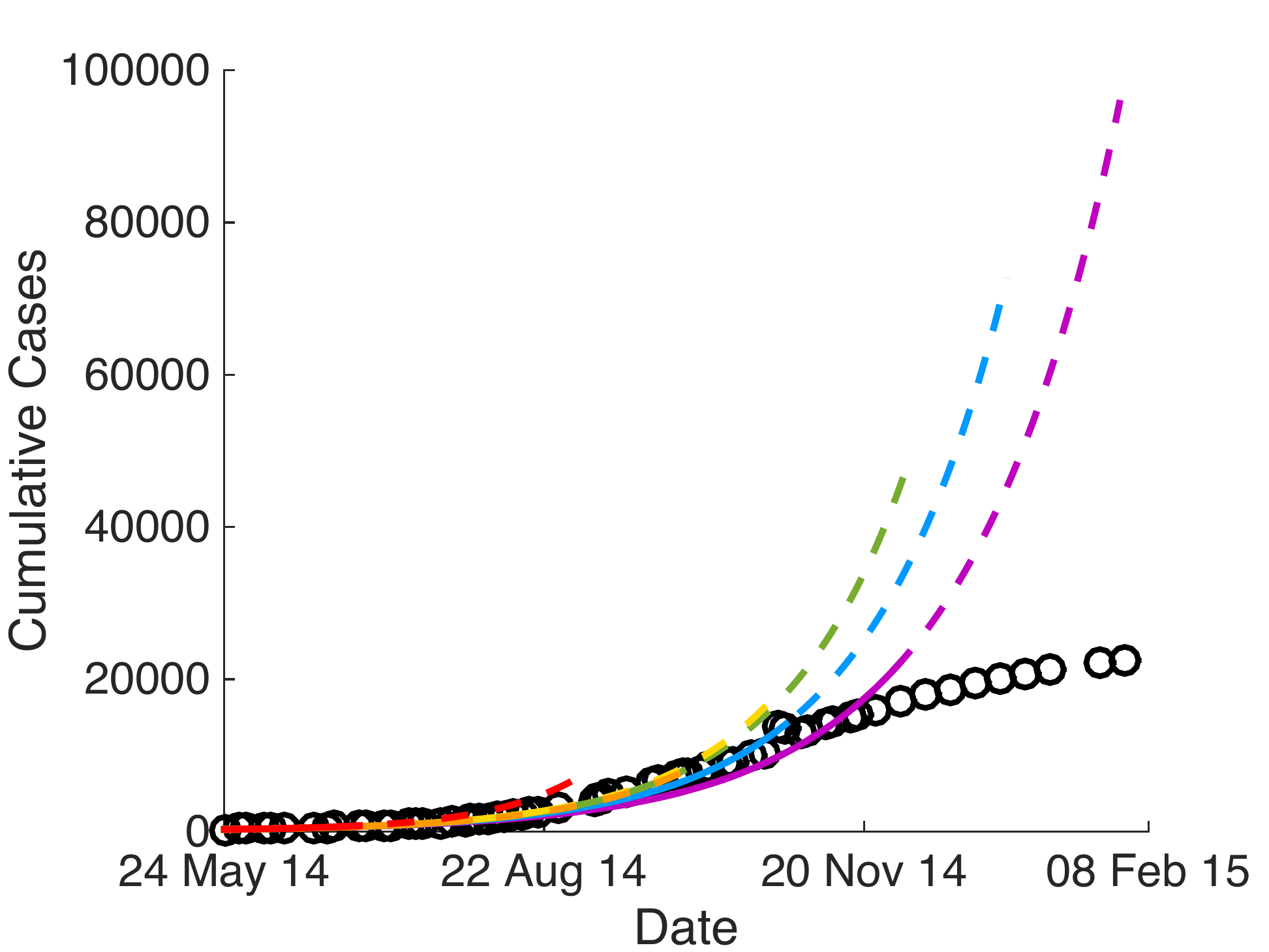}
\includegraphics[width=0.4\textwidth]{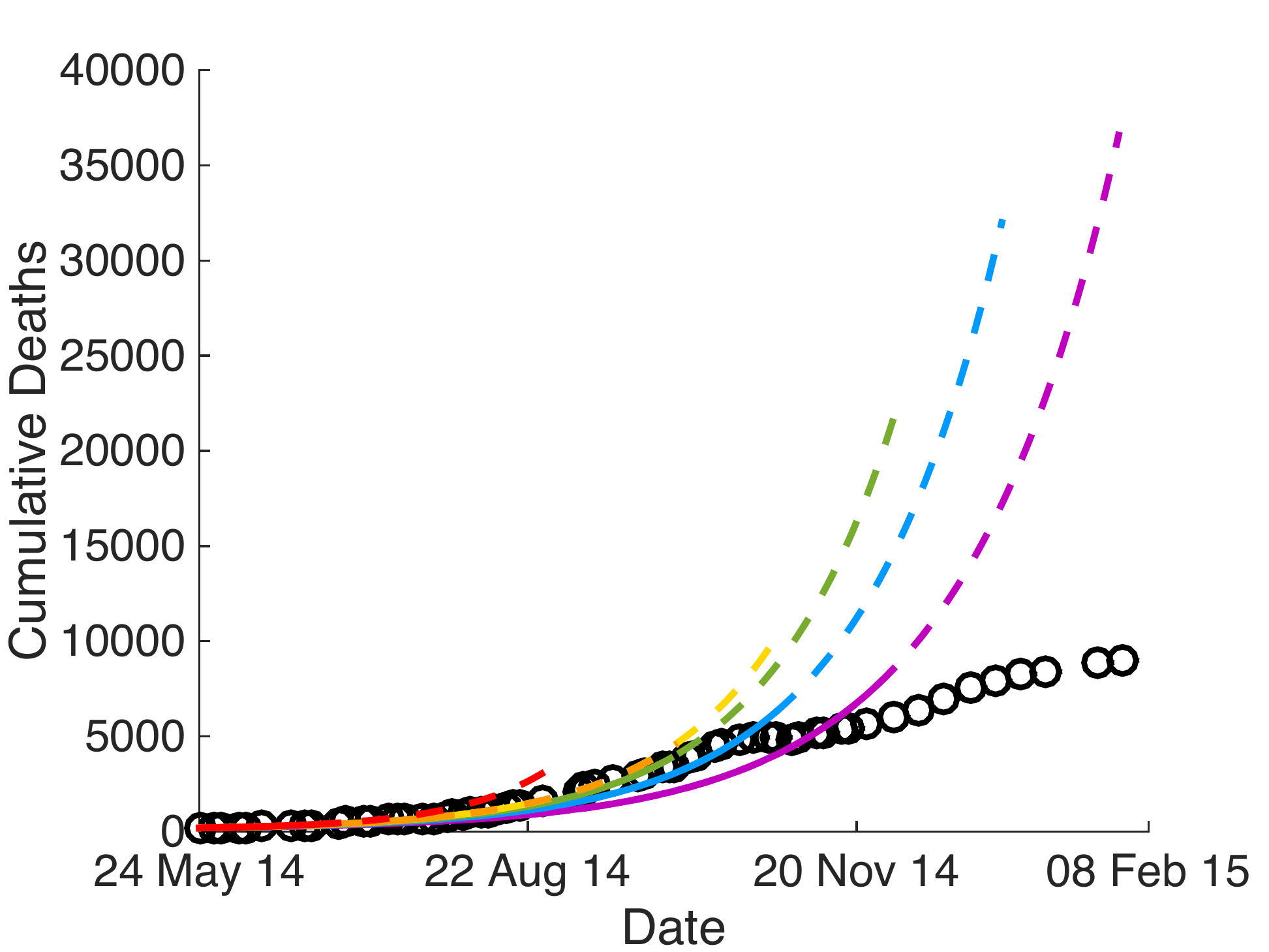}
\caption{Multiple model fits and forecasts generated either with $k$ included (top row) or not included (equivalent to fixing $k = 1$, shown bottom row). 
The model fits (solid lines) use the data up through July 1 (red), August 1 (orange), September 1 (yellow), October 1 (green), November 1 (blue) and December 1 (purple), with subsequent two months of forecasts shown as dashed lines in the same color. Both models with and without $k$ forecast similarly for the early data (red, orange, and yellow forecasts), after which point the model with $k$ performs better (green, blue, purple forecasts). Note the different y-axis scales in the top row vs. the bottom row.}
\label{fig:MultiForecastAll}
\end{figure}

As shown in Figure \ref{fig:MultiForecastAll} for all countries combined, the models with and without $k$ performed quite similarly until the data from September 2014 is included in the fit, at which point the model with $k$ outperformed the model without $k$ (i.e. where $k$ is fixed at $1$). In particular, the model with $k$ was able to predict and capture the downturn in cases and deaths, while the model without $k$ consistently over-forecasts the number of cases and deaths. The resulting parameter estimates (Table \ref{tab:MultiForecastAll}) show relatively consistent estimates for $\beta_1$ and $\delta$ across the length of data and value or inclusion of $k$ in the model. When included in the model, the fitted value of $k$ was larger during the period of exponential growth and then dropped by two orders of magnitude once the data began to show indications of epidemic slowdown.
Supplementary Figures \ref{fig:MultiForecast_k} and \ref{fig:MultiForecast_nok} and Supplementary Tables \ref{tab:MultiForecast_k} and \ref{tab:MultiForecast_nok} show similar results for each of the individual countries as well. While the model with $k$ sometimes over- or under-forecasted, it did not suffer from the consistent over-forecasting generated seen when $k$ was not included. We also tested a fixed value of $k = 1/2.5$ to account only for underreporting as estimated in \cite{Meltzer2014}, which we found yielded very similar results to the model without $k$ included ($k=1$), as shown in Supp. Figure \ref{fig:MultiForecast_k0pt4} and Supp Table \ref{fig:MultiForecast_k0pt4}. These results appeared to depend primarily on the inclusion of $k$ rather than the particular model structure used here, as preliminary tests using a simplified form of the model with only a single infectious stage (i.e. the SEIRD model \cite{weitz2015modeling}) generated similar forecasts when $k$ was included vs. not included (not shown).

It is notable that the data for Liberia, Guinea, and the all countries combined cases showed earlier epidemic slowdown in 2014, whereas the Sierra Leone data showed continuing exponential growth through all data used in fitting (through December 2014). The forecasts and parameter estimates for the model including $k$ reflected this pattern, with estimates for $k$ for Guinea, Liberia, and all countries combined dropping by two to three orders of magnitude once the data through October 1, 2014 was included (Tables \ref{tab:MultiForecastAll} and \ref{tab:MultiForecast_k}), corresponding to slowed forecasts as well. By contrast, the estimates for Sierra Leone stayed larger ($k$ ranged from 0.34 to 0.73) for all fits, corresponding to continuing projections of exponential growth. Because of these larger estimates due to continuing exponential growth, the forecasts for Sierra Leone were similar to those using fixed values of $k$ (Figures \ref{tab:MultiForecast_k}, \ref{fig:MultiForecast_nok}, and \ref{fig:MultiForecast_k0pt4}).

\section{Discussion}

Forecasting under conditions presented in the current Ebola outbreak will always be challenged by uncertainty in the data.  As much as we would like to insist on better quality data, the immediacy of public health demands will always be of primary importance for the health care workers on the ground.  Analytical approaches need to be able to address uncertainty and unidentifiability while still providing useful information to public health officials working to mitigate the outbreak.  In the case of the Ebola outbreak, uncertainties centered around reporting and lack of knowledge of numbers at risk necessitated an approach to account for these unknown factors.  In spite of large uncertainty and an absence of detailed data, model forecasts can still provide useful predictions of the behavior of the epidemic by using a simplified correction term.
\\

\begin{table*}
\centering
\def\arraystretch{1.4}
\begin{tabular}{c  c  c  c  c   c   c  c   c }
\hline
&& \multicolumn{3}{c}{Estimates with $k$} && \multicolumn{2}{c}{Without $k$}&\\
\cline{3-5} \cline{7-8}
Using data through: && $\beta_1$ & $\delta$ & $k$ && $\beta_1$ & $\delta$&\\
\hline
July 1, 2014 && 0.14  &  0.69  &  0.26 &&     0.14 &  0.69  &\\
Aug 1, 2014 && 0.12  &  0.67  &  0.31 &&     0.12  &  0.67  &\\
Sep 1, 2014 && 0.13 &  0.64  &  0.38 &&     0.13  &  0.64  &\\
Oct 1, 2014 && 0.14  &  0.57  &  0.0018 &&  0.13  &  0.58  &\\
Nov 1, 2014 && 0.14 & 0.51 &  0.0018 &&    0.13  & 0.53  &\\
Dec 1, 2014 && 0.16 & 0.43 &  0.0015 &&     0.14 &  0.45  &\\
\hline
\end{tabular}
\caption{Parameter estimates for each fit in Figure \ref{fig:MultiForecastAll}, both with $k$ included (left) and without $k$ (i.e. with $k$ fixed $=1$). Each model is fitted to the data up through the dates given on the left.}
\label{tab:MultiForecastAll}
\end{table*}

\noindent \textit{Forecasting and the Correction Factor $k$.} 
A key finding was that in spite of the large uncertainty, the best-fit models that included the reporting rate/population-at-risk parameter $k$ were able to successfully forecast the numbers of reported cases and deaths up to two months ahead (Figures \ref{fig:simplefit}, \ref{fig:MultiForecastAll}, and Supp. Figure \ref{fig:MultiForecast_k}), while models without $k$ (Figs. \ref{fig:MultiForecastAll} and \ref{fig:MultiForecast_nok}) or using a $k$ value based on reporting rate estimates alone (Fig. \ref{fig:MultiForecast_k0pt4}) consistently overestimated the incidence as they were unable to capture the slowing of the epidemic away from exponential growth.
We propose that $k$ provides an ad hoc way to represent the effects of changing conditions on the ground, shifts in behavior in the population, and ongoing intervention efforts. As interventions and changes in behavior reduce transmission, this may reduce the effective population at risk (decreasing $k$), so that the overall epidemic burns out earlier than it otherwise would (i.e. with increasing protection from transmission viewed as whole or partial removal from the population at risk). 
A lower value of $k$ thus results in a fitted model with a rapid increase matching the exponential growth in the data, followed by an earlier epidemic turn over/burn out due to the smaller effective population at risk, which would not be possible if the model treated the full population as at-risk (since the rapid early growth would correspond to a much larger epidemic).

The usefulness of $k$ in forecasting is particularly evident when comparing the patterns seen in the data for each country. The data for Liberia, Guinea, and all countries combined shows slowing of the epidemic from exponential growth much earlier (late 2014), while Sierra Leone continued to show exponential growth through the end of 2014 and into 2015 \cite{WHO_EbolaSitReps}. 
This pattern was difficult for many models to distinguish and predict \cite{TimeForecast, 538Forecast, Butler2014} (although we note that most models were worst-case estimates and were not intended to account for how changes in reporting or interventions would affect the epidemic trajectory, as discussed further by Rivers et al. \cite{Rivers2014Nature}). 
However, when $k$ was included, the model forecasted these qualitative differences between countries correctly. Correspondingly, the fitted and LH-sampled $k$ values for Guinea, Liberia, and all countries were two orders of magnitude smaller once the September data was included, but estimates for Sierra Leone remained large (Tables \ref{tab:MultiForecastAll} and \ref{tab:MultiForecast_k}). This suggests that even though the slow-down may not be immediately evident by eye as of October 1, 2014, the upcoming change in dynamics was preluded in the data, and subsequently captured by the best-fit models. 

Parameters similar to $k$ are often used to account for the reporting fraction \cite{hooker2010parameterizing, evans2005structural, chapman2009structural}; alternatively it is also common to re-scale or adjust the data to account for underreporting \cite{white2009estimation, pourbohloul2009initial}. However, correction parameters such as the $k$ used here are not often considered explicitly for representing other factors such as the population at risk and symptomatic fraction seen here (although similar issues of saturation have been noted and examined in other ways \cite{grassly2008mathematical}). A related modeling study \cite{Bellan2014} recently showed that forecasts will tend to overestimate incidence of EVD if asymptomatic infections and their resulting immunity are not taken into account, which illustrates a similar phenomenon to that observed here. 
It is of interest to note that one of the most commonly observed identifiable combinations in SIR-type models is between the reporting rate and total population size \cite{evans2005structural, chapman2009structural, eisenberg2013identifiability}, making such a combined parameter natural to consider. 
In previous work we explicitly used $k$ to correct for the population at risk in the context of cholera \cite{eisenberg2013identifiability, eisenberg2013examining}, but here we extend these results to evaluate the effects of this correction parameter on forecasting have been examined.

There are a wide range of factors which may alter the value of $k$ and the apparent population at risk, including underreporting, asymptomatic infections, pre-existing immunity, social contact structures, spatial spread, behavior patterns, and ongoing interventions. As many of these factors are likely to be at work in outbreak settings we would expect that, without any adjustment or accounting for these issues, most models will tend to over-forecast. 
Our results show that even the extremely simplified and agglomerated adjustment used here can make a significant difference the accuracy of the model forecasts. 
However, it is important to note that this approach does not mechanistically account for these factors, so that if there is no evidence of saturation of the epidemic in the data, even models including $k$ can over-estimate, as illustrated in the last forecast for Sierra Leone in Supplementary Figure \ref{fig:MultiForecast_k}. 
All of these factors would also be expected to change over time, making it difficult to generate long-term predictions---we were able to accurately forecast up to two months ahead using this constant factor, but more explicitly mechanistic models might be helpful for more accurate longer-term predictions. In general, some factors may change over time in ways which a simple correction term cannot account for---the trajectory of the Guinea data over 2014-2015 shows an extended, nearly-linear growth period, and occasional plateaus, which this approach may not completely capture depending on what data is used for forecasting. Indeed, the the forecasts for Guinea seen in Supplementary Figure \ref{fig:MultiForecast_k} show some cases of both over- and under-prediction, although the forecasts remain significantly better than those without $k$ or using $k=1/2.5$. 

Additional data on the factors making up $k$ would allow us to understand their relative importance in the epidemic dynamics, and may help further improve forecasting by understanding how these components are likely to change. 
Forecasting efforts are often useful for comparing outcomes from alternative scenarios with varying interventions. Understanding the uncertainty inherent in these forecasts can itself be useful in evaluating the range of intervention efforts that may be needed based on current data, and could potentially help guide additional data collection efforts to reduce this uncertainty \cite{lofgren2014opinion}. 
Comparative modeling approaches may also be helpful in this process \cite{lofgren2014opinion}, as different model structures and approaches may result in further uncertainty, or potentially provide some consensus across a range of different model assumptions. 
\\

\noindent \textit{Uncertainty \& Unidentifiability.} That the full range of the sampled parameters fit the data similarly well but yielded a wide range of forecasted trajectories (shaded regions in Figure \ref{fig:simplefit}) illustrates the expected uncertainty and parameter unidentifiability associated with forecasting from exponential growth phase data, in which a range of parameter values can yield the same initial growth rate, but may differ as to when their resulting trajectories begin to slow from exponential growth. This idea has been used to develop simple two-parameter predictive models of the epidemic, as demonstrated by Fisman et al. \cite{fisman2014early}. This practical unidentifiability of the model results in identifiable combinations of parameters, so that increasing the transmission parameter of one stage can be compensated by decreasing that of another stage (as shown in the approximate plane of best fit parameter estimates in  Fig. \ref{fig:Beta12FCombo}). This issue has also been noted for infected and funeral transmission by Weitz and Dushoff \cite{weitz2015modeling}, using the SEIRD model, wherein they also note that the unidentifiability issues are likely to persist more broadly in models fitted to exponential growth data, as is indeed the case here. Unidentifiability of transmission parameters has also been noted to be a common issue more generally \cite{brookhart2002statistical, tien2010multiple, eisenberg2013identifiability, cortez2013distinguishing}.
Transmission parameters are often difficult to measure empirically, so that a lack of identifiability may hinder estimates of the magnitudes of different transmission pathways or terms of $\Ro$ (Supp. Fig. \ref{fig:BestR0s} and Supp. Table \ref{tab:stageR0}). This can make it difficult to inform decision-making between alternative intervention strategies focused on different transmission pathways (e.g. increasing safe burials vs. building ETUs). Additional data collection that distinguishes transmission pathways (e.g. on behavioral and contact patterns, burial practices, etc.) is thus important to improve parameter identifiability and make modeling efforts more practically applicable to decision-making. 
\\

\noindent \textit{Limitations \& Future work.} The model does not account for factors such dynamic changes in reporting, differences in reporting between cases and deaths, or effects of spatial spread and mobility (all likely changing with time). Changing conditions on the ground made forecasting further than 1-2 months ahead a difficult task (as illustrated in Figures \ref{fig:simplefit} and \ref{fig:MultiForecastAll}), and further work to capture these time-varying parameters is warranted. Similarly, while the parameter $k$ may capture the effects of behavior change on the population at-risk in a rough way, deeper exploration of how behavior and contact patterns shifted over the course of the outbreak (particularly related to hygiene and funeral practices) is important, both for modeling and data collection efforts. Another direction for future work is in examining the stage structure of the clinical course of EVD in more detail. This model includes a simplified representation of the natural history of EVD using two stages. However, some descriptions of EVD break the clinical course into three or more stages, for example, as described by Beeching et al. \cite{beeching2014ebola}: ``a few days of non-specific fever, headache, and myalgia, followed by a gastrointestinal phase in which diarrhoea and vomiting, abdominal symptoms, and dehydration are prominent. In the second week, the patient may recover or deteriorate, with a third phase of illness including collapse, neurological manifestations, and bleeding, which is often fatal.'' Our model likely includes part of this gastrointestinal phase in each of the two stages, and further work expanding the stage structure would be useful, particularly in evaluating the potential for stage-driven interventions. The complex dynamics of the viral load and immune response have recently begun to be explored in modeling efforts \cite{yamin2014effect}, and further work in this area would be particularly useful for uncovering clinical features that can be used to optimize treatment approaches. 

\subsection*{Conclusions} 
Explicitly accounting for under-reporting and fraction of population at risk can improve forecasting even when the uncertainty in predictions and parameter estimates are high.  Specifically, we found that a scalar correction factor was identifiable and correctly forecasted the discrimination between continued exponential growth of cases in Sierra Leone through 2014 versus early leveling off in Liberia and Guinea.

\section*{Acknowledgements} 
This work was supported by the National Institute of General Medical Sciences of the National Institutes of Health under Award Number U01GM110712 (supporting JNSE, MCE, RM, and EVW), as part of the Models of Infectious Disease Agent Study (MIDAS) Network. The content is solely the responsibility of the authors and does not necessarily represent the official views of the National Institutes of Health.

\section*{Competing interests}
The authors declare that we have no competing interests. 

\section*{Authors' Contributions}
MCE designed the study, performed all model simulations and analyses, contributed to the collection of online data sources, and drafted and revised the manuscript. MCE and JPD developed the initial model, with substantial revision by all authors to form the final model. JPD led the collection of online data sources, and provided critical revisions to the manuscript for intellectual content. JNSE, EVW, SC, YK, and RM provided critical revisions to the manuscript for intellectual content. All authors gave final approval for publication.

{\small
\bibliographystyle{ieeetr}
\bibliography{EbolaRefs.bib}
}

\pagebreak
\beginsupplement
\section{Supplementary Information}

\subsection{Model Initial Conditions}
The model initial conditions were determined as follows: for the measurement equations (Eq. \eqref{eq:meas}), $y_C(0)$ and $y_D(0)$ were both set equal to the initial values of the data set being fitted; $I_2(0)$ was approximated by examining the number of new deaths reported in the next two days after the first data point (as these would have been in $I_2$ when $t=0$), and then re-scaling this quantity from numbers of individuals to fraction of the population using $k$ and $N$; $I_1(0)$ was approximated by evaluating the number of cases that had been reported in the past six days (as these would, on average, still be in $I_1$) and then re-scaling as for $I_2$; $F(0)$ was approximated by the re-scaled number of deaths reported within the previous two days; $R(0)$ was approximated roughly by taking the number of incident cases in the last 15 days, subtracting the number of dead in the last 15 days, and then re-scaling to fractions of the population; $E(0)$ was taken as twice the initial values for infectious individuals. 
$S(0)$ was then calculated as 1 minus the remaining model variable initial values. These initial conditions were quite rough (reflecting the uncertainty in the data as well), but we tested a wide range of initial conditions based on differing assumptions and also sampled within realistic ranges for the initial conditions; we found that the model fit well to the data (with similar residual sum of squares values) for all initial condition values we set.

\subsection{Additional Information on Parameter Estimates and Uncertainty}
\begin{figure}[h]
\centering
\includegraphics[width=0.4\textwidth]{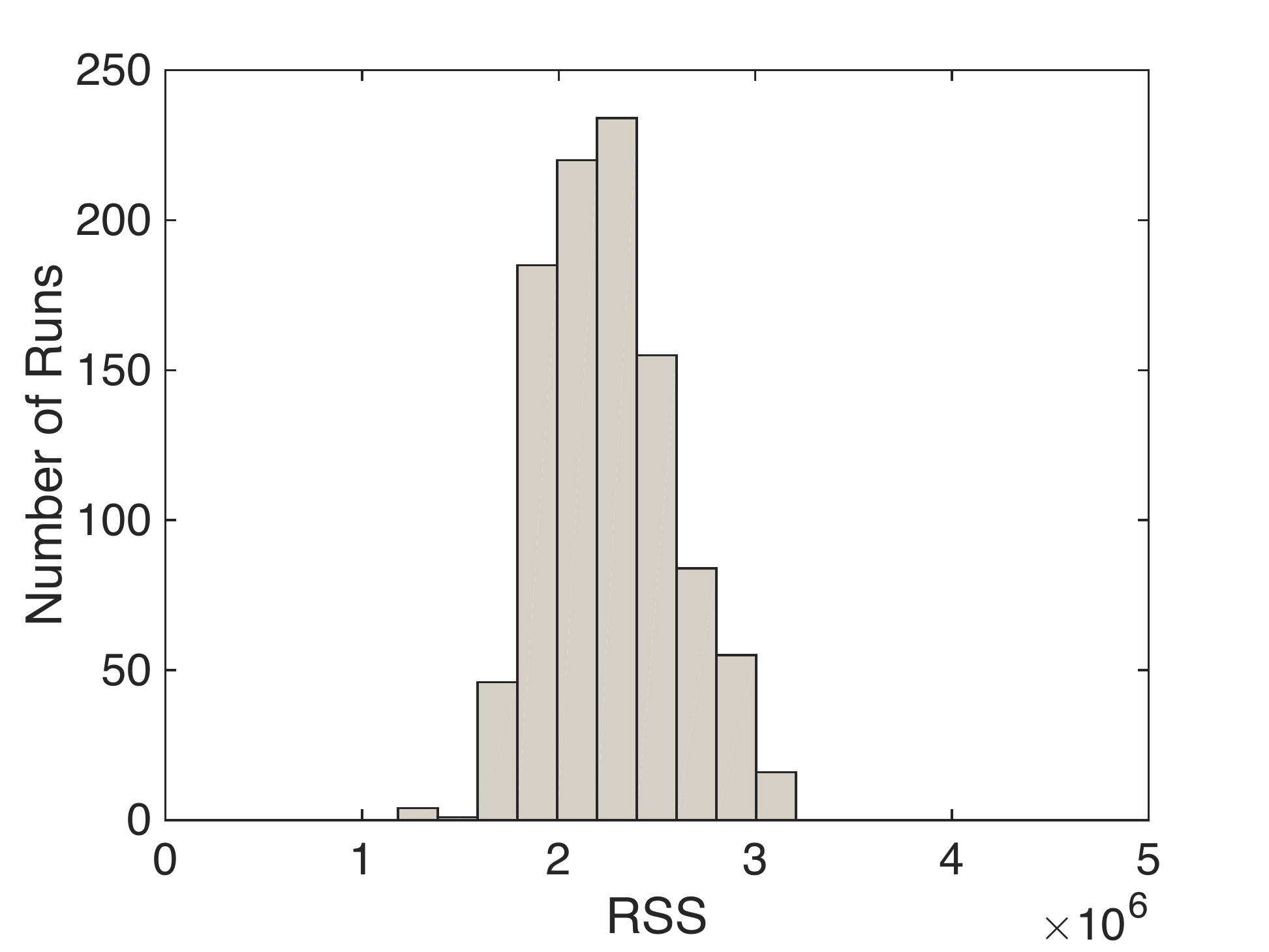}
\includegraphics[width=0.4\textwidth]{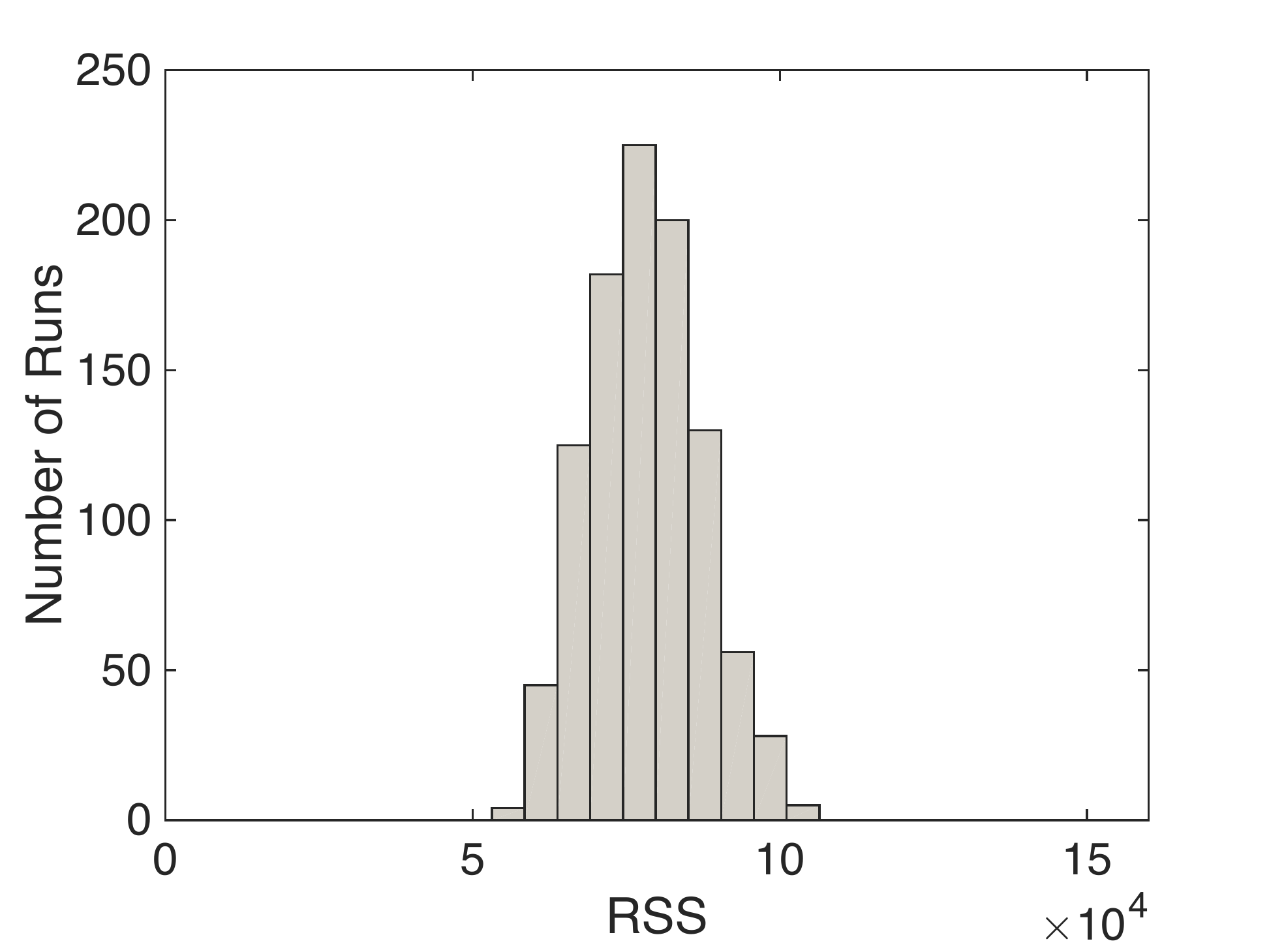}
\includegraphics[width=0.4\textwidth]{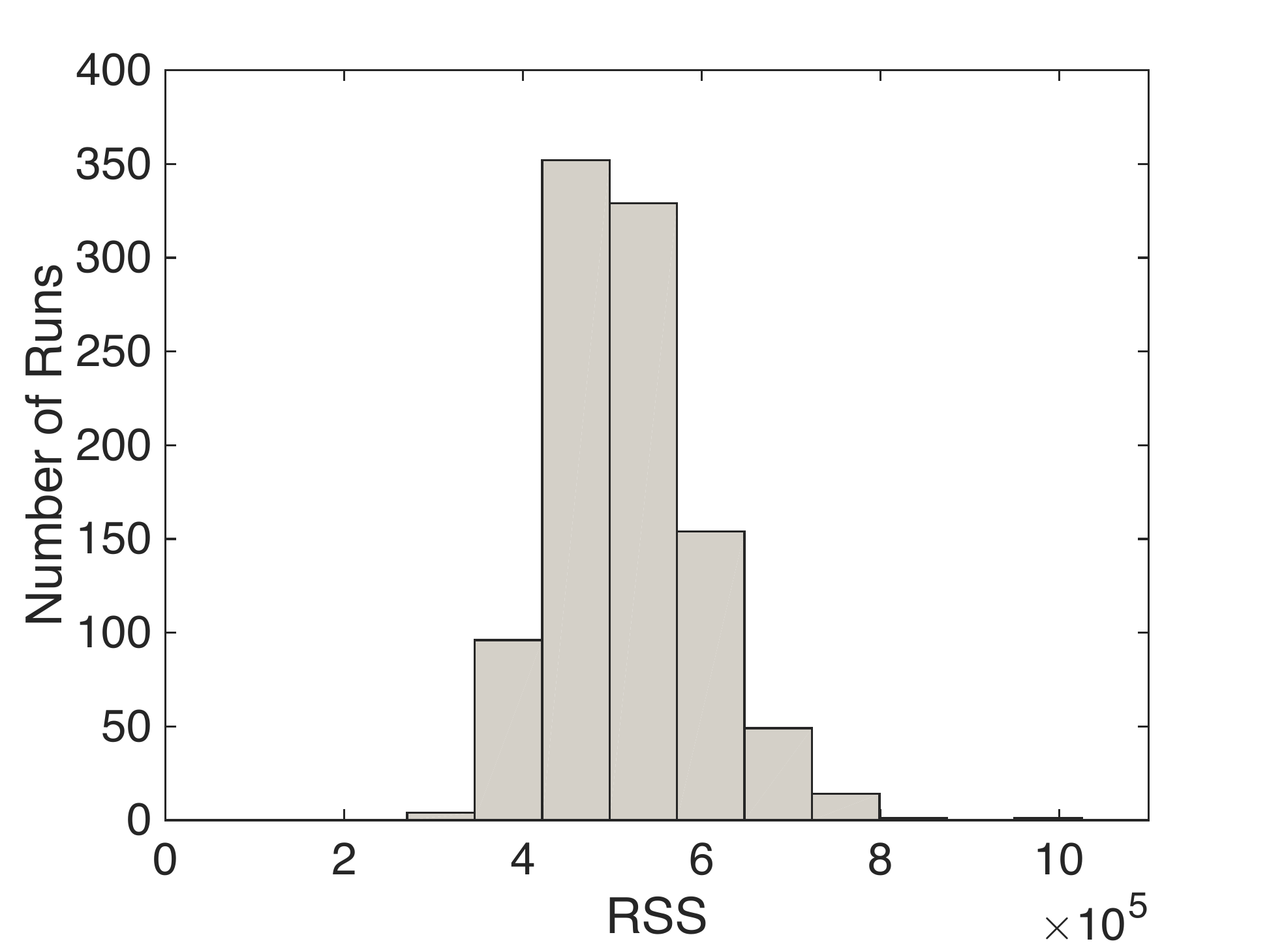}
\includegraphics[width=0.4\textwidth]{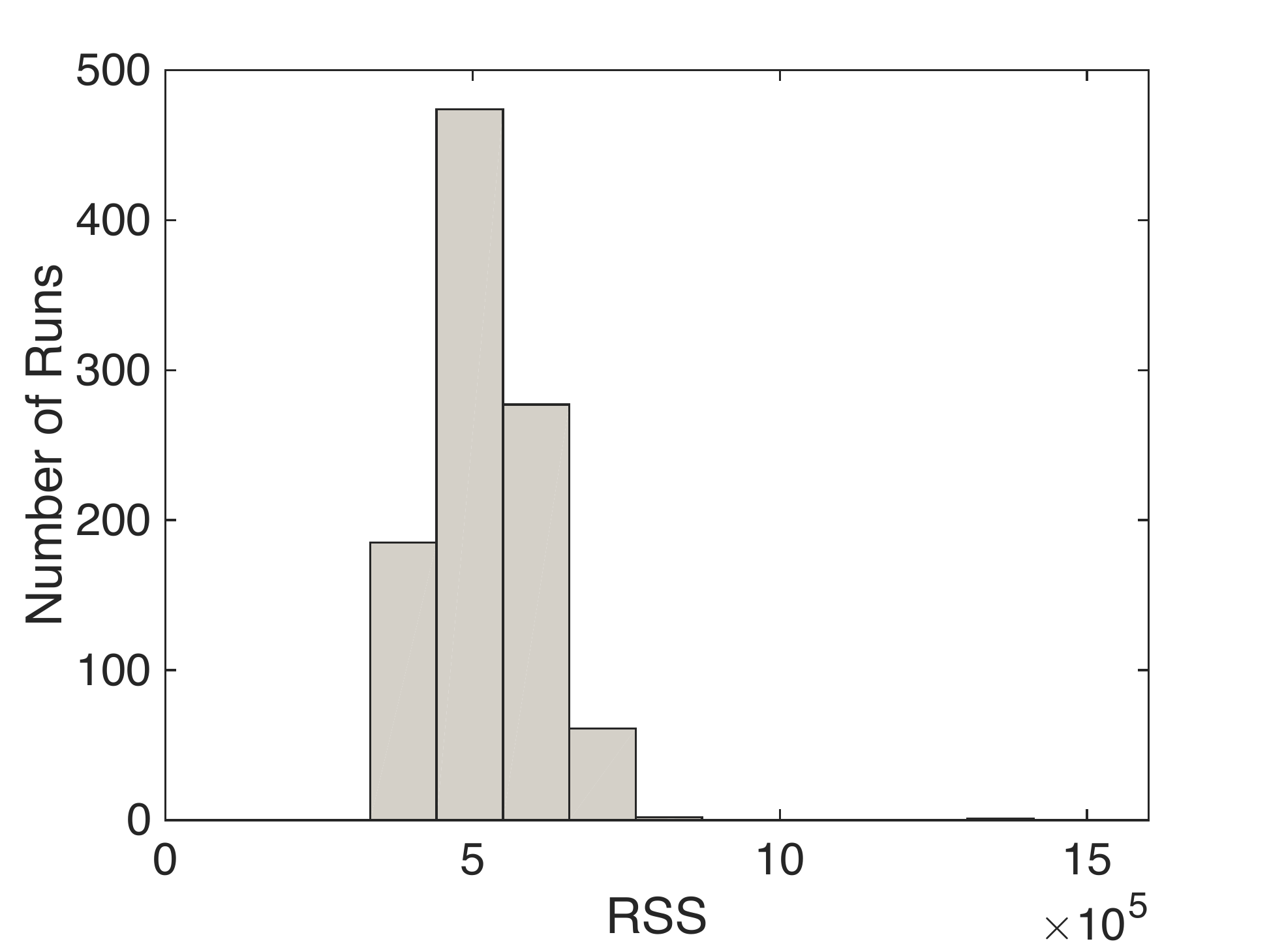}
\caption{Histogram of residual sum of squares values for the simplified model fits to data from (left to right): all countries combined, Guinea, Liberia, and Sierra Leone (corresponding to Figure \ref{fig:simplefit}). }
\label{fig:GofHist}
\end{figure}

The non-fitted, LH-sampled parameter values associated with the best-fit trajectories in Figure \ref{fig:simplefit} were:
\begin{itemize}
\item All countries: $\alpha = 0.1$, $\beta_2/\beta_1 = 3.38$, $\beta_F/\beta_1 = 3.44$, $\delta_2 = 0.99$, $\gamma_F = 0.36$, $\gamma_2 = 0.53$, $\gamma_1 = 0.19$, $k = 0.0026$
\item Guinea: $\alpha = 0.11$, $\beta_2/\beta_1 = 2.04$, $\beta_F/\beta_1 = 2.63$, $\delta_2 = 0.99$, 
$\gamma_F = 0.40$, $\gamma_2 = 0.92$, $\gamma_1 = 0.19$, $k = 0.0006$
\item Liberia: $\alpha = 0.11$, $\beta_2/\beta_1 = 4.90$, $\beta_F/\beta_1 = 3.84$, $\delta_2 = 0.93$, $\gamma_F = 0.35$, $\gamma_2 = 0.96$, $\gamma_1 = 0.20$, $k = 0.0024$
\item Sierra Leone: $\alpha = 0.10$, $\beta_2/\beta_1 = 1.73$, $\beta_F/\beta_1 = 2.97$, $\delta_2 = 0.92$, $\gamma_F = 0.38$, $\gamma_2 = 0.73$, $\gamma_1 = 0.15$, $k = 0.68$
\end{itemize}
where the corresponding fitted values for $\beta_1$ and $\delta$ are given in Table \ref{tab:simplefit}. We note that as these parameters were LH-sampled rather than estimated there may be many other parameter values which yield the same fit to the data. Figure \ref{fig:GofHist} shows histograms of the goodness of fit scores associated with all trajectories in Figure \ref{fig:simplefit}. 

\begin{figure}
\centering
\includegraphics[width=0.32\textwidth]{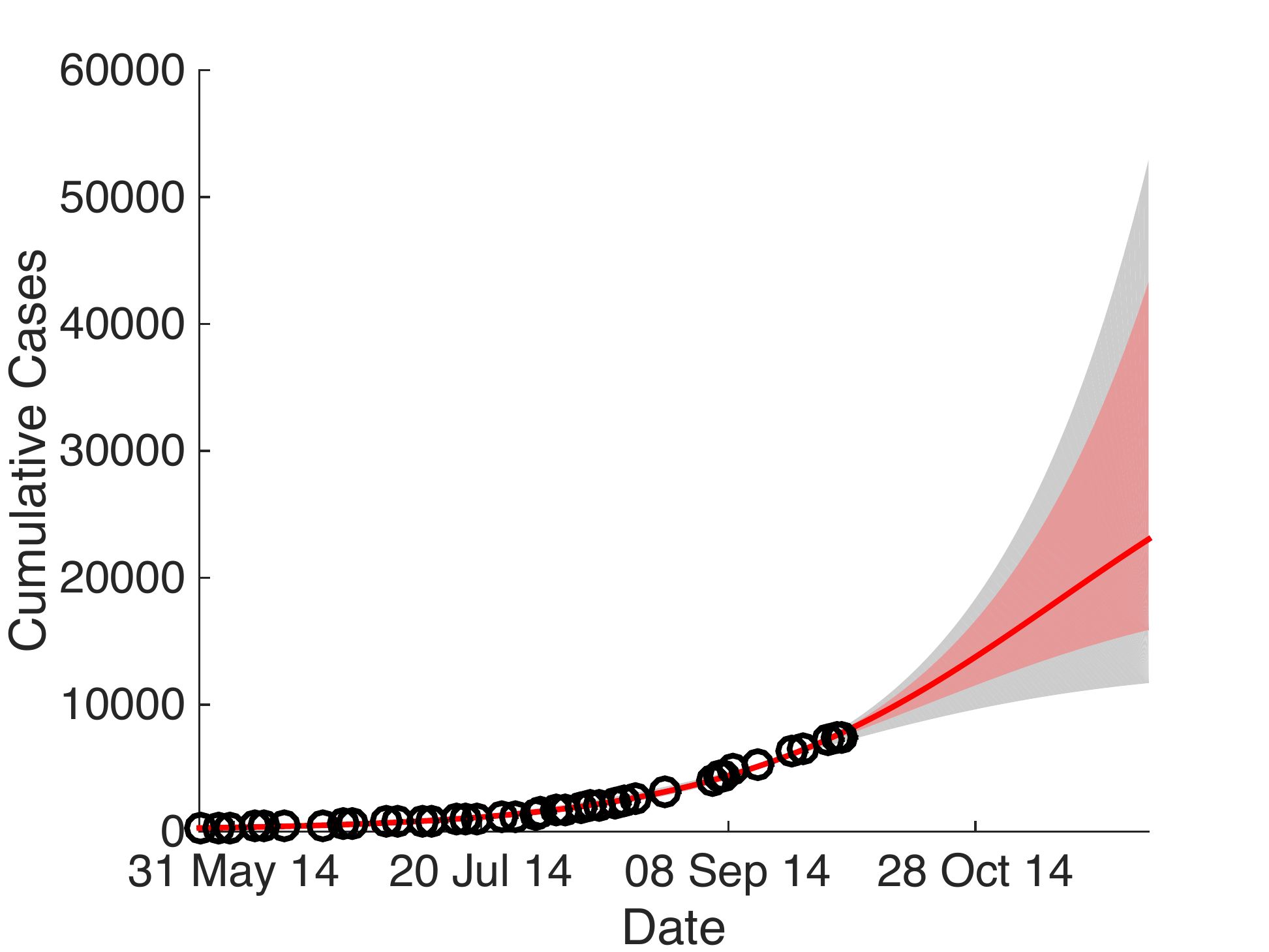}
\includegraphics[width=0.32\textwidth]{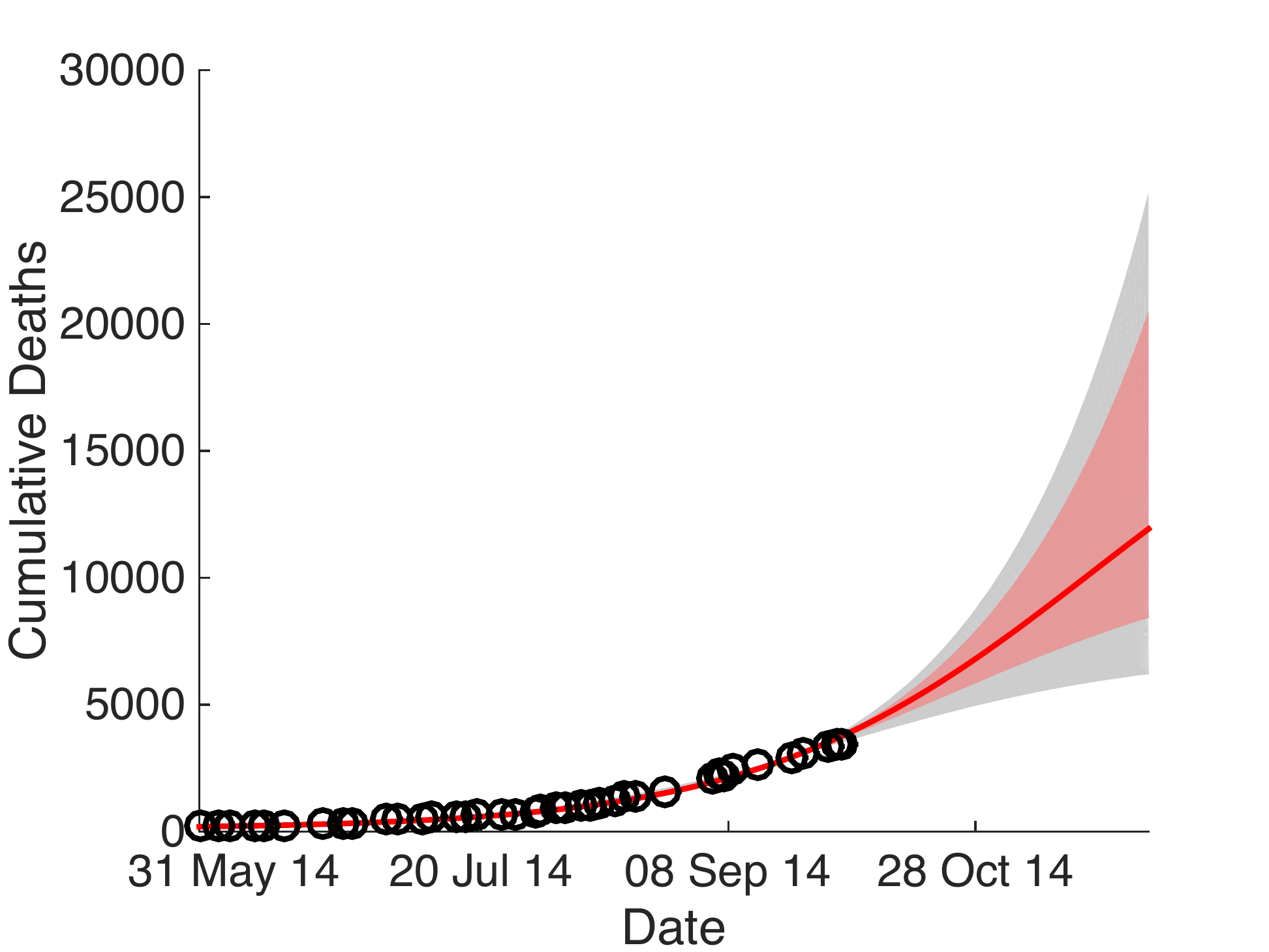}
\includegraphics[width=0.32\textwidth]{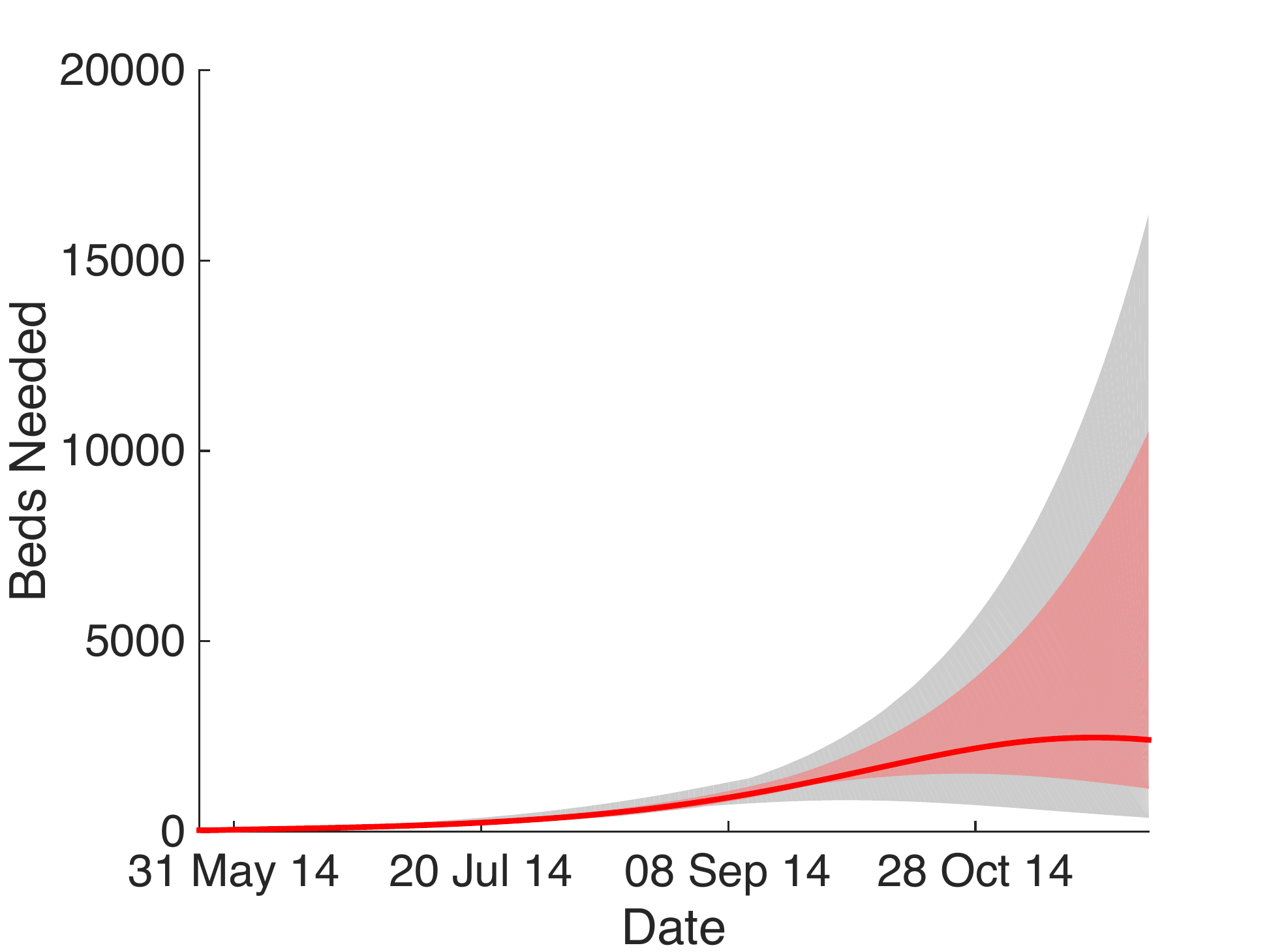}
\caption{Model fits to data from all countries combined (as in Figure \ref{fig:simplefit}), with the region covered by the best 10\% of all fits (lowest residual sum of squares) highlighted with red shading.}
\label{fig:BestTraj}
\end{figure}

\begin{figure}
\centering
\includegraphics[width=0.4\textwidth]{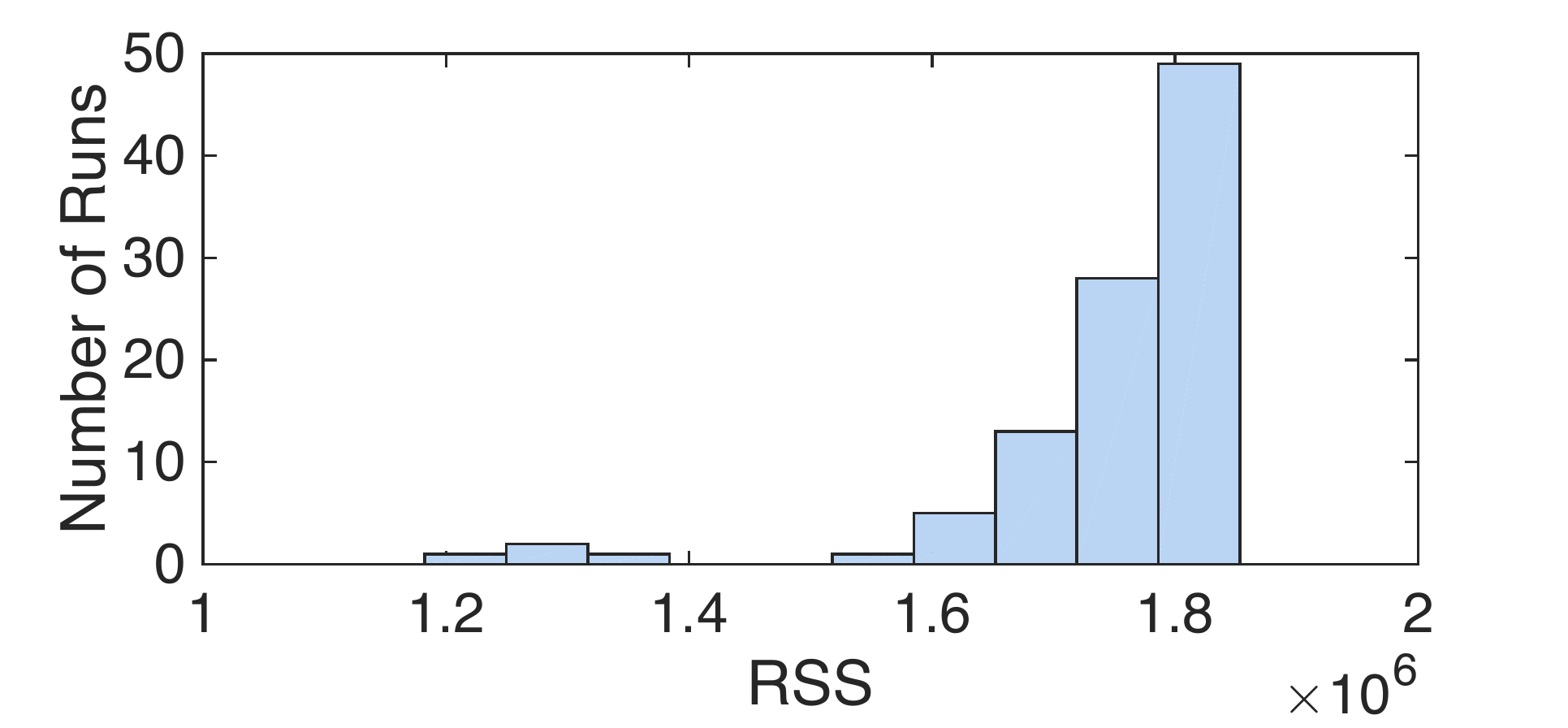}
\includegraphics[width=\textwidth]{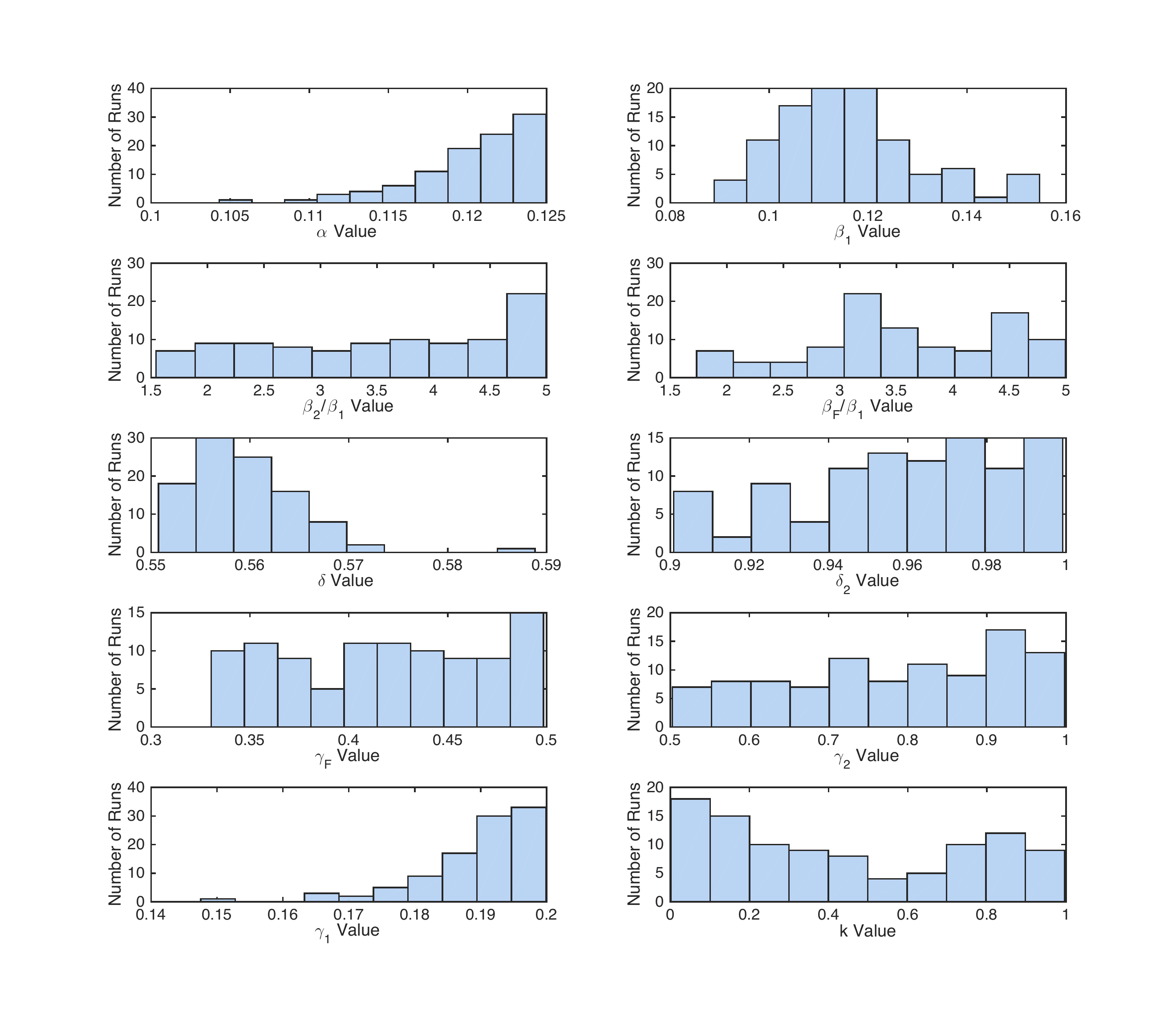}
\caption{Histograms of the residual sum of squares (RSS, top) and parameter values (bottom) from the best 10\% of all fits (lowest residual sum of squares) for all countries combined (Fig. \ref{fig:simplefit}). While in some cases, the best-fit parameters tend to be clustered at a particular values, others span the full range of realistic values given in Table \ref{tab:params}. 
}
\label{fig:BestHist}
\end{figure}

For the all countries combined case, we also examined the best $10\%$ of fits (evaluated by their sum of squares value), with the range (min to max) of these best fits shown in red shading in Figure \ref{fig:BestTraj}. These best $10\%$ of fits tended to cluster around the overall best fit line (red line in Fig. \ref{fig:BestTraj}). Figure \ref{fig:BestHist} shows histograms of the fitted ($\beta_1$ and $\delta$) and LH-sampled parameter values corresponding to the best $10\%$ of fits, and a histogram of the corresponding sum-of-squares values. These histograms can help to give a sense of what parameter values yield the overall best fits to the data. As shown in Figure \ref{fig:BestHist}, most non-fitted parameters span the full range of biologically reasonable values in Table \ref{tab:params}, even when only the best $10\%$ of fits are considered, suggesting they are likely to be practically unidentifiable. However, some parameters do appear to be primarily drawn from specific regions, e.g. $\alpha$, $\gamma_1$, and the fitted parameters $\beta_1$ and $\delta$. We note that while $k$ shows only a weak tendency towards the small values seen in the best fit, the very lowest sum of squares values consistently had low values for $k$, while the remaining parameters tended to cover the same ranges seen in Figure \ref{fig:BestHist}.

\clearpage

\subsection{Transmission by Stage of Infection}
\begin{figure}
\centering
\includegraphics[width=0.9\textwidth]{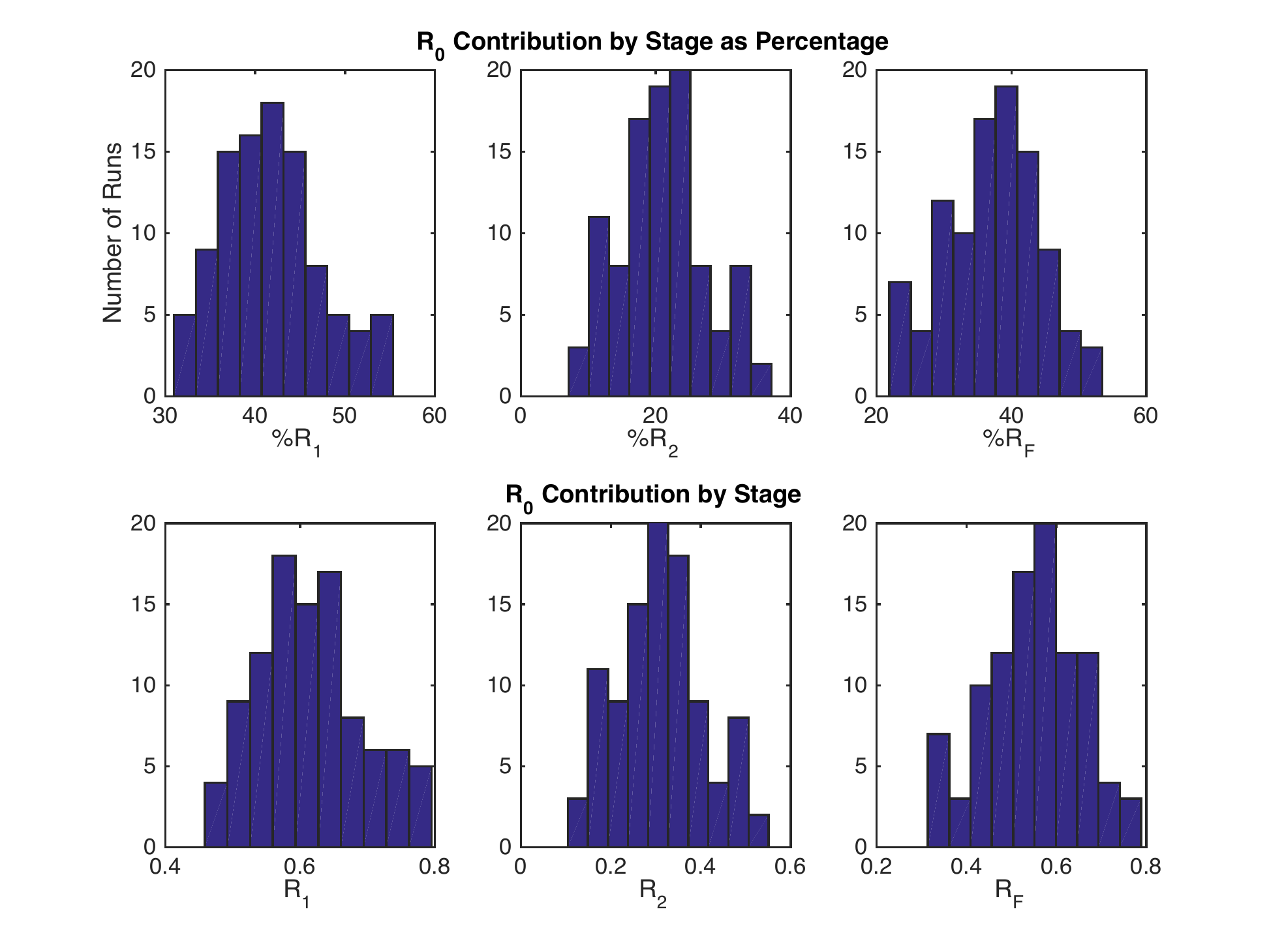}
\caption{Histograms of $\Ro$ contributions by stage from the best 10\% of all fits (lowest residual sum of squares) for all countries combined (Figure \ref{fig:simplefit}). The top row shows contributions by stage as a percentage of overall $\Ro$, while the bottom row shows the magnitudes of $\mathcal{R}_1$, $\mathcal{R}_2$, and $\mathcal{R}_F$ (the contributions of $I_1$, $I_2$, and $F$ respectively). Even among the best fit estimates, the contributions by stage to $\Ro$ span from quite low to the majority of transmission for any given stage, making it difficult to estimate the true contribution of each stage to transmission from incidence data alone.}
\label{fig:BestR0s}
\end{figure}

We also examined the contributions of each stage of EVD to transmission. The model $\Ro$ breaks up naturally into terms for each of the three stages (with each term in Eq. \eqref{eq:R0} corresponding to $I_1$, $I_2$, and $F$), representing the relative contributions of each transmission stage to $\Ro$. Each term can be interpreted as the average number of secondary cases generated by an infectious individual while in a particular stage of transmission. We denote each of these terms as $\mathcal{R}_1$, $\mathcal{R}_2$, and $\mathcal{R}_F$ respectively. Table \ref{tab:stageR0} shows the contributions to $\Ro$ by each stage. 
In all cases, all three transmission stages contributed to $\Ro$, with $I_1$ and $F$ tending to make the largest contributions to $\Ro$. The estimated ranges of contribution by each stage are quite wide, likely due to the unidentifiability issues for the transmission parameters by stage as noted above. Indeed, even if only the best $10\%$ of fits are considered, we still found wide ranges of contributions by stage, as shown for all countries combined in Figure \ref{fig:BestR0s}. 

\begin{table*}
\centering
\def\arraystretch{1.4}

\begin{tabular}{| c | c || c | c | c |}

\hline
\rowcolor{LightGray} & $\Ro$ & $\mathcal{R}_1$ & $\mathcal{R}_2$ & $\mathcal{R}_F$  \\
\hline
\multirow{2}{*}{All countries}  & \multirow{2}{*}{1.6 (1.43 - 1.64)} & 0.58 (0.45 - 1.0) & 0.41 (0.10 - 0.61) & 0.61 (0.22 - 0.86)\\
 &  & 36\% (30\% - 69\%) & 26\% (7\% - 40\%) & 38\% (14\% - 56\%)\\
\hline
\multirow{2}{*}{Guinea} & \multirow{2}{*}{1.79 (1.47 - 1.79)} &  0.86 (0.41 - 1.0) & 0.23 (0.12 - 0.70) & 0.69 (0.26 - 0.97) \\
 &  & 48\% (26\% -  63\%) & 13\% (7\% - 43\%) & 39\% (16\% - 57\%)\\
\hline  
\multirow{2}{*}{Liberia} & \multirow{2}{*}{1.81 (1.34 - 2.75)} & 0.60 (0.38 - 0.99) & 0.41 (0.11 - 0.70) & 0.81 (0.24 - 1.1)\\
 &  & 33\% (28\% - 62\%) & 22\% (8\% - 47\%) & 45\% (16\% - 57\%)\\
\hline
\multirow{2}{*}{Sierra Leone }& \multirow{2}{*}{1.32 (1.19 - 1.37)} &  0.84 (0.49 - 0.96) & 0.12 (0.065 - 0.41) & 0.36 (0.15 - 0.55)\\
 &  &  64\% (39\% - 75\%) & 9\% (5\% - 34\%) & 27\% (12\% - 44\%)\\
\hline
\end{tabular}
\caption{Overall best estimates and ranges for $\Ro$, and the contribution to $\Ro$ by $I_1$, $I_2$, and $F$ (denoted $\mathcal{R}_1$, $\mathcal{R}_2$, and $\mathcal{R}_F$ respectively), as both magnitude and percentage of overall $\Ro$.  
}
\label{tab:stageR0}
\end{table*}

The result that the first stage may contribute strongly to $\Ro$ is somewhat counterintuitive given the lower excretion of body fluids and lower viral load in the first few days of symptoms compared to late-stage infection. Studies during previous outbreaks have observed cases occurring after exposure during the first stage of disease \cite{dowell1999transmission}, and have shown significant transmission risk in the first stage \cite{Francesconi2003}. 
In this model, the first stage is 1.5-5 times less transmissible, however it
lasts a longer period of time, and at each subsequent stage some fraction of people recover (and so don't continue to transmit), which lends the first stage a stronger impact on the overall $\Ro$. Moreover, as individuals may be more likely to seek care as symptoms progress, late-stage individuals may generate fewer cases due to reduced contact once hospitalized as well. However, the relative contribution of the first stage to transmission would likely drop considerably if the ranges for $\beta_2/\beta_1$ or $\beta_F/\beta_1$ were extended above the upper limit of 5 in Table \ref{tab:params}, since similar fits and forecasts were obtained across the full range of values. Because of the unidentifiabilities in the transmission parameters of our model, without additional biological knowledge or data, it is difficult or impossible to determine what the bounds of the biological ranges should be, or estimate precisely where within the ranges shown in Table \ref{tab:stageR0} the contribution of each stage to $\Ro$ actually sits. 

Nonetheless, in many of our simulations across these ranges, even complete elimination of any single stage of transmission was not enough to reduce $\Ro$ below one, suggesting that interventions should target more than just the second stage to be fully effective (or indeed any other single stage), and highlighting the importance of earlier case-finding and reducing funeral transmission. 
These results underscore the importance of a wide range of intervention strategies, including behavior and community-based interventions such as improving hygiene, sanitation, contact reduction with early stage infected, and safe burials, as well as collection of additional data on each infection stage to transmission.

%

\subsection{Expanded Forecasting Simulations}
\begin{figure}[h]
\centering
\textbf{Fitted $k$}\\
\vspace{0.25cm}
\rule{2cm}{0.4pt} . Guinea . \rule{2cm}{0.4pt}\\
\includegraphics[width=0.32\textwidth]{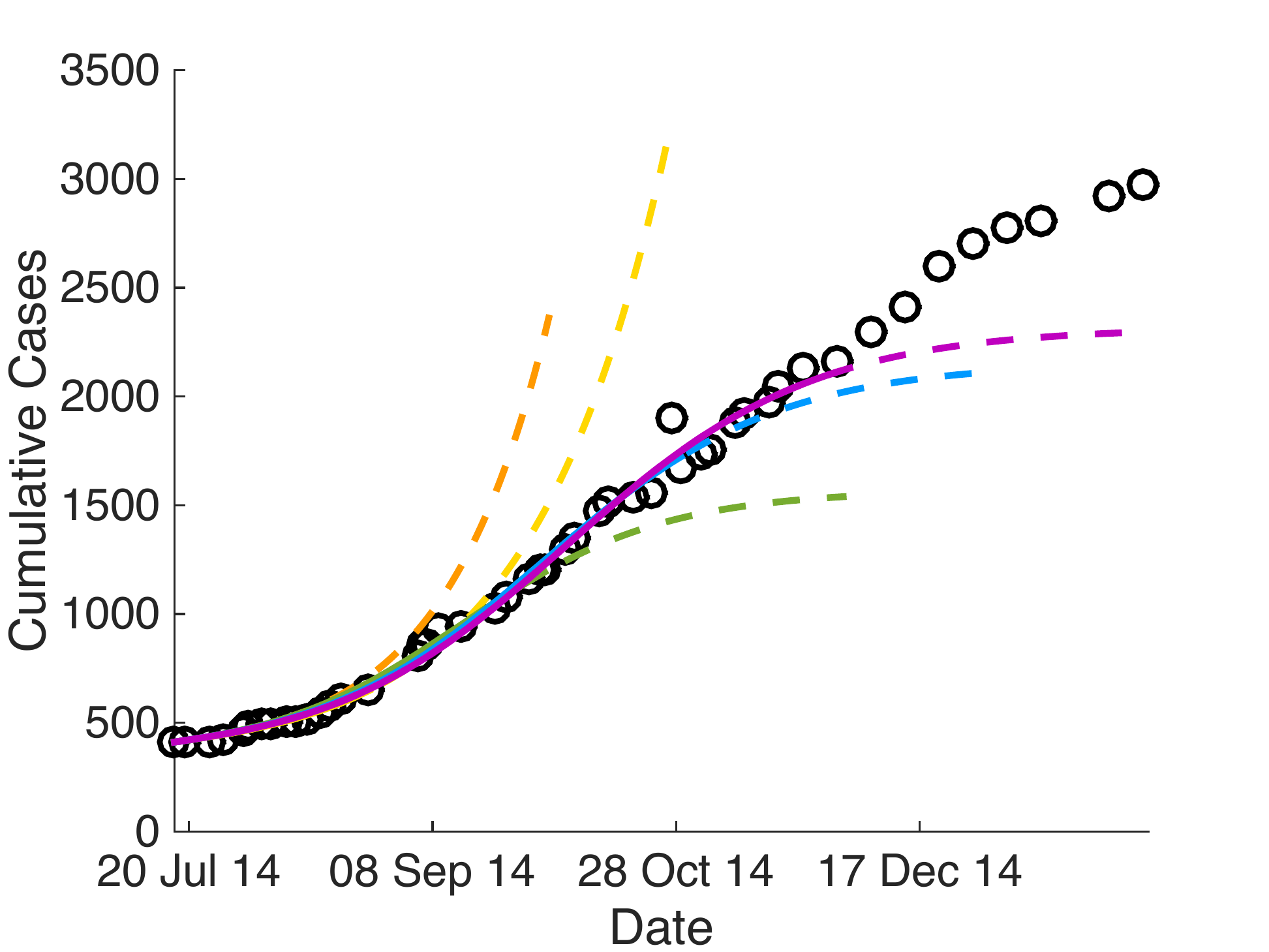}
\includegraphics[width=0.32\textwidth]{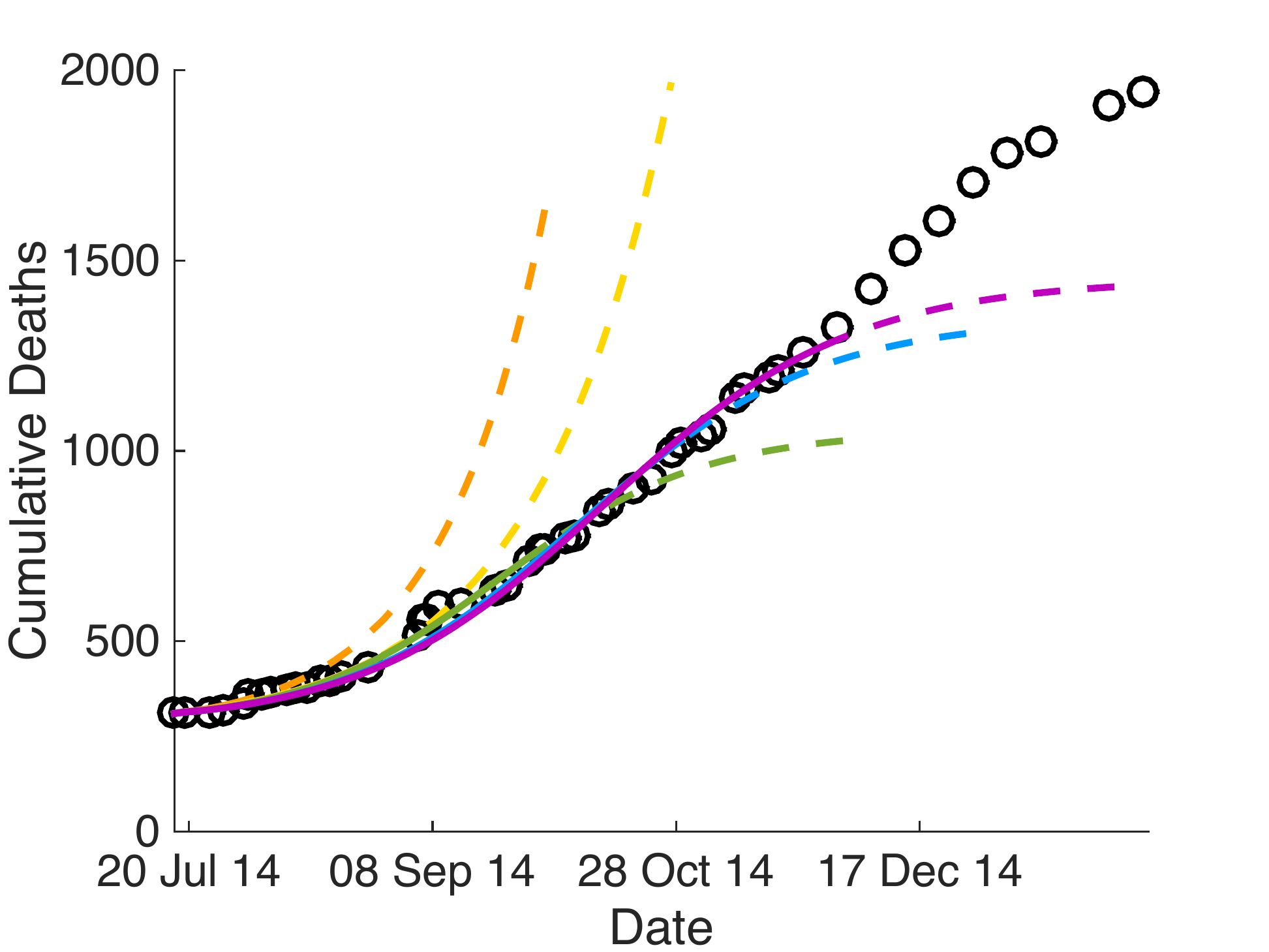}
\includegraphics[width=0.32\textwidth]{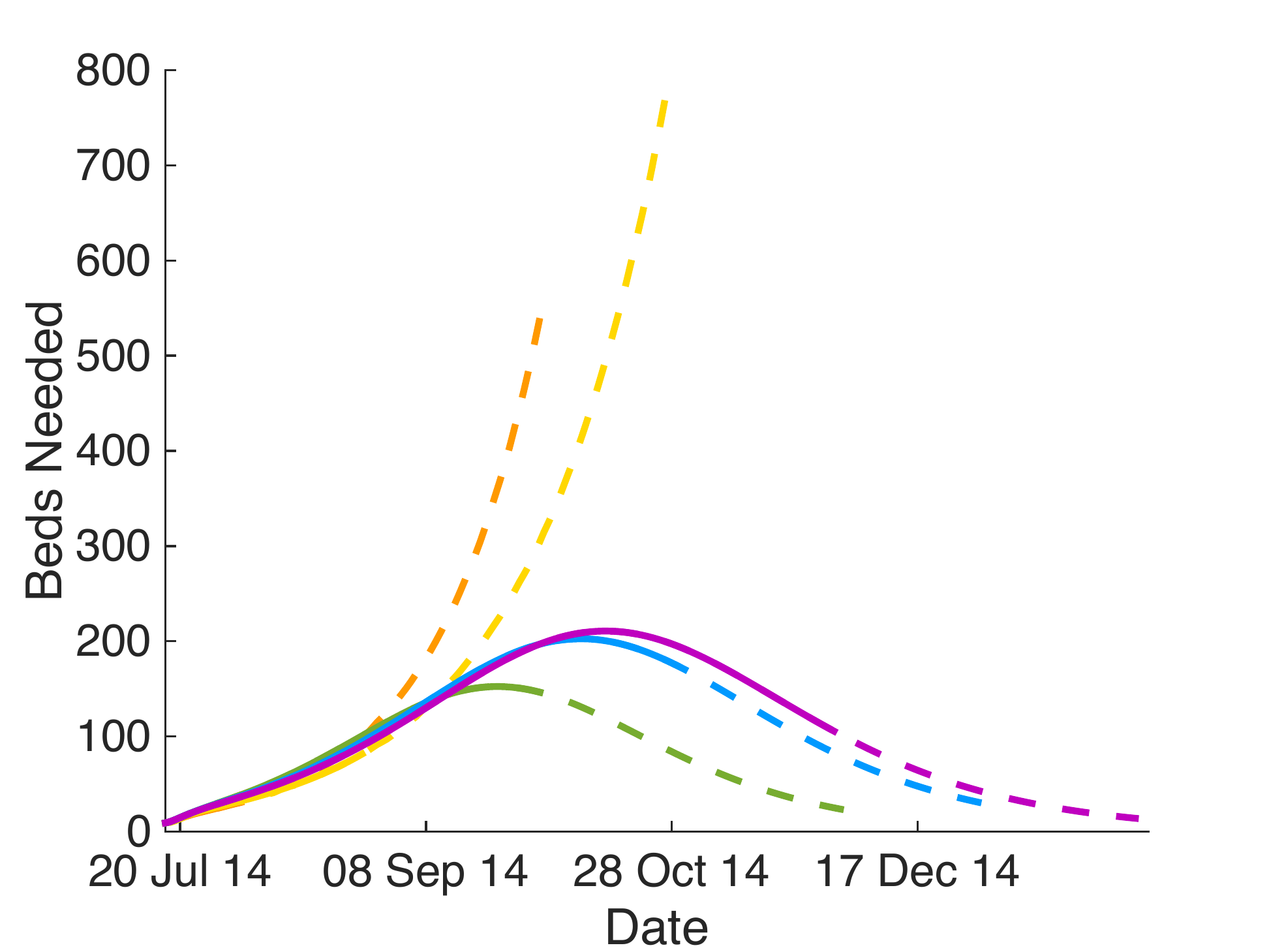}\\
\vspace{0.25cm}
\rule{2cm}{0.4pt} . Liberia . \rule{2cm}{0.4pt}\\
\includegraphics[width=0.32\textwidth]{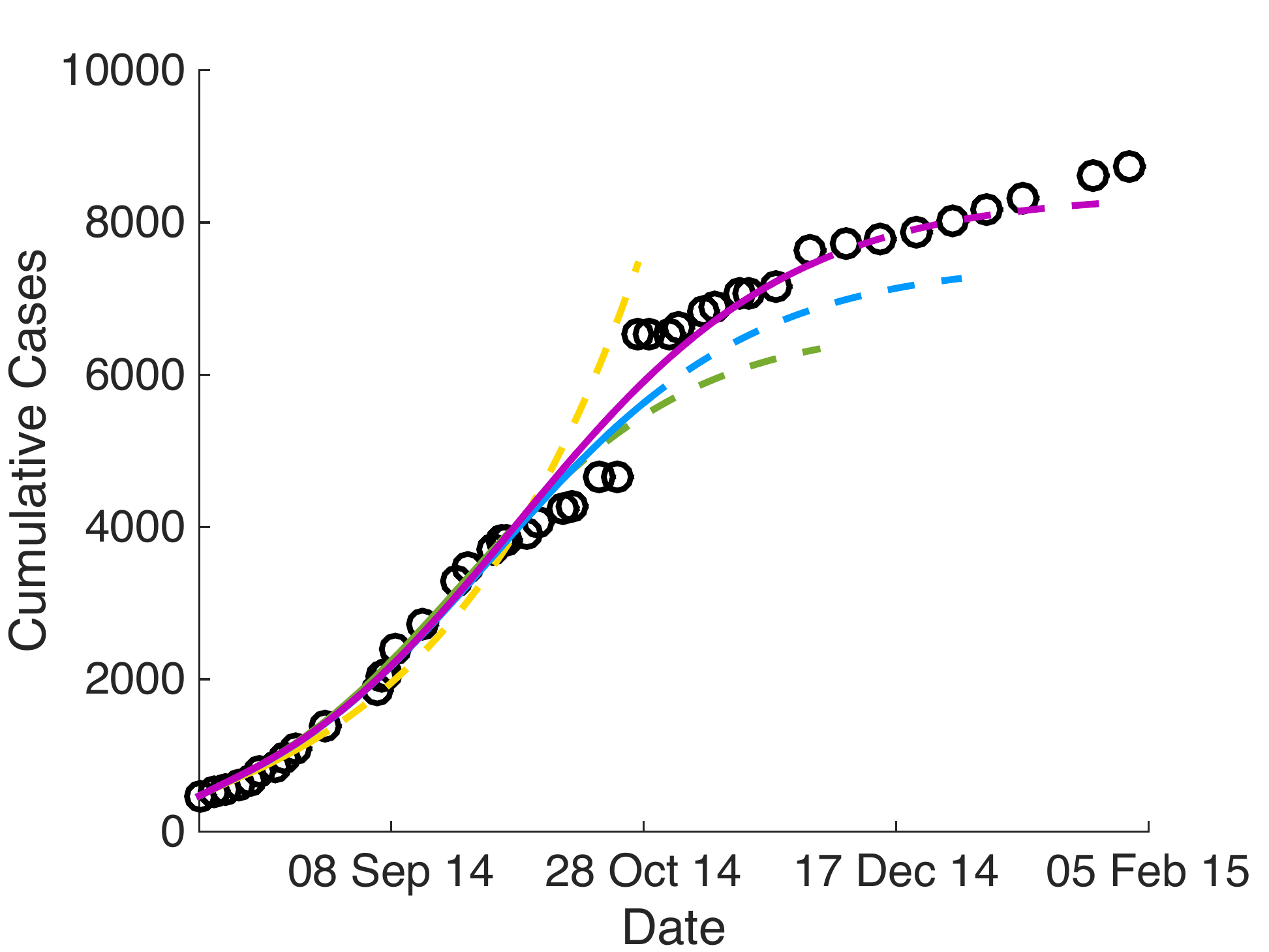}
\includegraphics[width=0.32\textwidth]{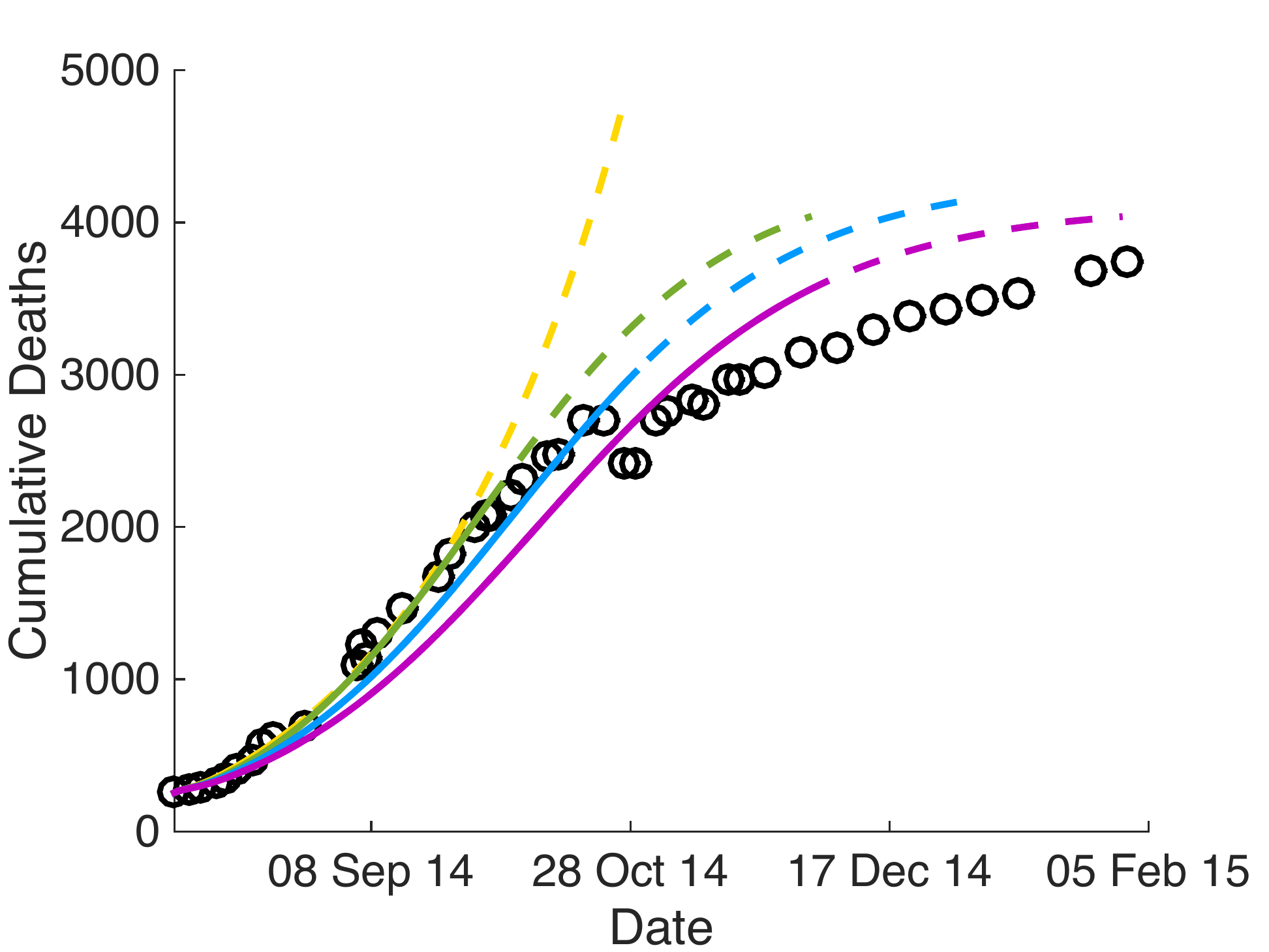}
\includegraphics[width=0.32\textwidth]{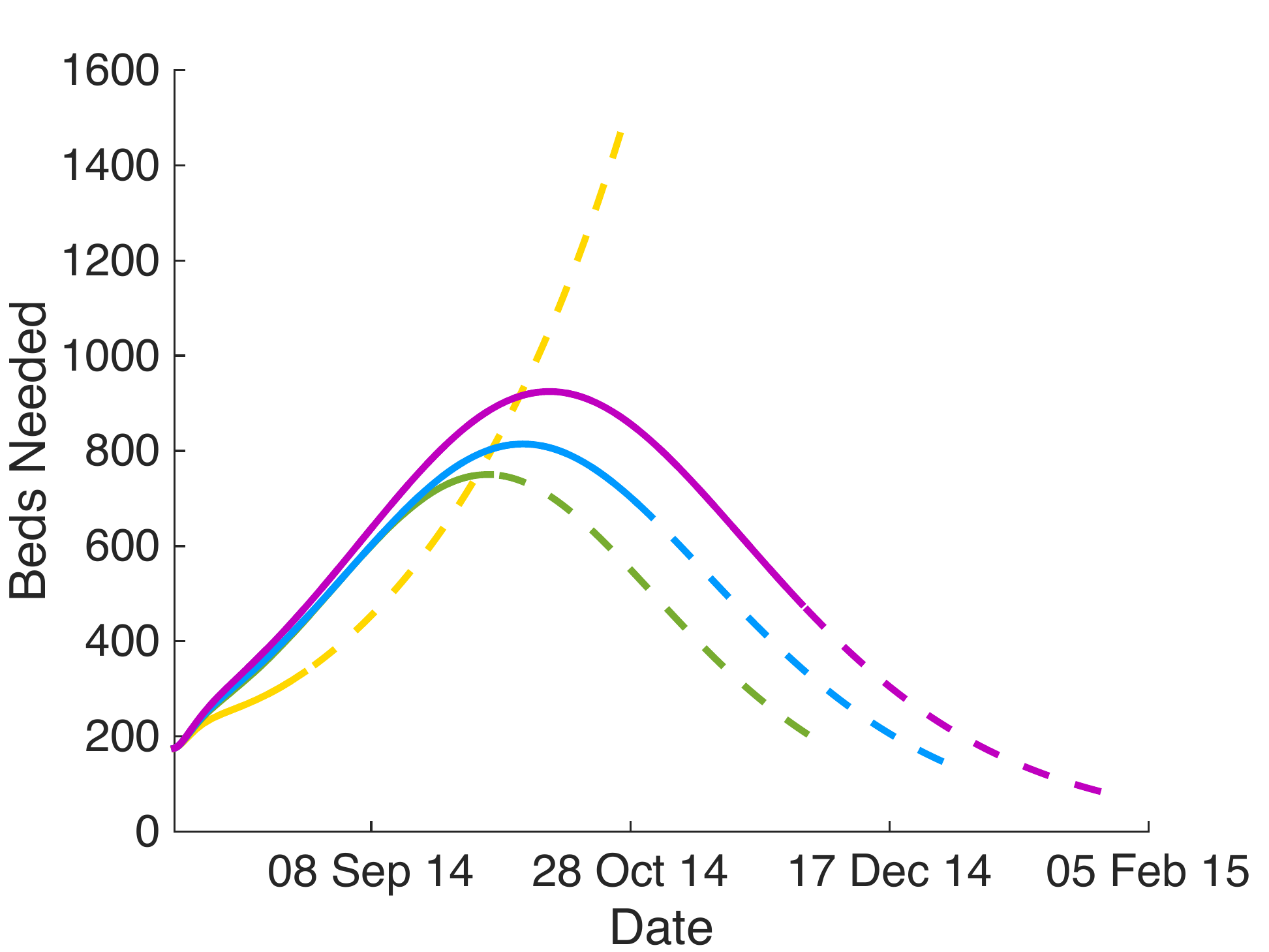}\\
\vspace{0.25cm}
\rule{2cm}{0.4pt} . Sierra Leone . \rule{2cm}{0.4pt}\\
\includegraphics[width=0.32\textwidth]{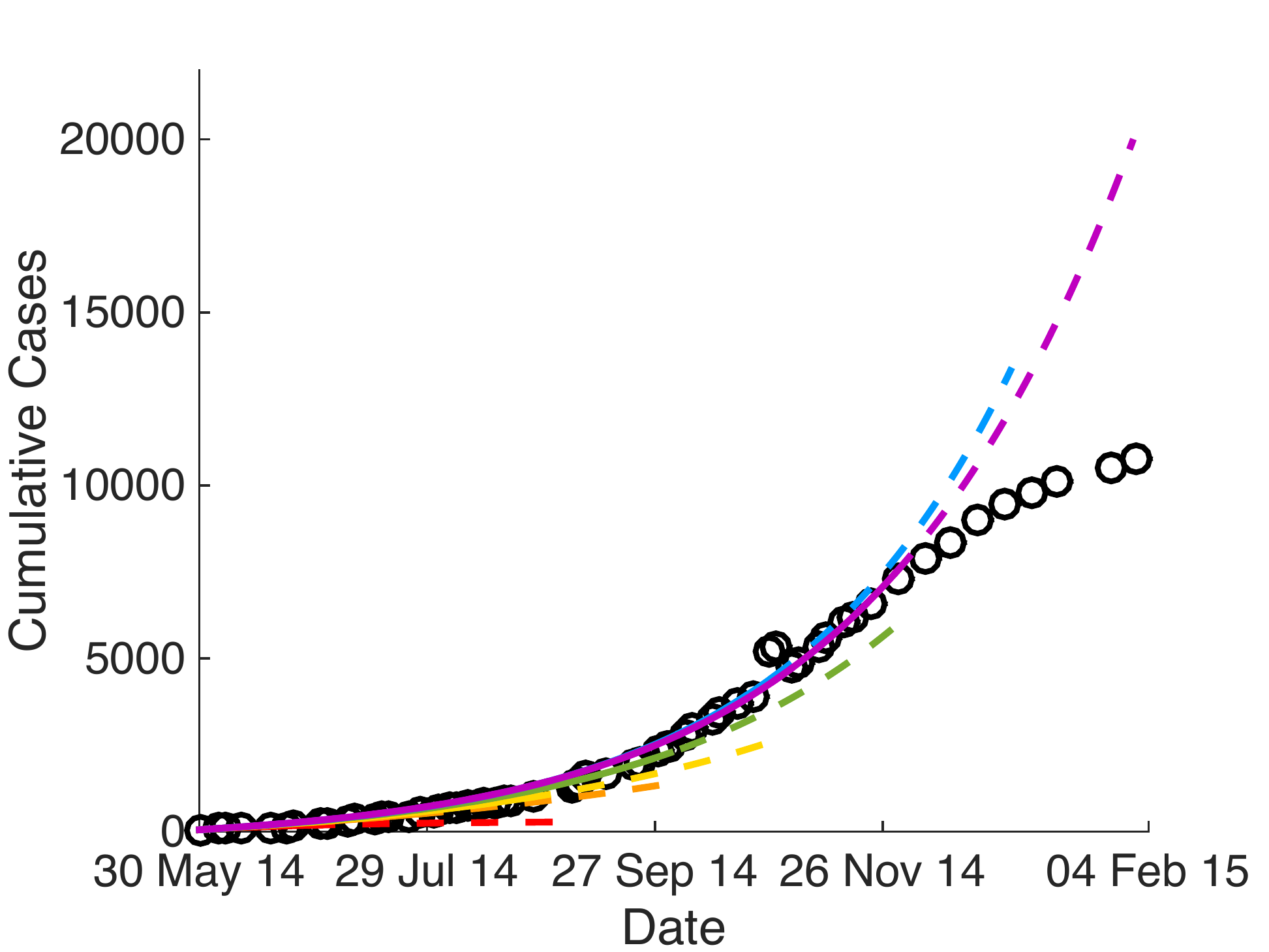}
\includegraphics[width=0.32\textwidth]{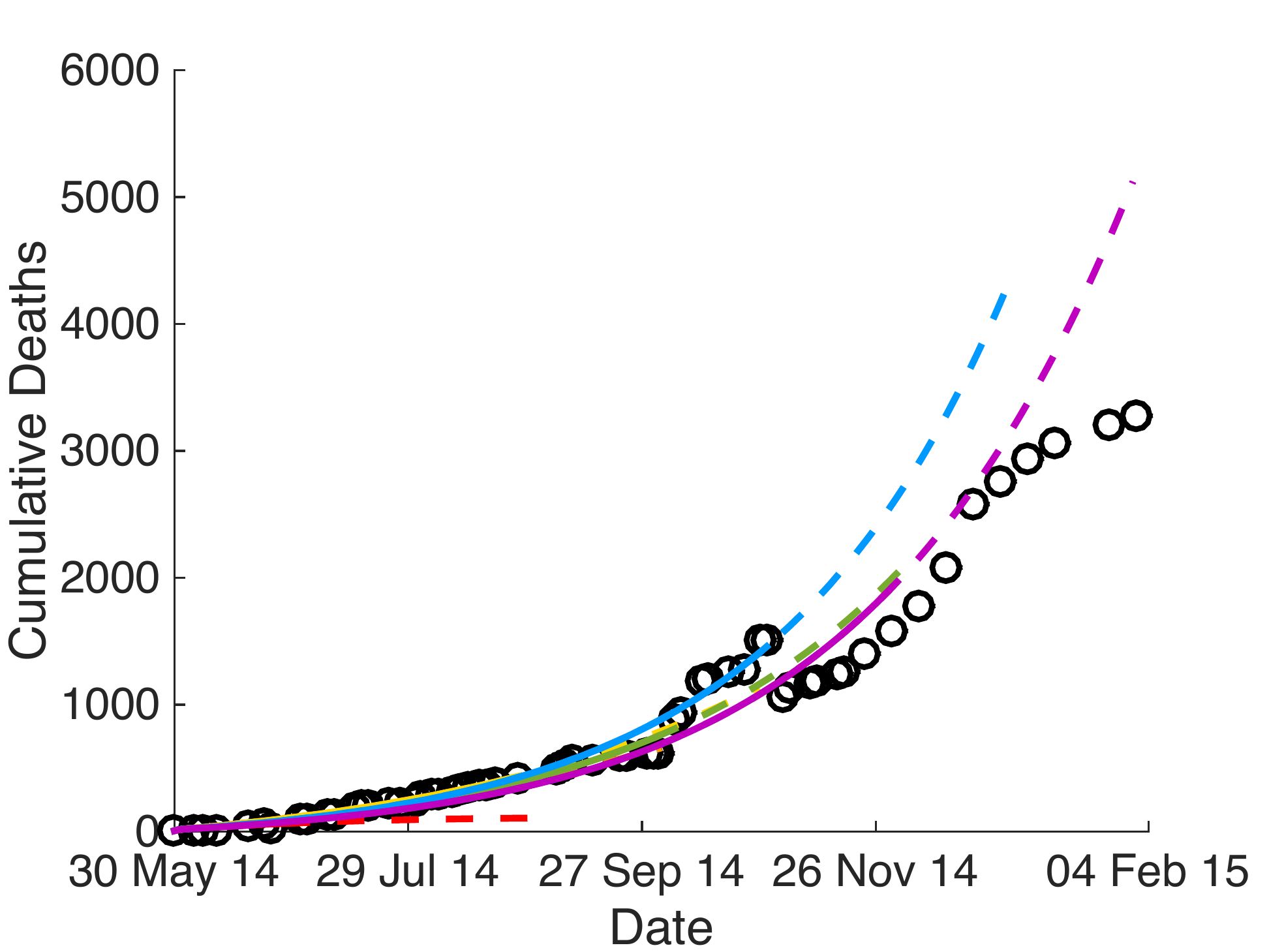}
\includegraphics[width=0.32\textwidth]{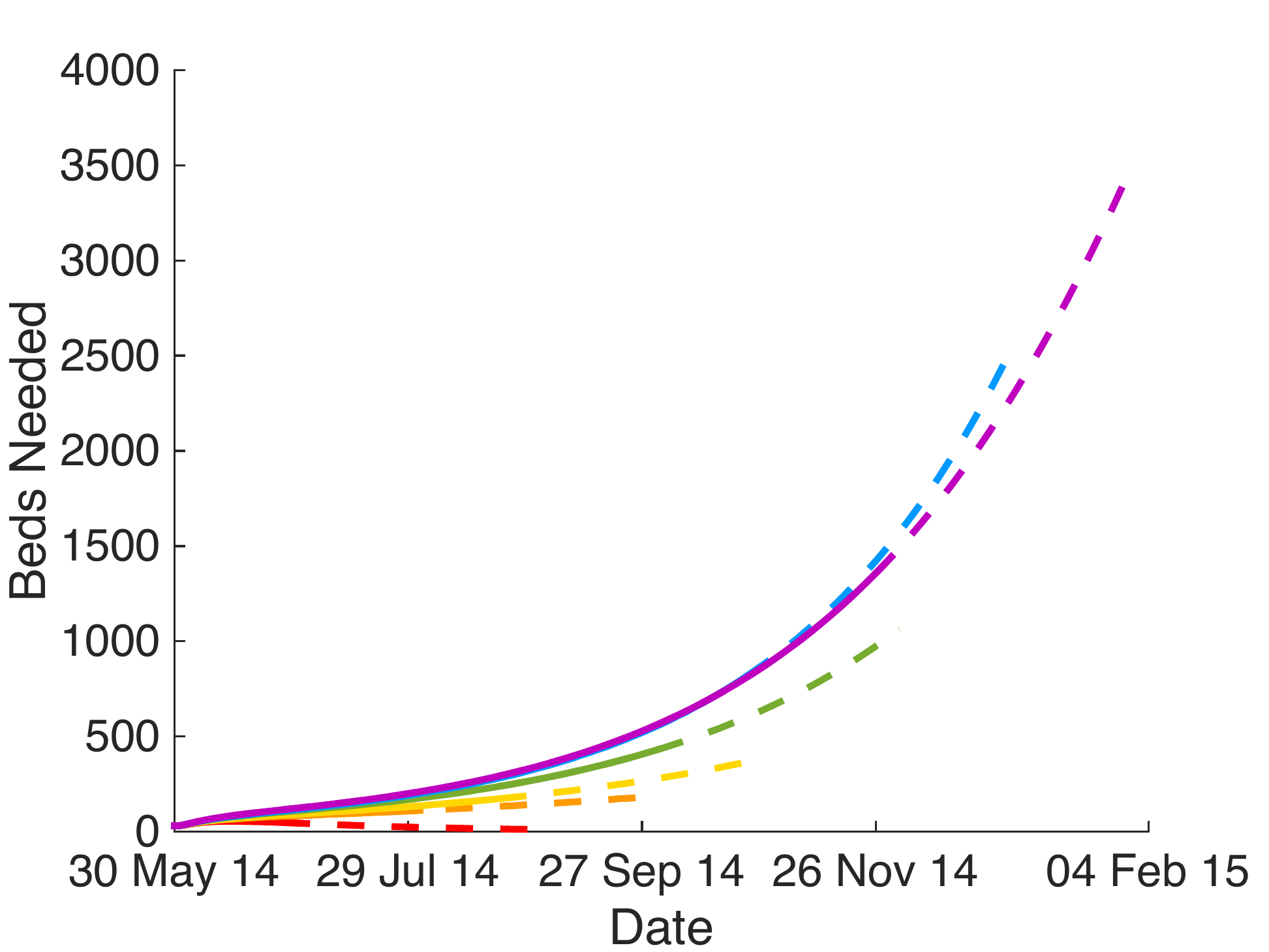}\\
\vspace{0.25cm}
\rule{2cm}{0.4pt} . All Countries Combined . \rule{2cm}{0.4pt}\\
\includegraphics[width=0.32\textwidth]{\figpath/MultiForecastCasesAll_k_flip_dates}
\includegraphics[width=0.32\textwidth]{\figpath/MultiForecastDeathsAll_k_flip_dates}
\includegraphics[width=0.32\textwidth]{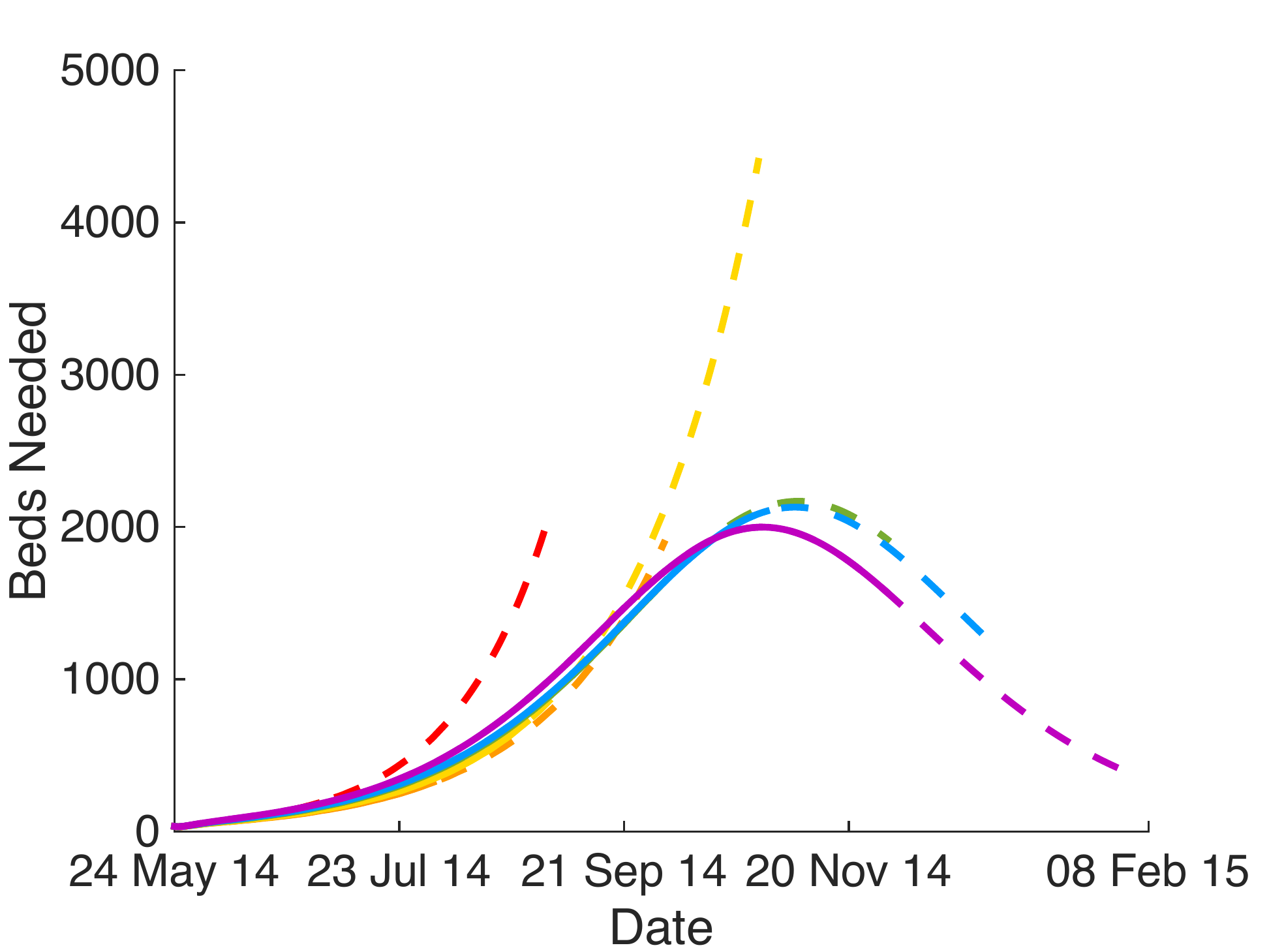}\\
\caption{Multiple model fits and forecasts for each country individually and all countries combined, with $\beta_1$, $\delta$, and $k$ fitted to data, and the remaining parameters fixed to the midpoints of the ranges in Table \ref{tab:params}. 
The model fits (solid lines) use the data up through July 1 (red), August 1 (orange), September 1 (yellow), October 1 (green), November 1 (blue) and December 1 (purple), with subsequent two months of forecasts shown as dashed lines in the same color. }
\label{fig:MultiForecast_k}
\end{figure}

\begin{figure}[h]
\centering
\textbf{Without $k$ (i.e. $k = 1$)}\\
\vspace{0.25cm}
\rule{2cm}{0.4pt} . Guinea . \rule{2cm}{0.4pt}\\
\includegraphics[width=0.32\textwidth]{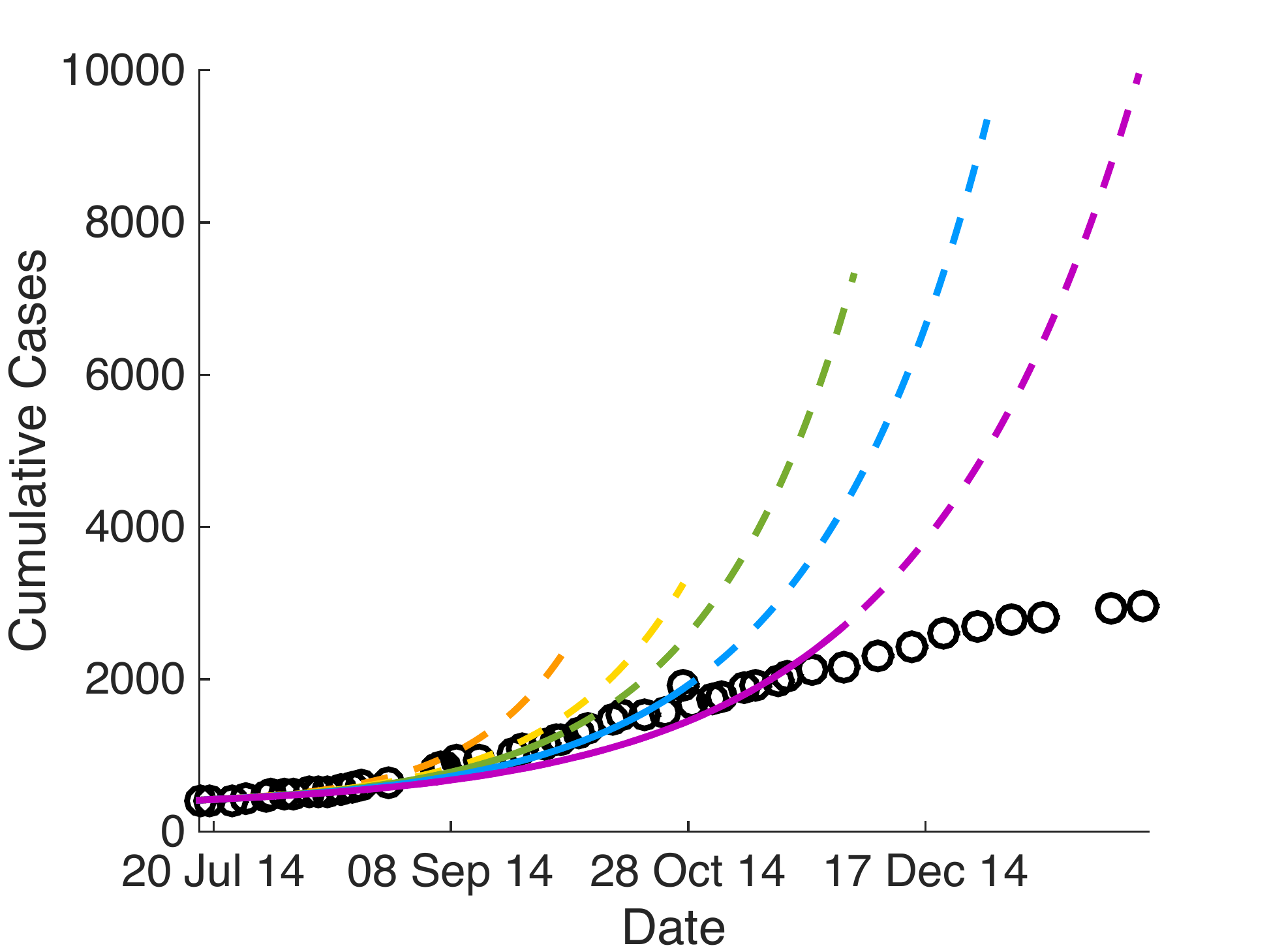}
\includegraphics[width=0.32\textwidth]{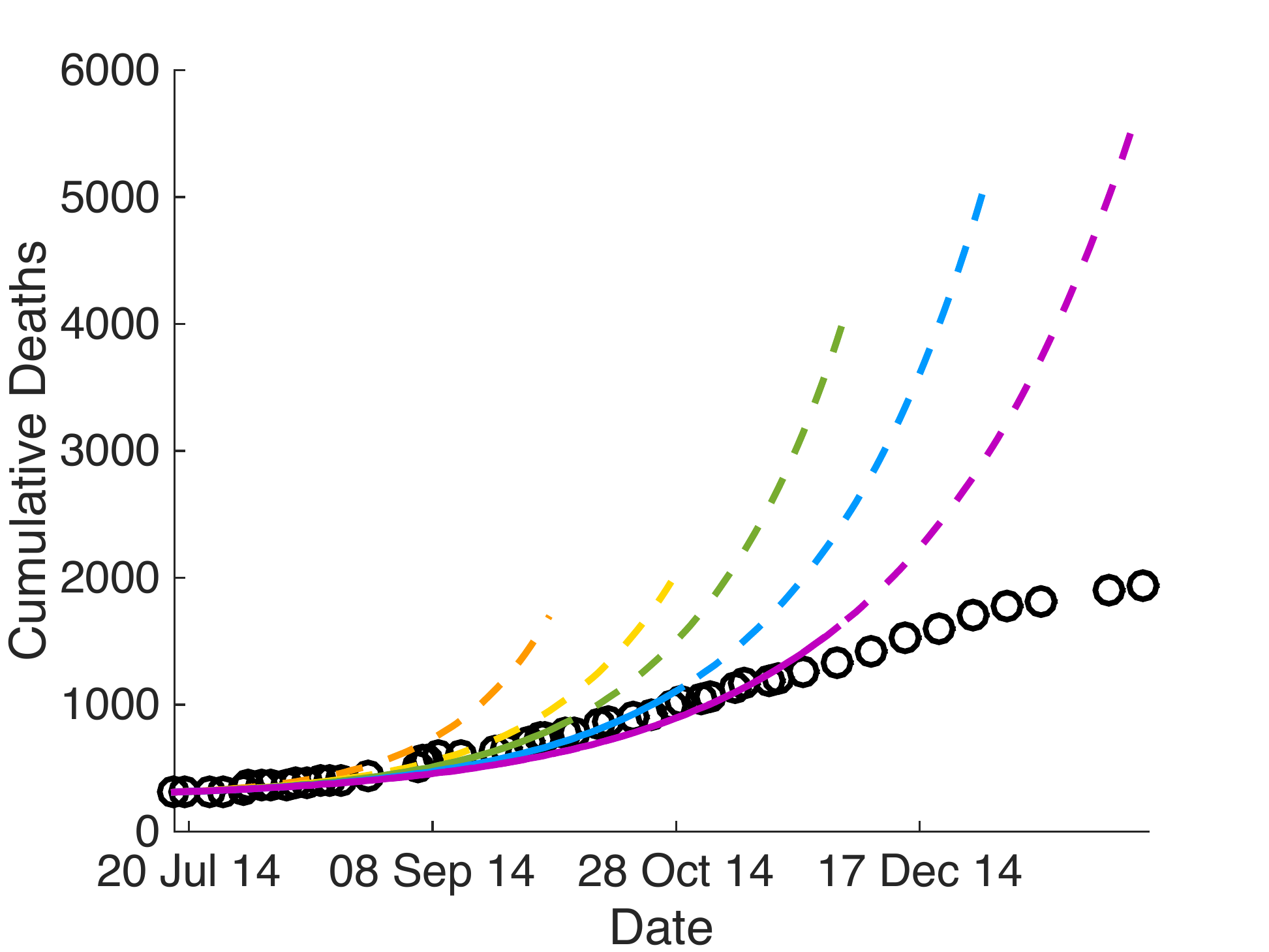}
\includegraphics[width=0.32\textwidth]{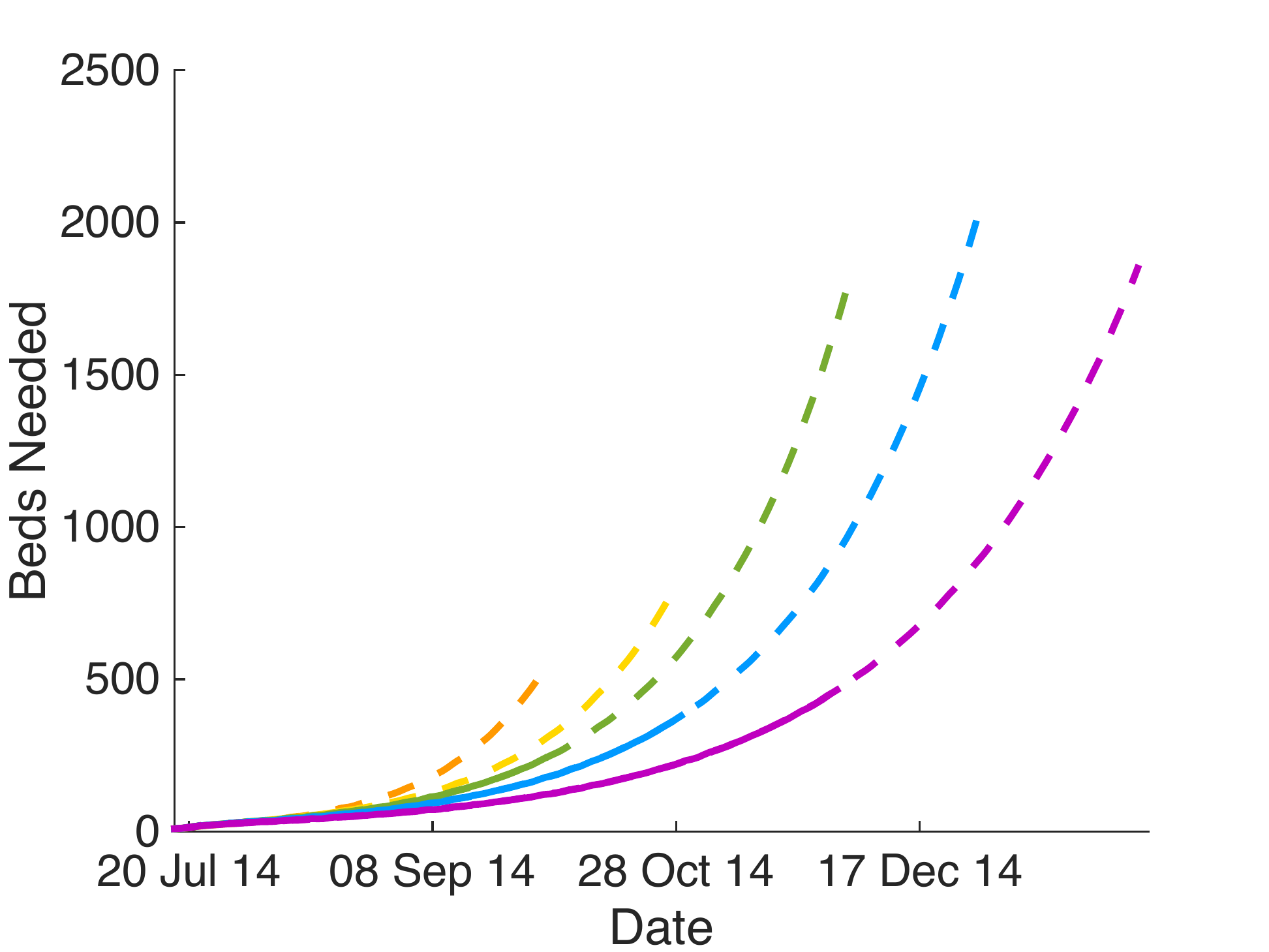}\\
\vspace{0.25cm}
\rule{2cm}{0.4pt} . Liberia . \rule{2cm}{0.4pt}\\
\includegraphics[width=0.32\textwidth]{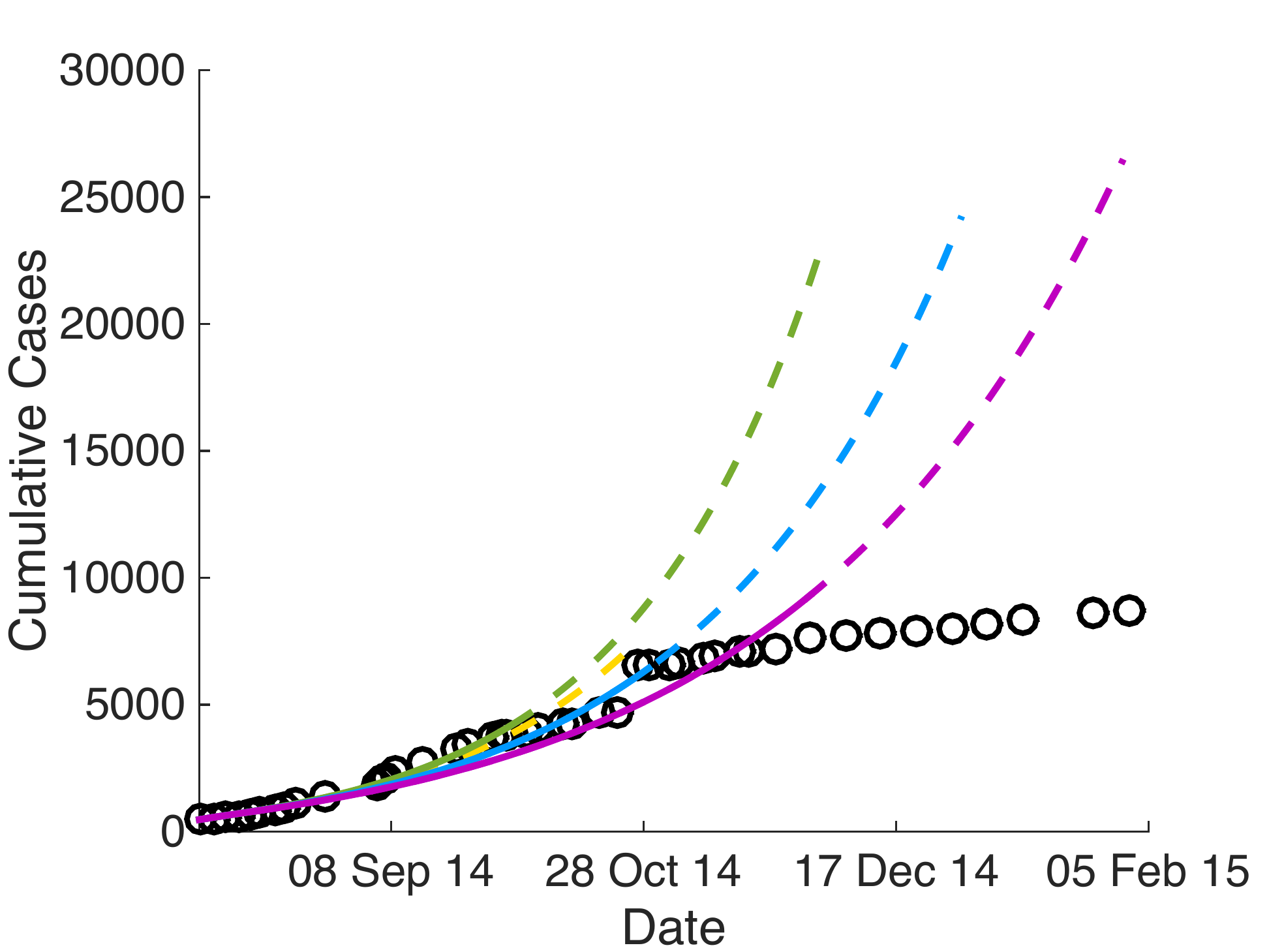}
\includegraphics[width=0.32\textwidth]{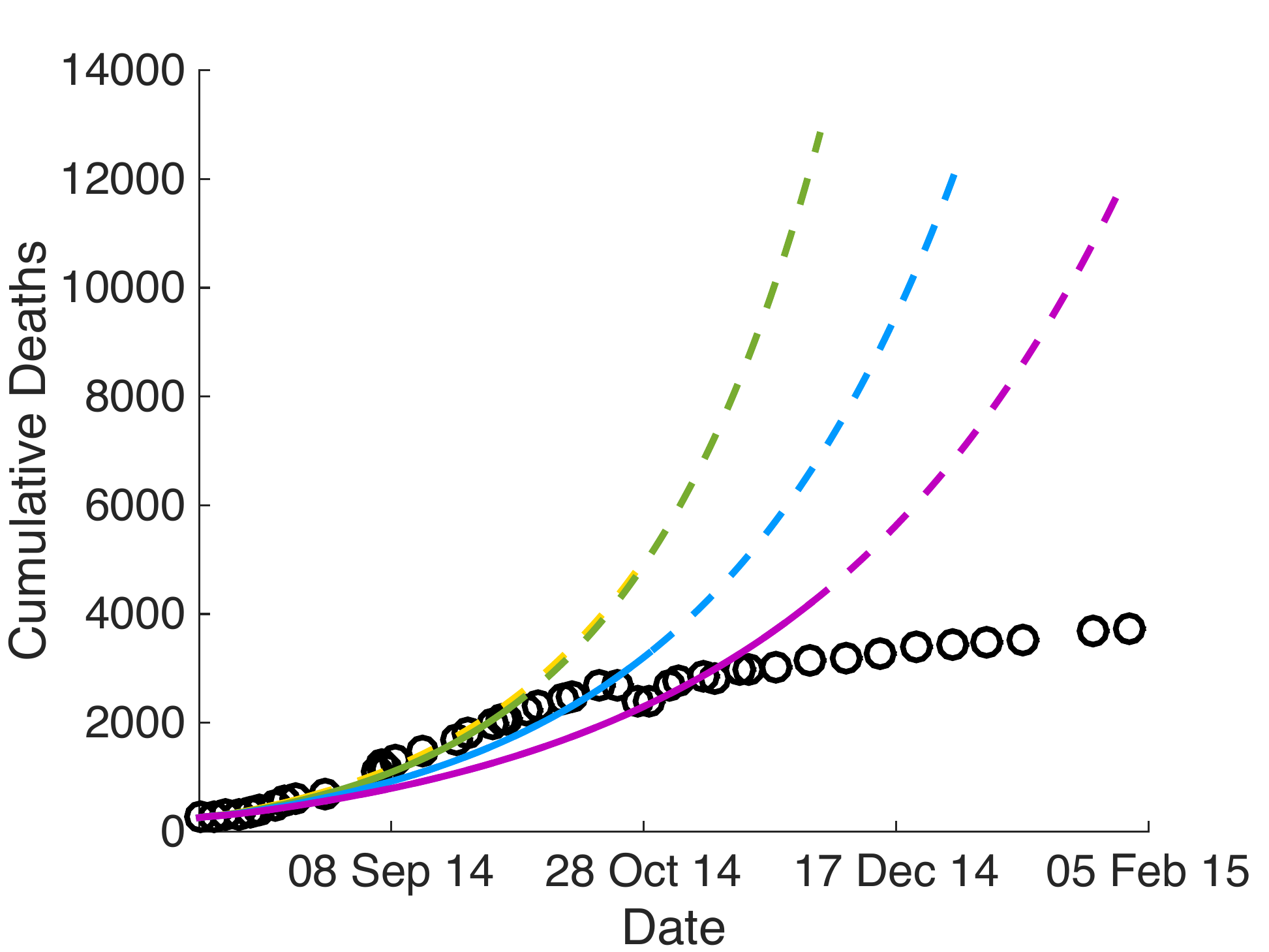}
\includegraphics[width=0.32\textwidth]{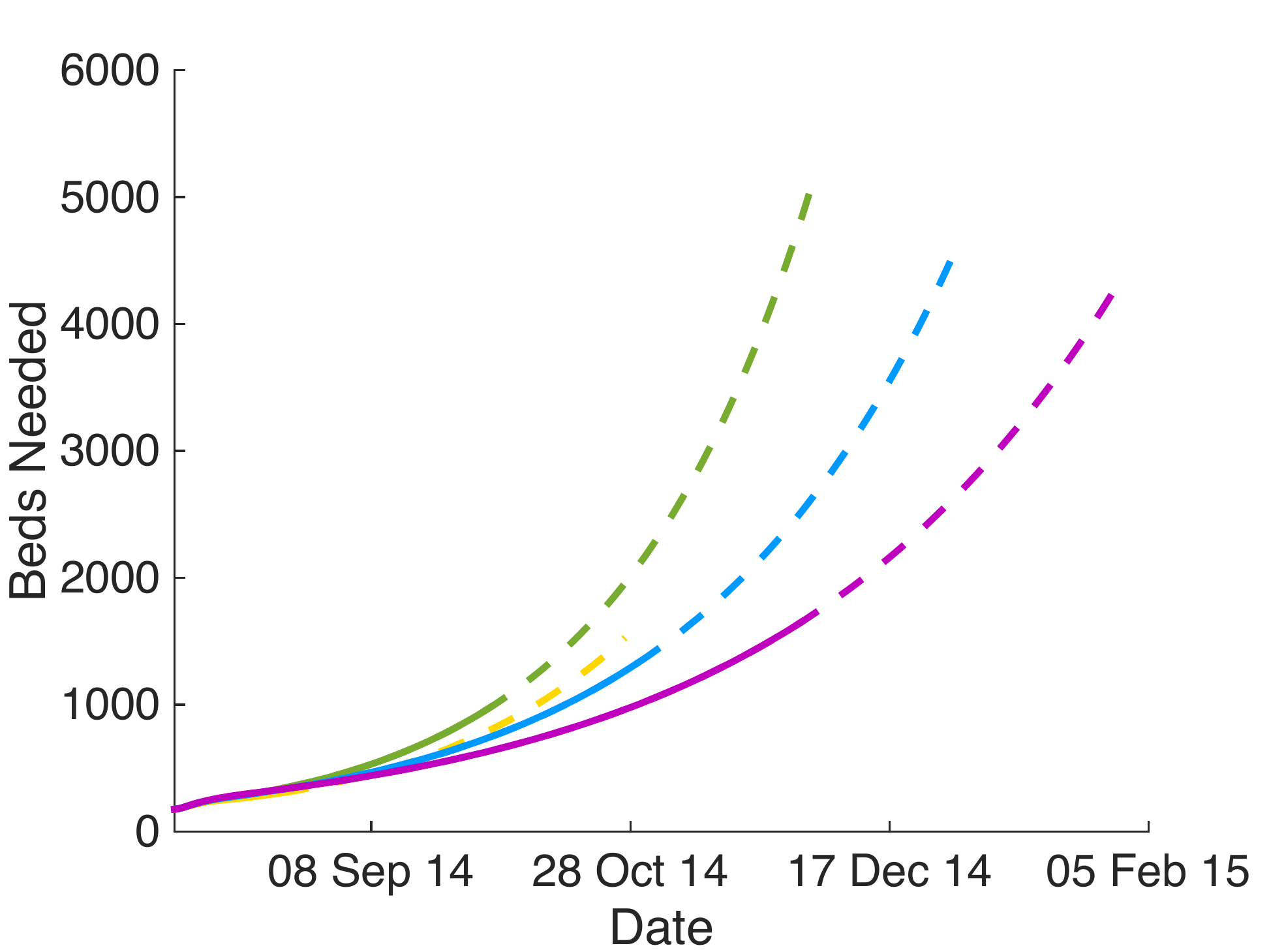}\\
\vspace{0.25cm}
\rule{2cm}{0.4pt} . Sierra Leone . \rule{2cm}{0.4pt}\\
\includegraphics[width=0.32\textwidth]{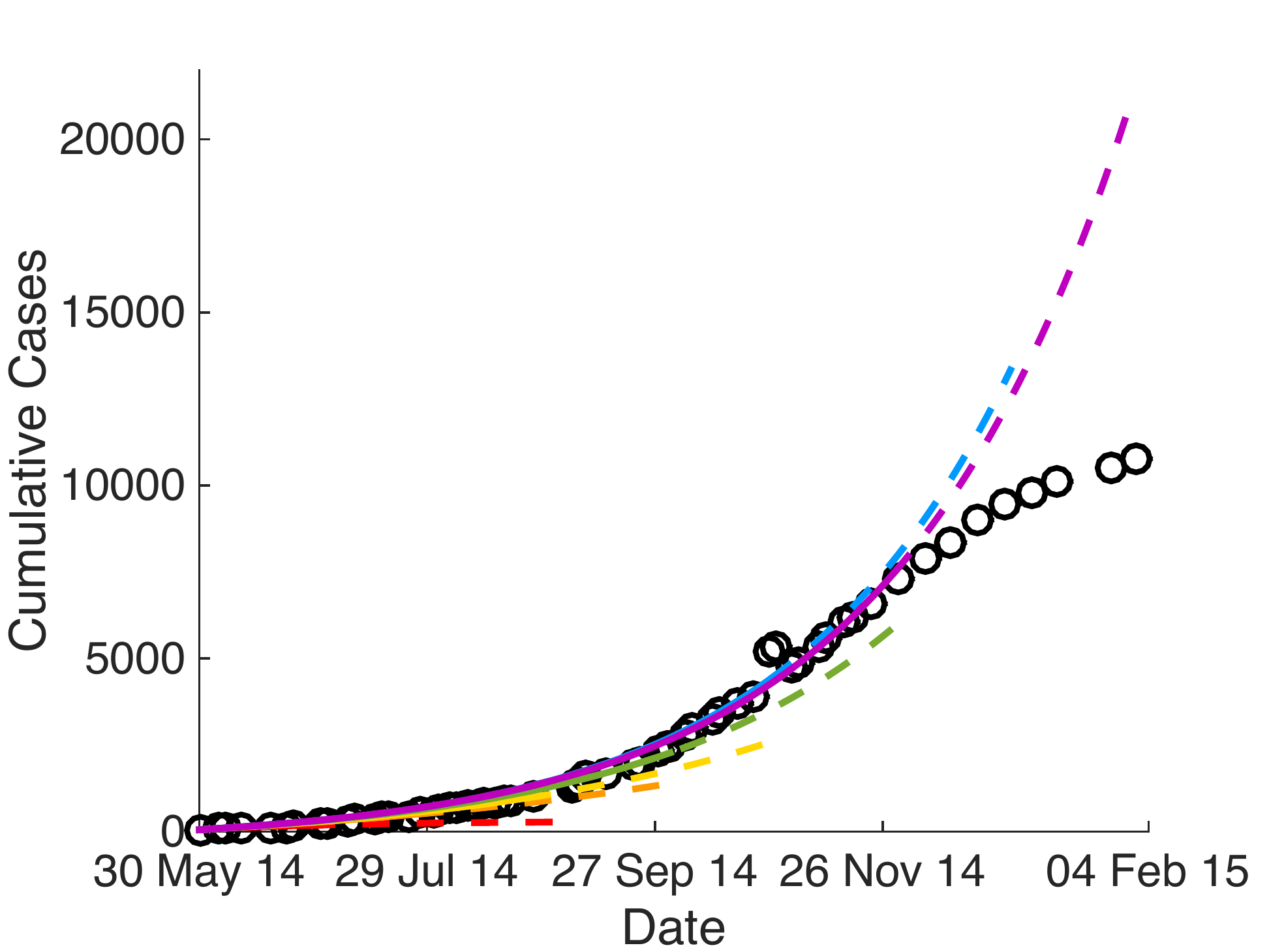}
\includegraphics[width=0.32\textwidth]{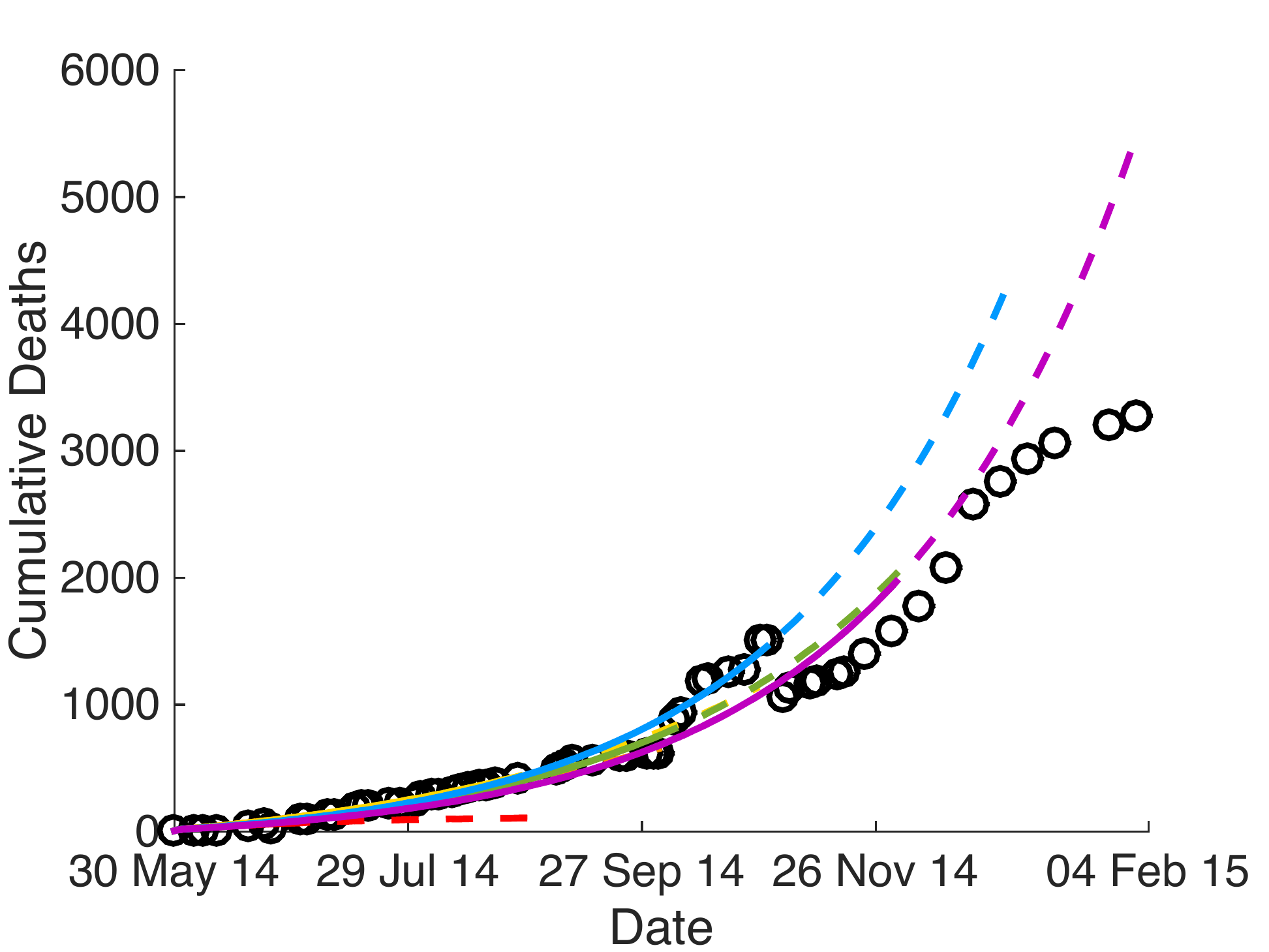}
\includegraphics[width=0.32\textwidth]{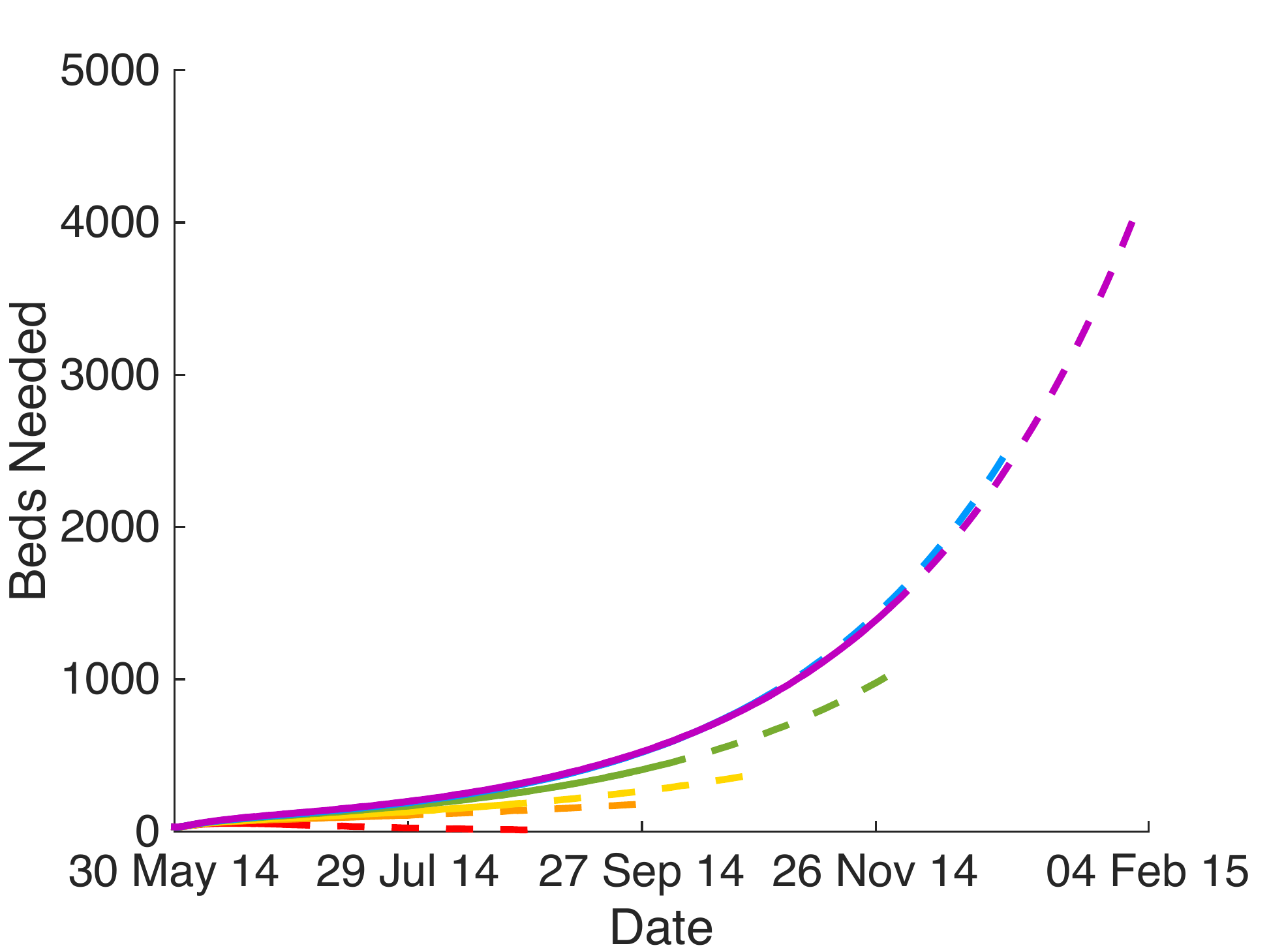}\\
\vspace{0.25cm}
\rule{2cm}{0.4pt} . All Countries Combined . \rule{2cm}{0.4pt}\\
\includegraphics[width=0.32\textwidth]{\figpath/MultiForecastCasesAll_nok_flip_dates}
\includegraphics[width=0.32\textwidth]{\figpath/MultiForecastDeathsAll_nok_flip_dates}
\includegraphics[width=0.32\textwidth]{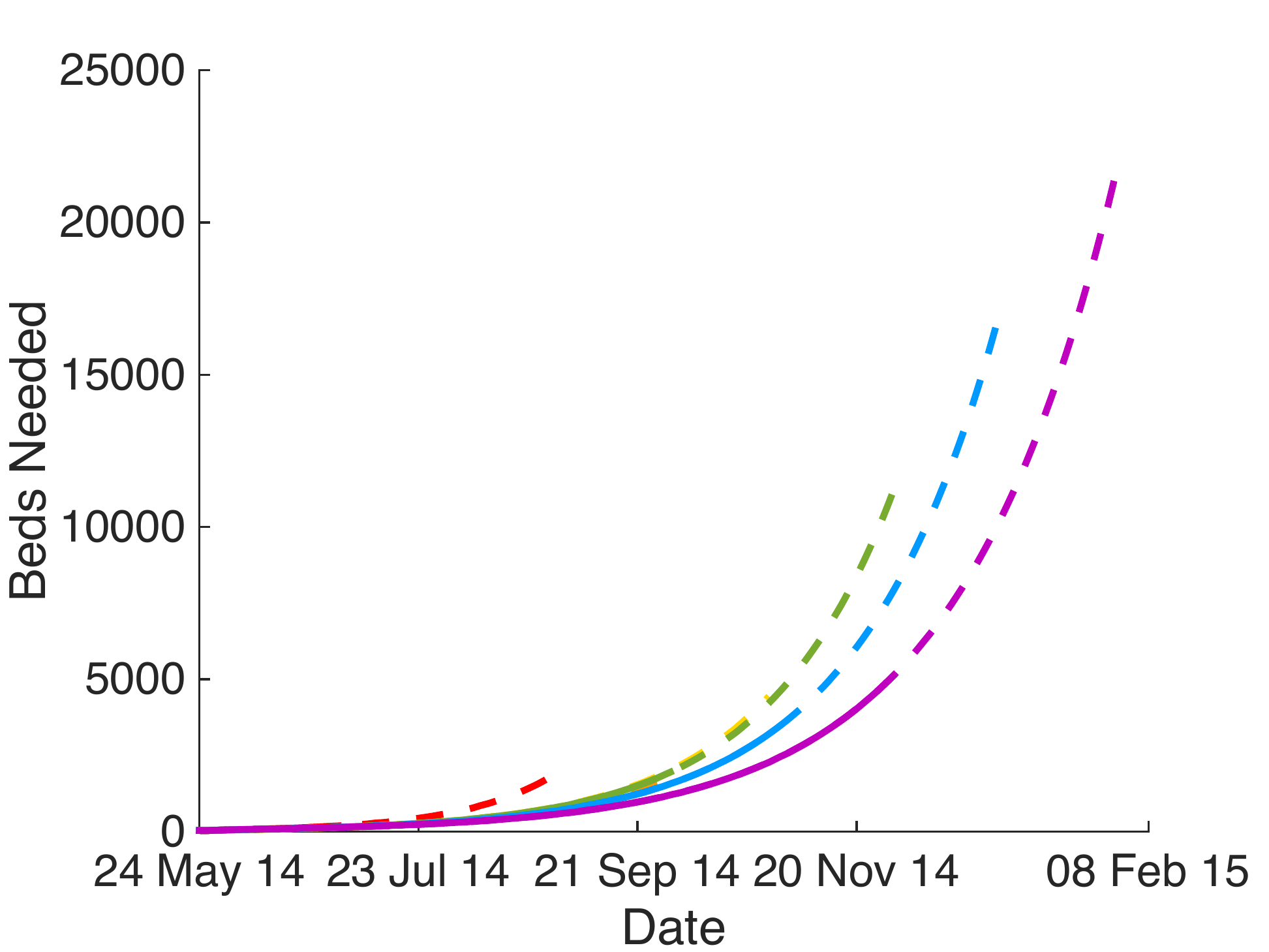}\\
\caption{Multiple model fits and forecasts for each country individually and all countries combined, with $\beta_1$ and $\delta$ fitted to data, and $k$ fixed to 1 (equivalent to not including $k$ in the model). The remaining parameters fixed to the midpoints of the ranges in Table \ref{tab:params}. 
The model fits (solid lines) use the data up through July 1 (red), August 1 (orange), September 1 (yellow), October 1 (green), November 1 (blue) and December 1 (purple), with subsequent two months of forecasts shown as dashed lines in the same color. Note the difference in y-axis scale from Figure \ref{fig:MultiForecast_k}.}
\label{fig:MultiForecast_nok}
\end{figure}

\begin{figure}[h]
\centering
\textbf{With $k$ fixed to $1/2.5$}\\
\vspace{0.25cm}
\rule{2cm}{0.4pt} . Guinea . \rule{2cm}{0.4pt}\\
\includegraphics[width=0.32\textwidth]{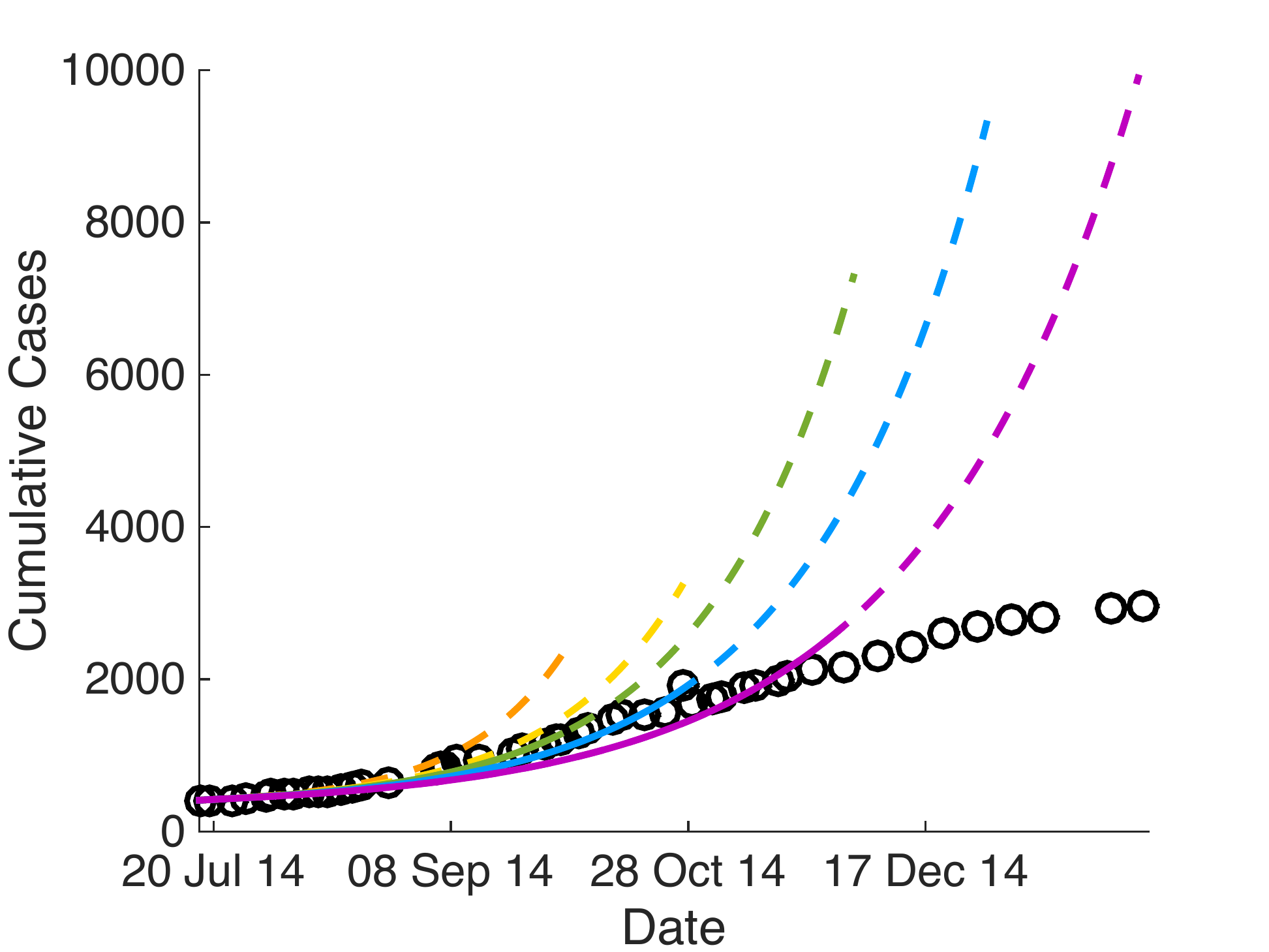}
\includegraphics[width=0.32\textwidth]{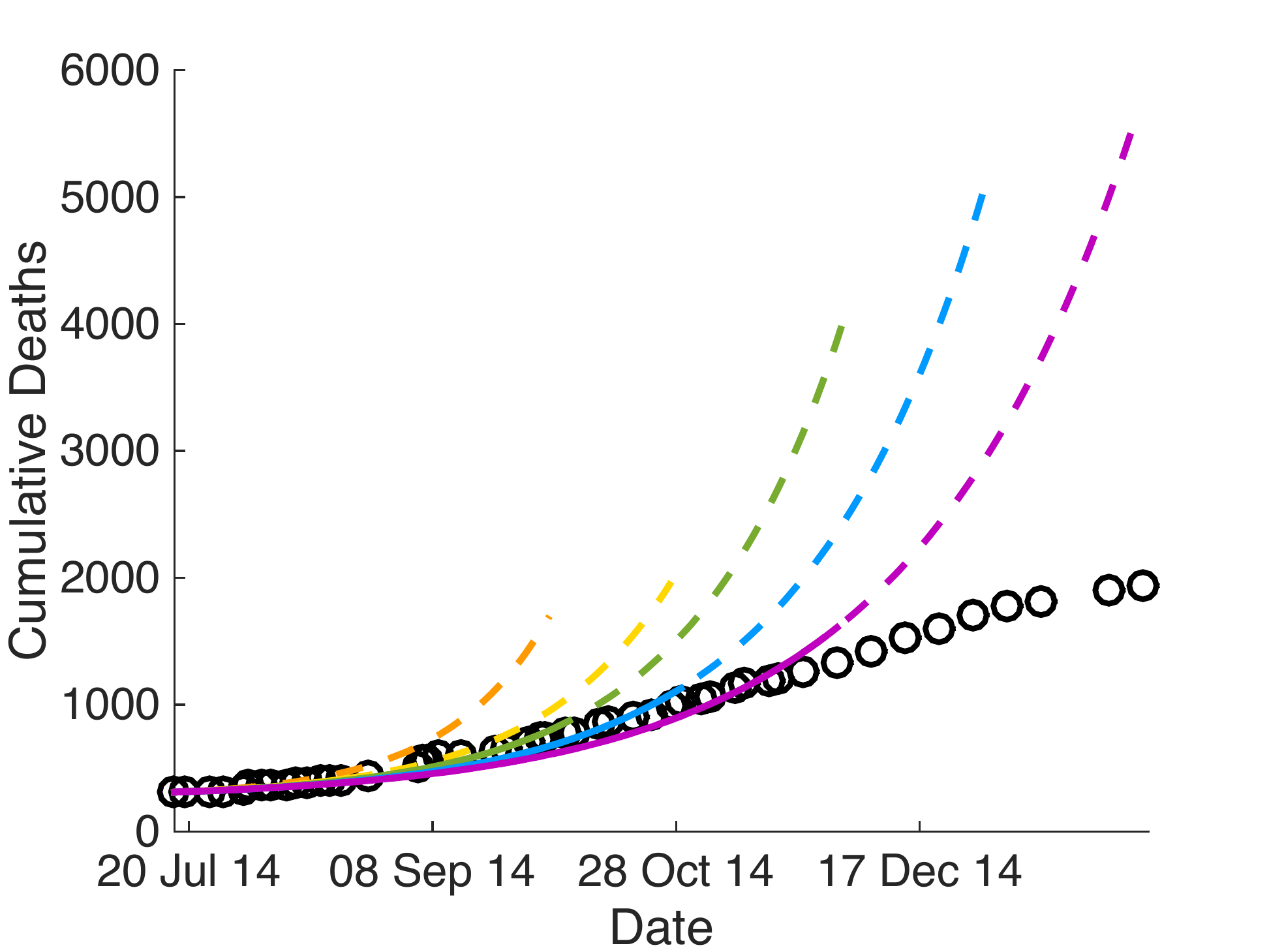}
\includegraphics[width=0.32\textwidth]{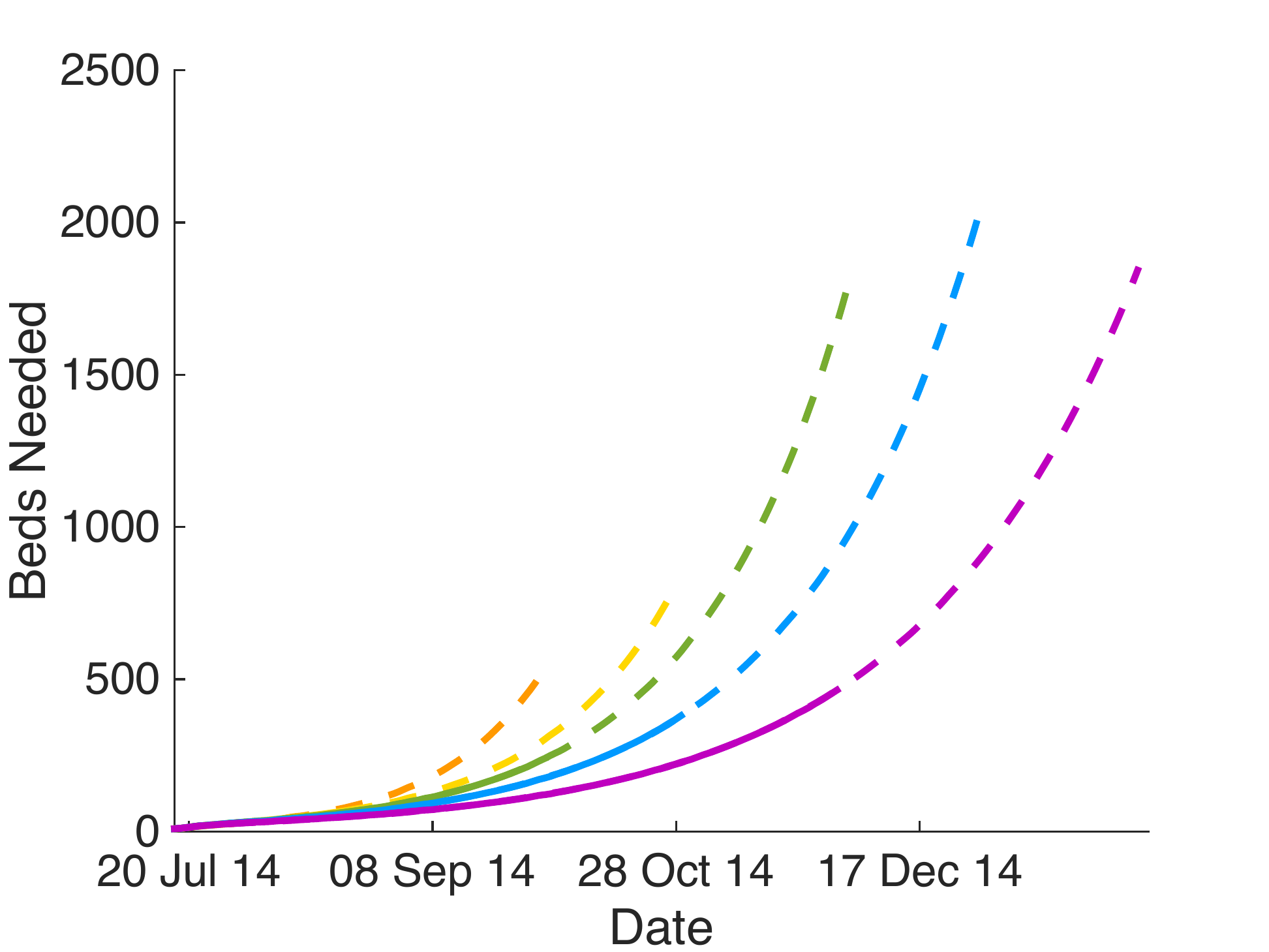}\\
\vspace{0.25cm}
\rule{2cm}{0.4pt} . Liberia . \rule{2cm}{0.4pt}\\
\includegraphics[width=0.32\textwidth]{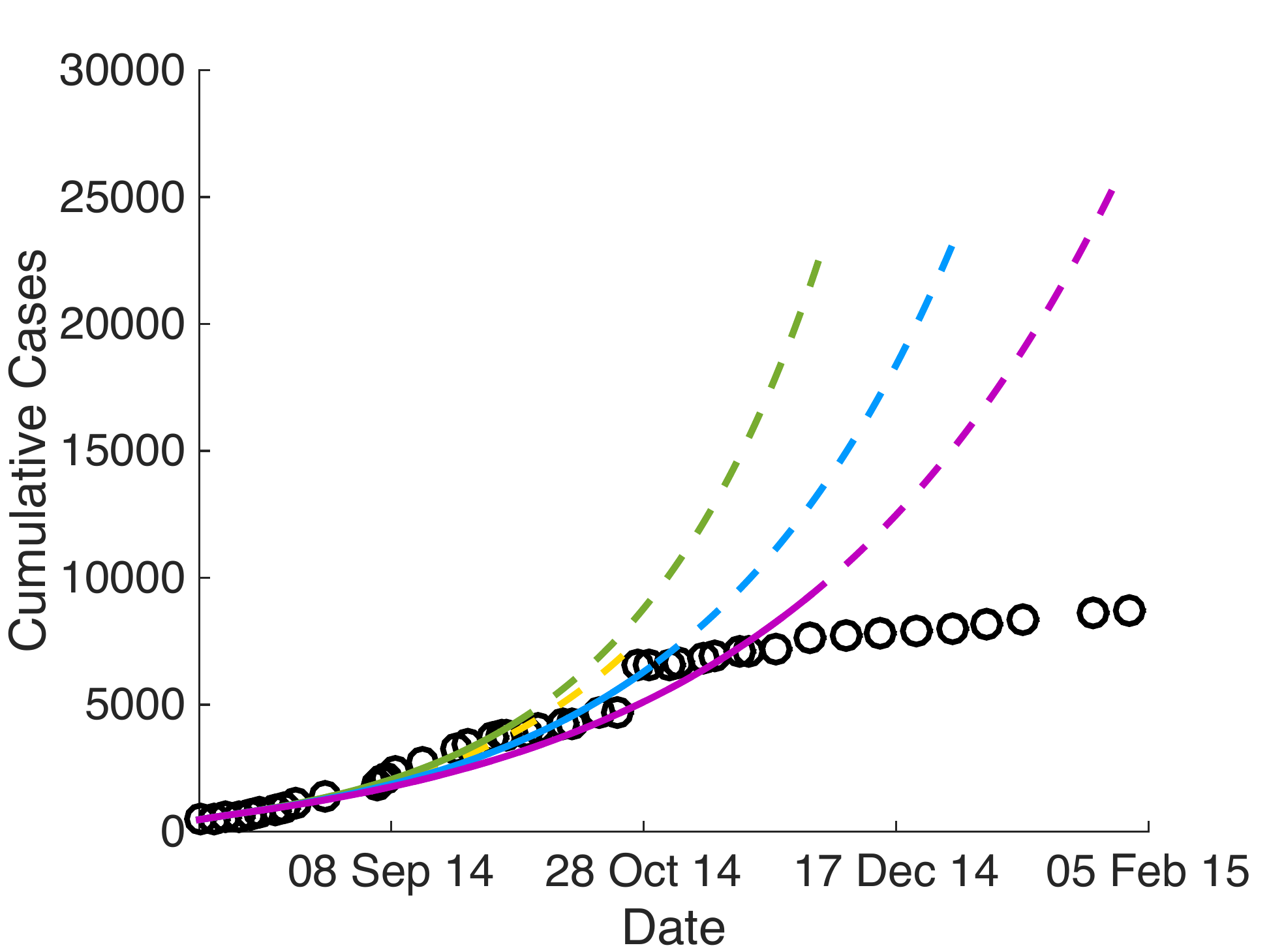}
\includegraphics[width=0.32\textwidth]{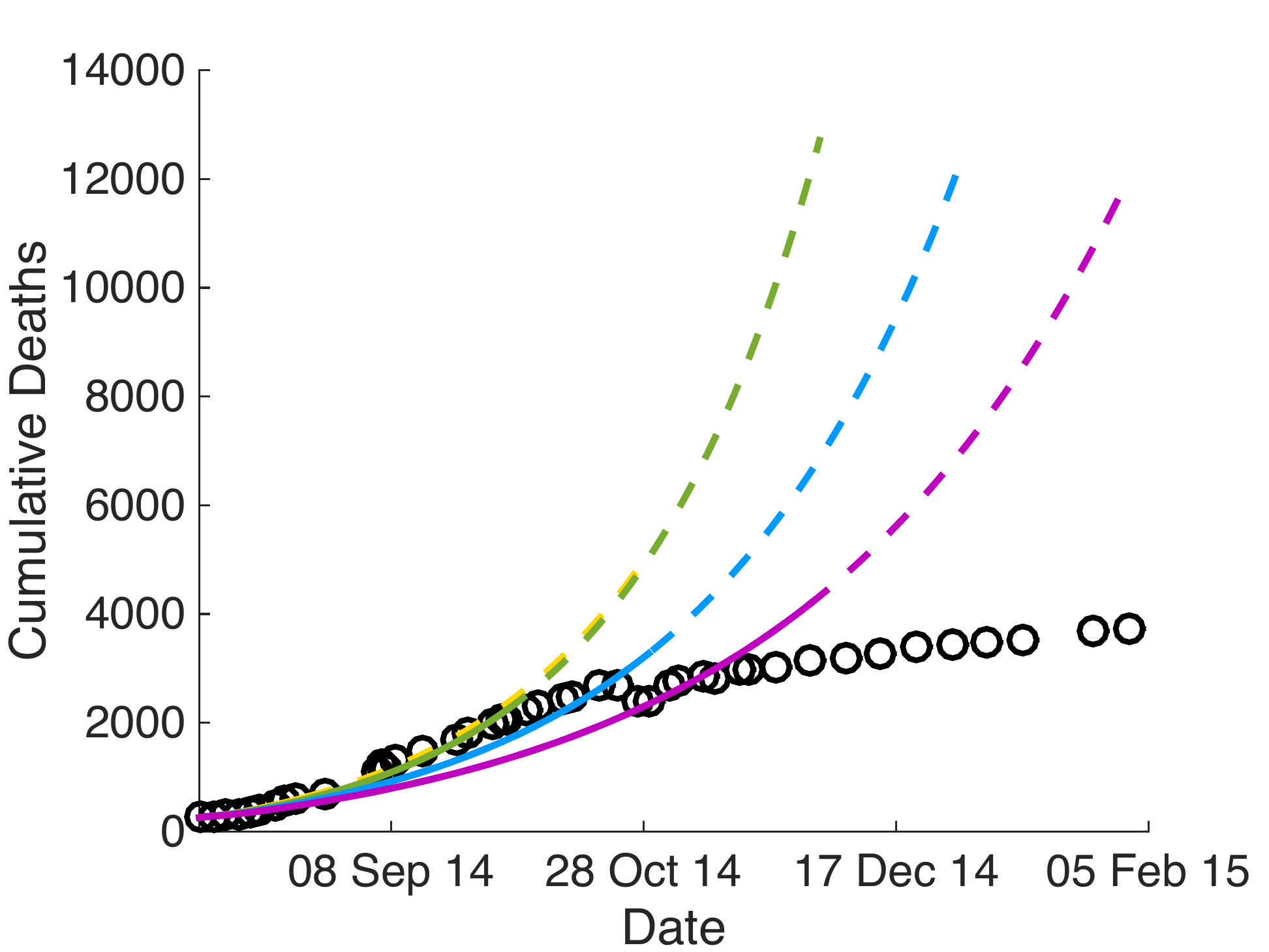}
\includegraphics[width=0.32\textwidth]{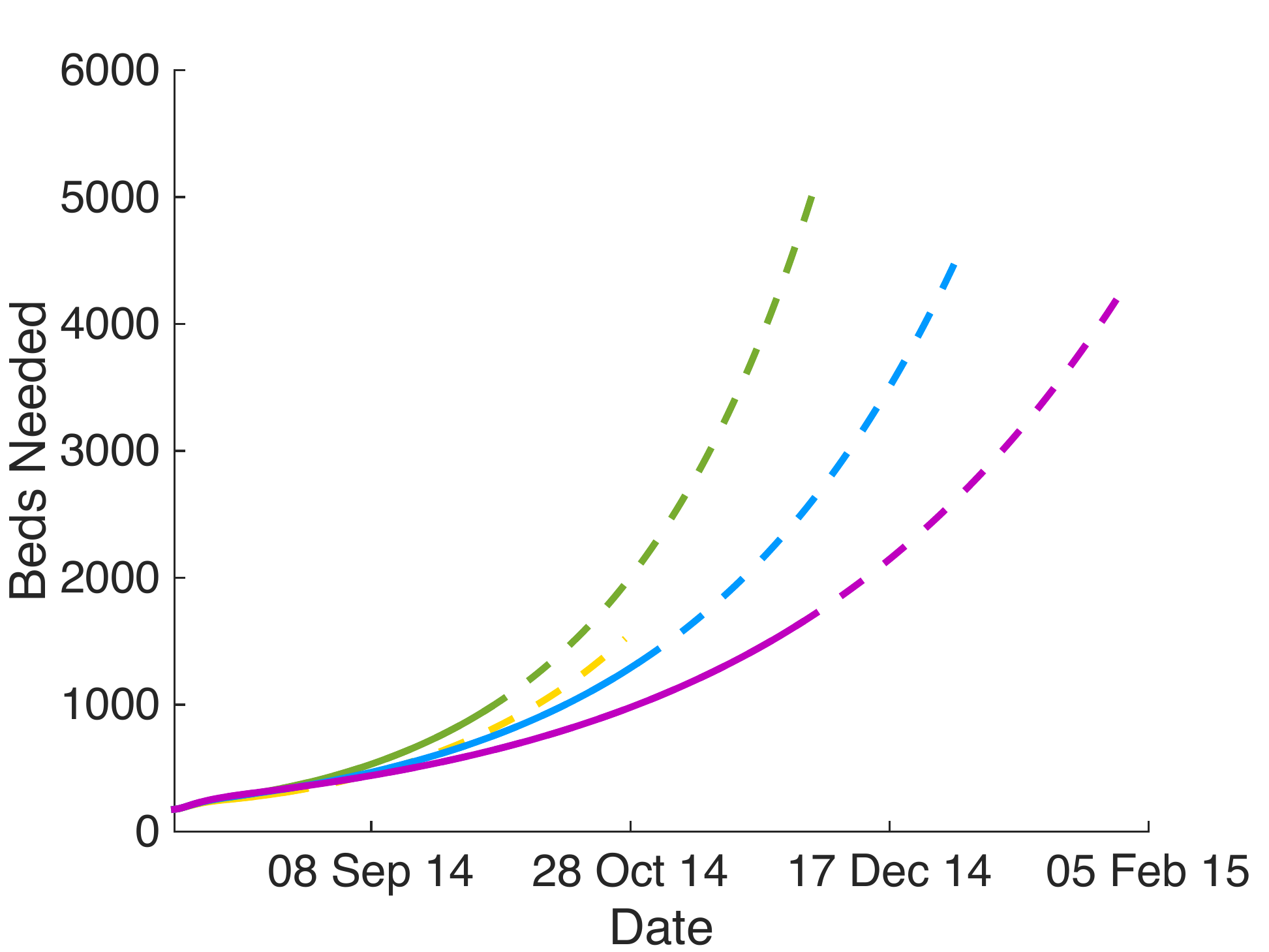}\\
\vspace{0.25cm}
\rule{2cm}{0.4pt} . Sierra Leone . \rule{2cm}{0.4pt}\\
\includegraphics[width=0.32\textwidth]{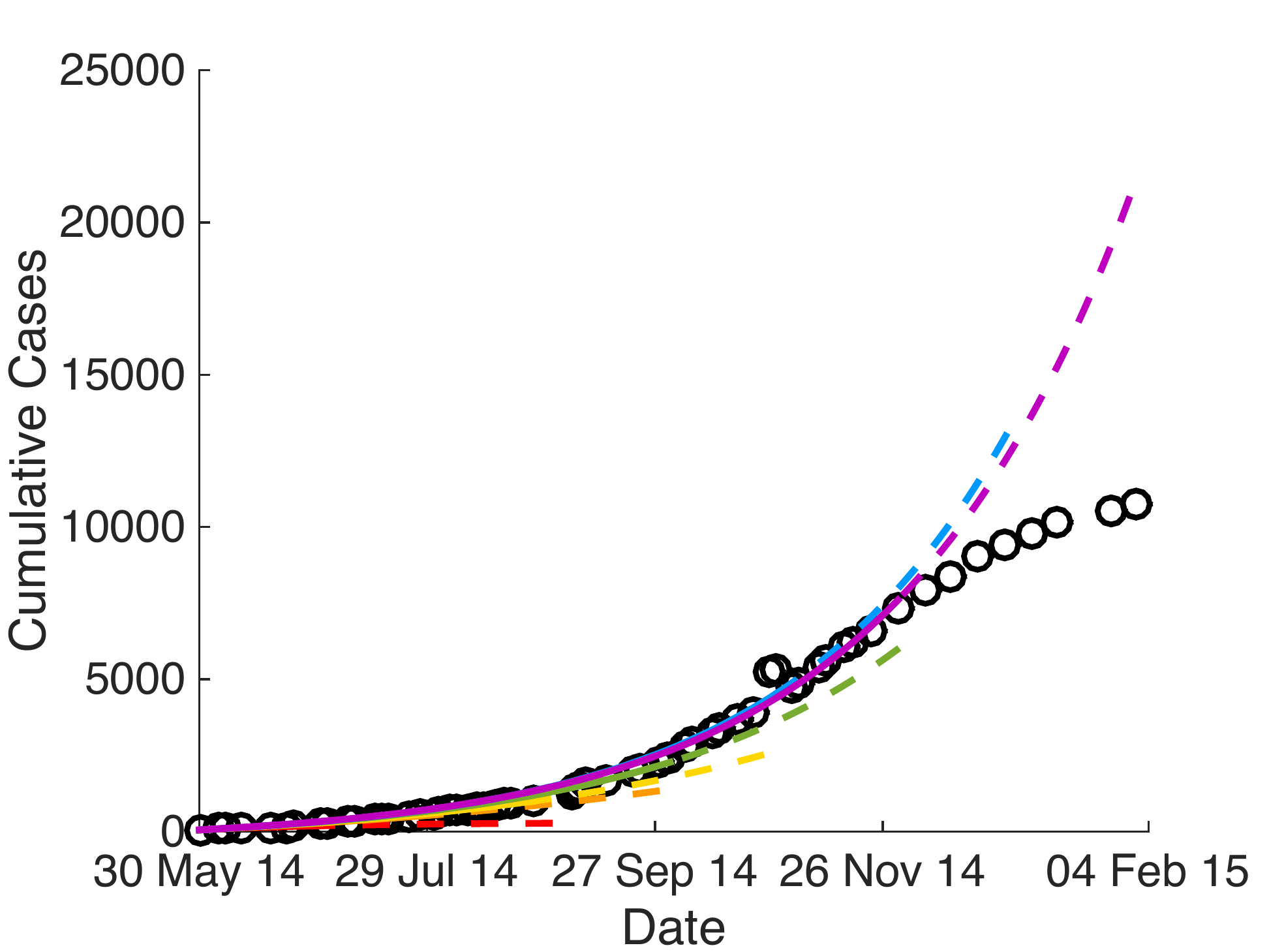}
\includegraphics[width=0.32\textwidth]{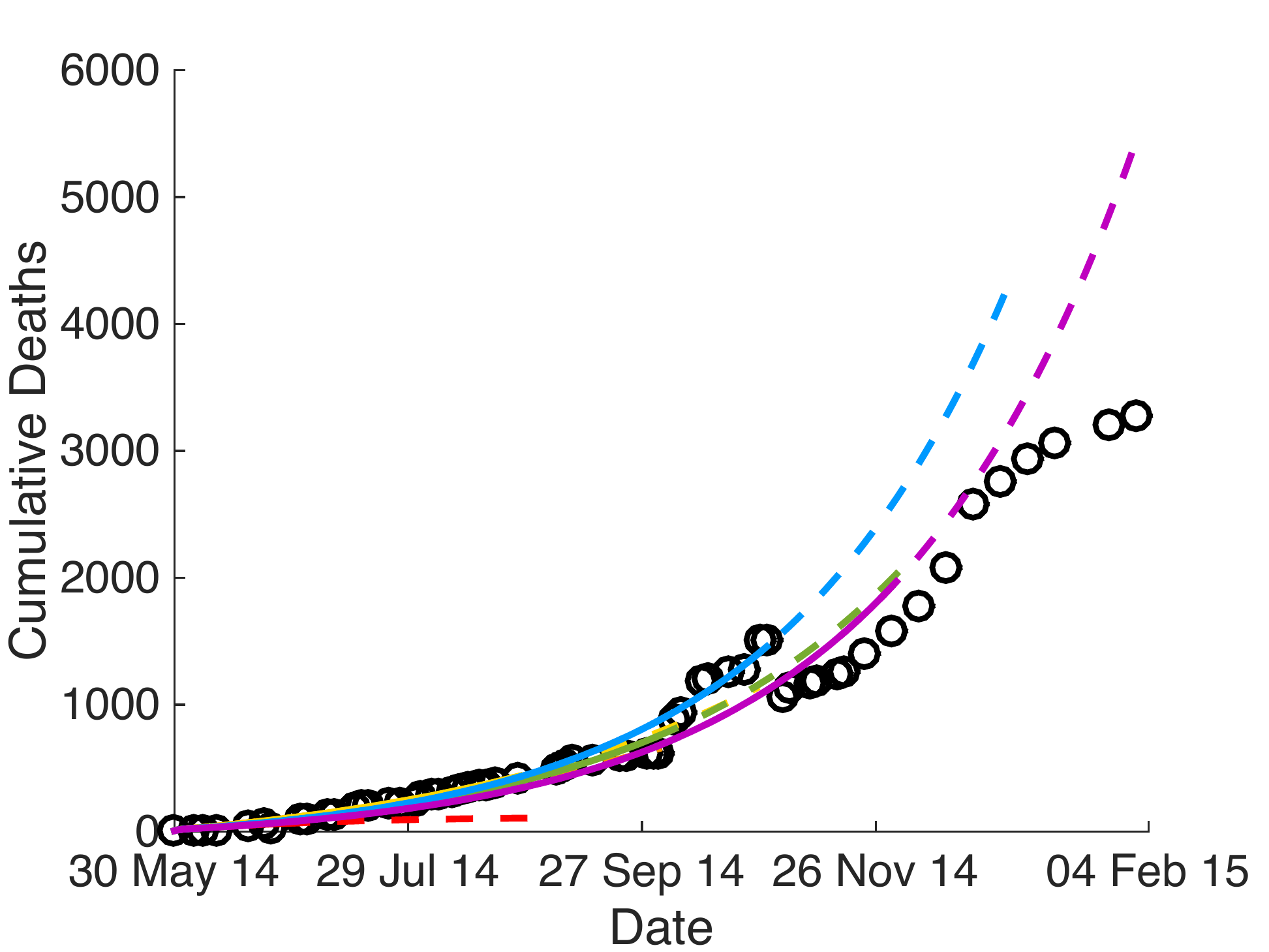}
\includegraphics[width=0.32\textwidth]{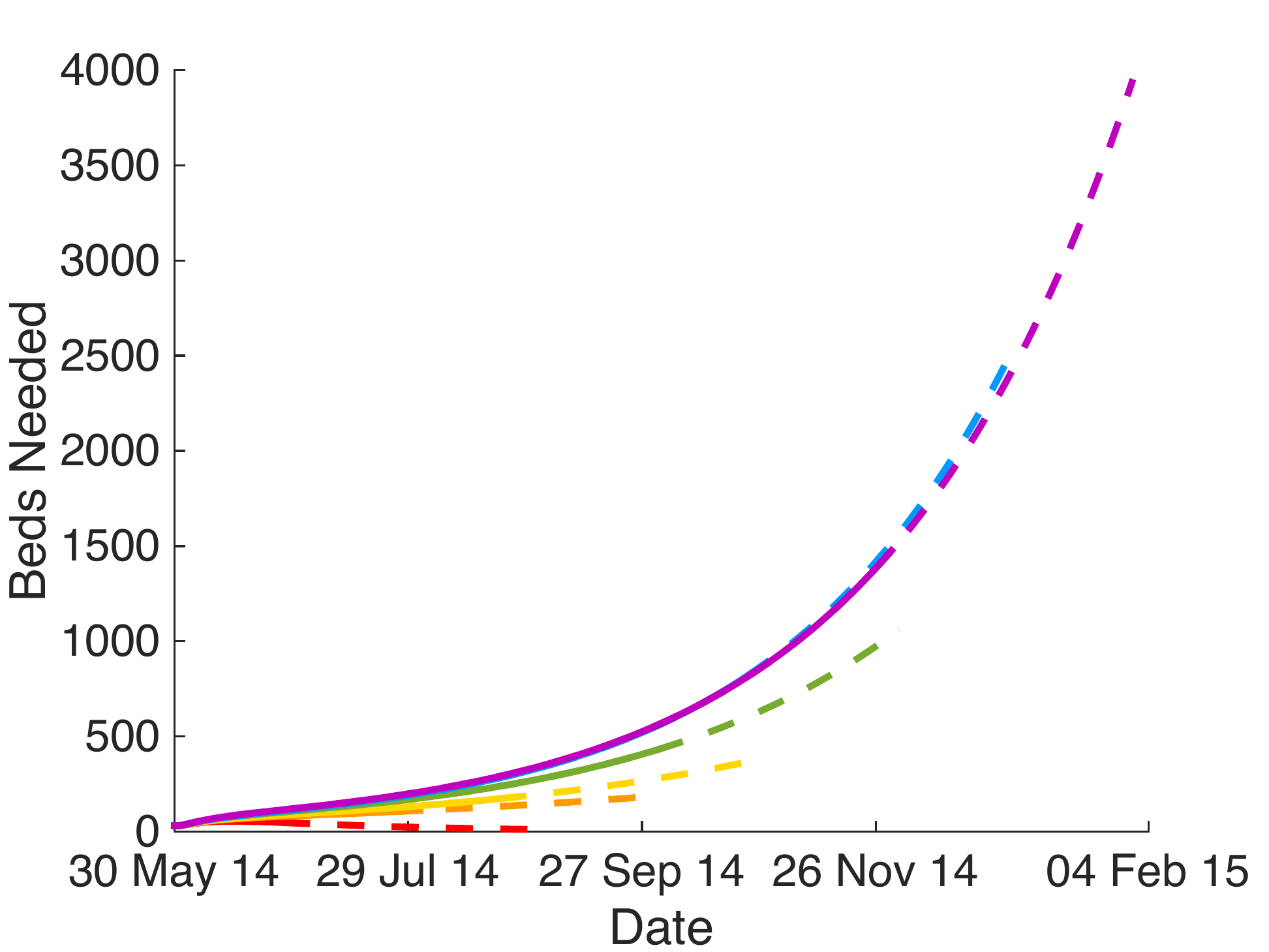}\\
\vspace{0.25cm}
\rule{2cm}{0.4pt} . All Countries Combined . \rule{2cm}{0.4pt}\\
\includegraphics[width=0.32\textwidth]{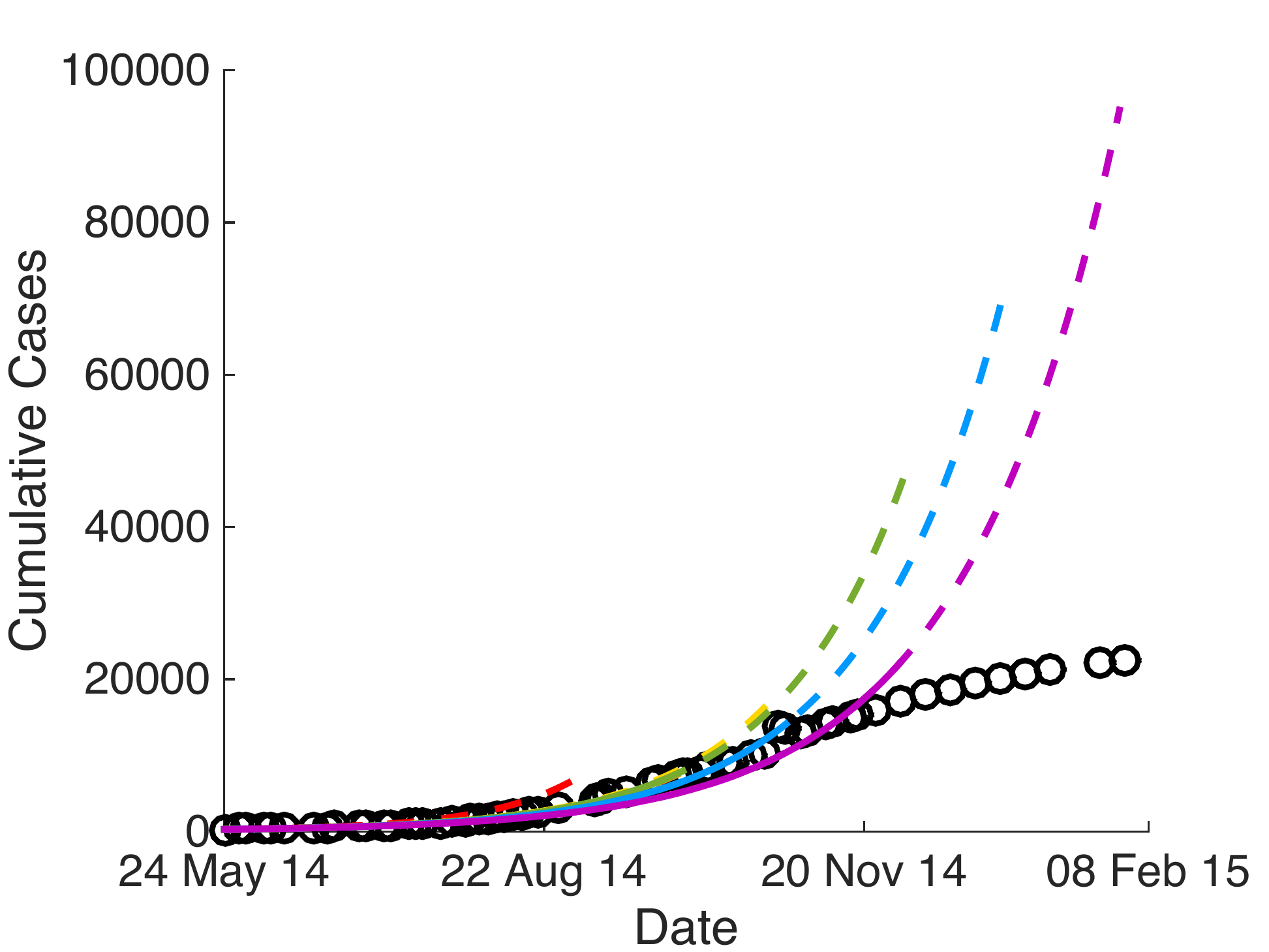}
\includegraphics[width=0.32\textwidth]{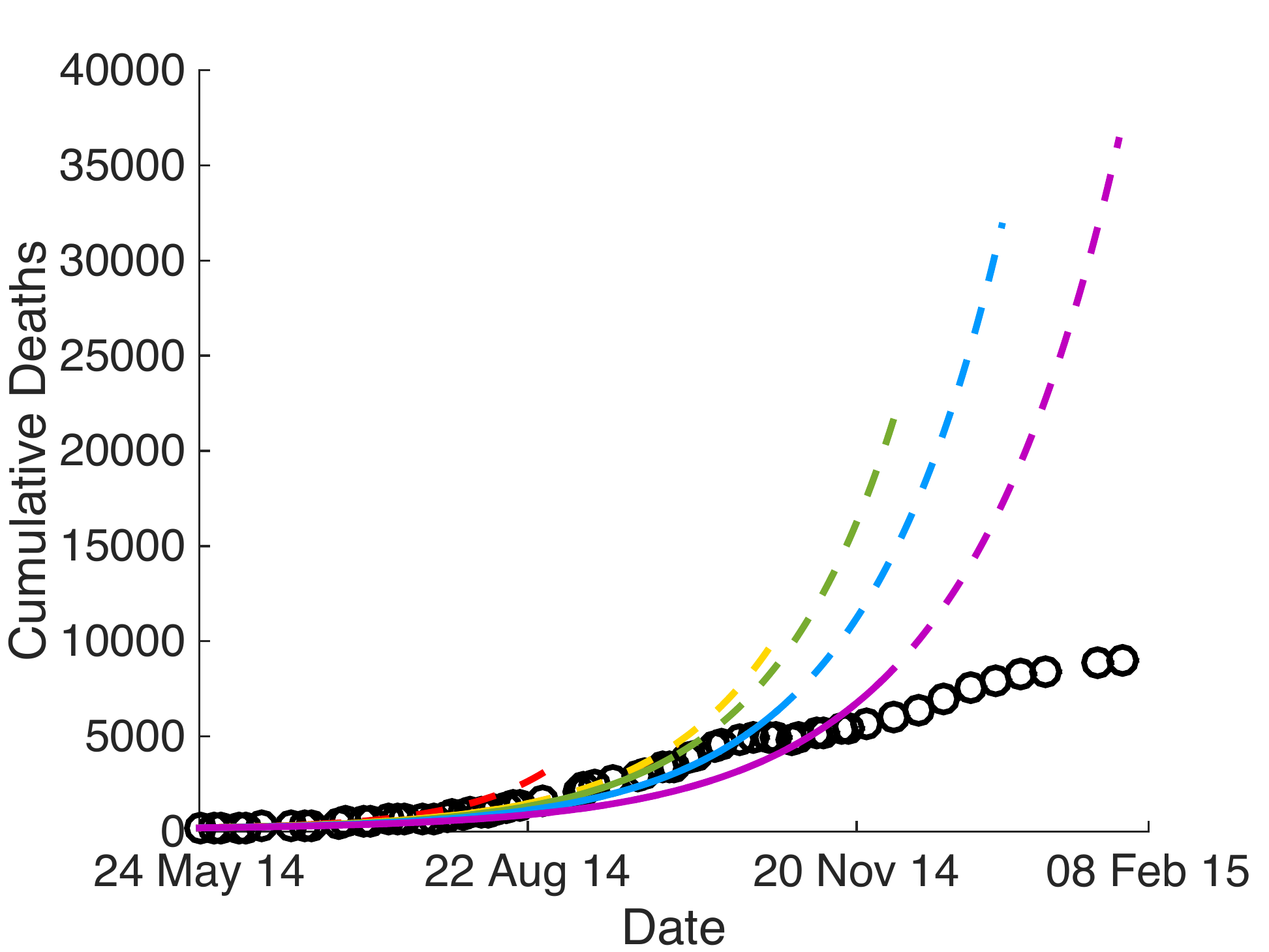}
\includegraphics[width=0.32\textwidth]{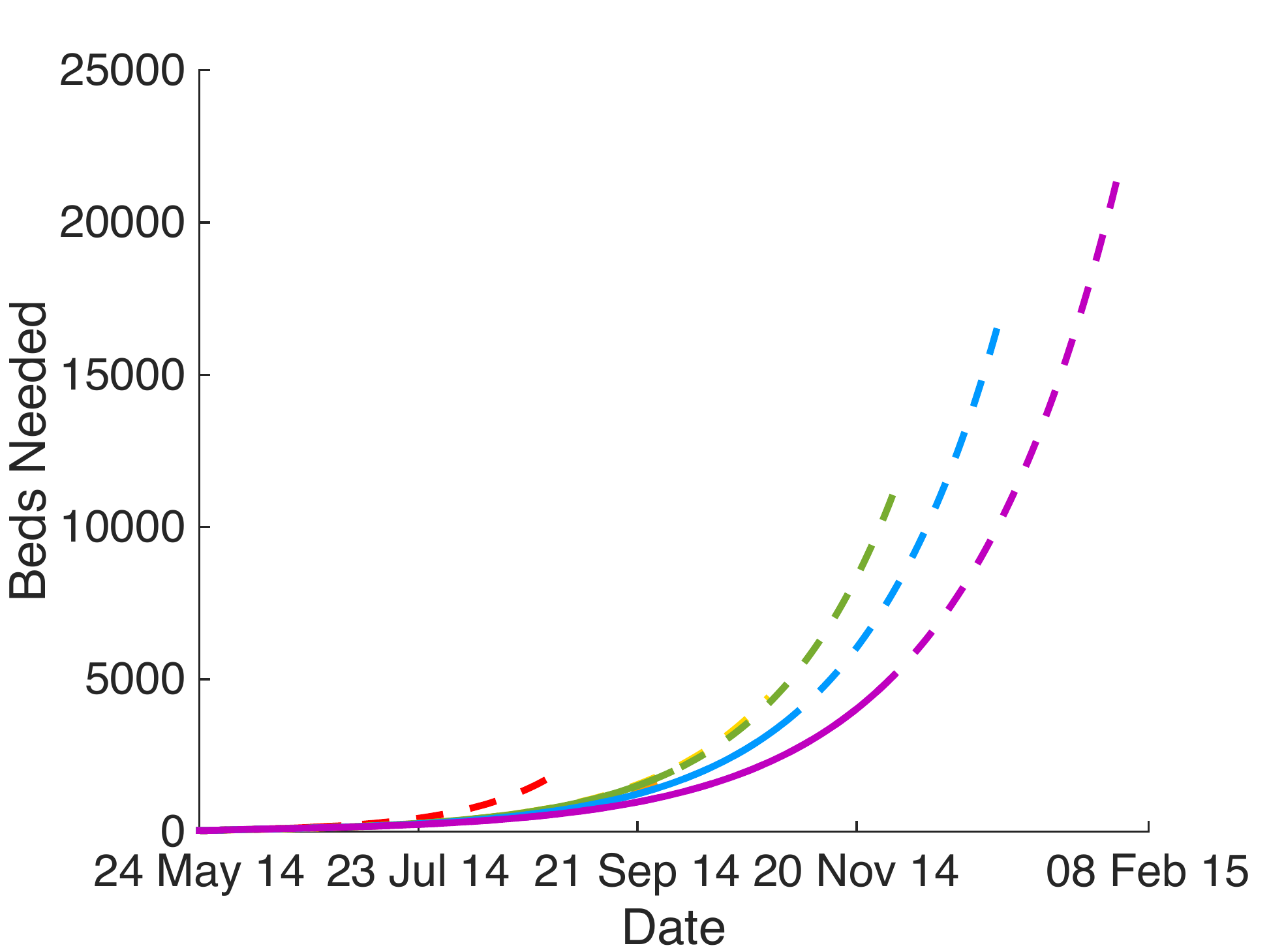}\\
\caption{Multiple model fits and forecasts for each country individually and all countries combined, with $\beta_1$ and $\delta$ fitted to data, and $k$ fixed to $1/2.5$ (based on the reporting rate estimate in \cite{Meltzer2014}). The remaining parameters fixed to the midpoints of the ranges in Table \ref{tab:params}. 
The model fits (solid lines) use the data up through July 1 (red), August 1 (orange), September 1 (yellow), October 1 (green), November 1 (blue) and December 1 (purple), with subsequent two months of forecasts shown as dashed lines in the same color. Note the difference in y-axis scale from Figure \ref{fig:MultiForecast_k}.}
\label{fig:MultiForecast_k0pt4}
\end{figure}

Figure \ref{fig:MultiForecast_k} shows expanded forecasting simulations for each country and all countries combined, using fixed parameter values at the midpoints of the ranges in Table \ref{tab:params}, with $\beta_1$, $\delta$ and $k$ fitted to the data using least squares. Figures \ref{fig:MultiForecast_nok} and \ref{fig:MultiForecast_k0pt4} show analogous forecasts but with $k$ fixed equal to 1 or $1/2.5$, to represent either perfect reporting and complete population at risk, or a correction only for underreporting using the estimates in \cite{Meltzer2014}. The corresponding 
parameter estimates for each fitted trajectory are given in Tables \ref{tab:MultiForecast_k}, \ref{tab:MultiForecast_nok}, and \ref{tab:MultiForecast_k0pt4}. 

\clearpage

\begin{table*}[h!]
\centering
\def\arraystretch{1.4}
\begin{tabular}{| c | c  c  c    c    c  c  c    c    c  c  c    c    c  c  c |}
\hline
Using data & \multicolumn{3}{c}{All countries}  && \multicolumn{3}{c}{Guinea} && \multicolumn{3}{c}{Liberia} && \multicolumn{3}{c|}{Sierra Leone}\\
\cline{2-4}\cline{6-8} \cline{10-12} \cline{14-16}
through:  & $\beta_1$ & $\delta$ & $k$ && $\beta_1$ & $\delta$ & $k$ && $\beta_1$ & $\delta$ & $k$ && $\beta_1$ & $\delta$ & $k$\\
\hline
July 1, 2014 & 
0.14  &  0.69  &  0.26 && 
-- & -- & -- &&
-- & -- & -- &&
0.067 &  0.37 & 0.34\\

Aug 1, 2014 & 
0.12  &  0.67  &  0.31 &&
0.13 & 0.97 & 0.24 &&
-- & -- & -- &&
0.10 & 0.51 & 0.63\\

Sep 1, 2014 & 
0.13 &  0.64  &  0.38 &&
0.13 &  0.75 & 0.46 &&
0.11 &  0.79 & 0.17 &&
0.11 & 0.49 &  0.46\\

Oct 1, 2014 & 
0.14  &  0.57  &  0.0018 &&
0.24 & 0.64 & 0.0002 &&
0.16 & 0.65 & 0.0021&&
0.13 & 0.37 & 0.43 \\

Nov 1, 2014 & 
0.14 & 0.51 &  0.0018 &&
0.20 &  0.59 & 0.0003 &&
0.16 & 0.57 & 0.0025 &&
0.13 & 0.36 & 0.71\\

Dec 1, 2014 & 
0.16 & 0.43 &  0.0015 &&
0.19 & 0.60 &  0.0003 &&
0.16 & 0.48 & 0.0029&&
0.15 & 0.28 & 0.073\\
\hline
\end{tabular}
\caption{Parameter estimates for each fit in Figure \ref{fig:MultiForecast_k}, with $k$ included in the fitted parameters. Each model is fitted to the data up through the dates given on the left.}
\label{tab:MultiForecast_k}
\end{table*}

\begin{table*}[h!]
\centering
\def\arraystretch{1.4}
\begin{tabular}{| c | c  c     c    c  c     c    c  c     c    c  c |}
\hline
Using data & \multicolumn{2}{c}{All countries}  && \multicolumn{2}{c}{Guinea} && \multicolumn{2}{c}{Liberia} && \multicolumn{2}{c|}{Sierra Leone}\\
\cline{2-3}\cline{5-6} \cline{8-9} \cline{11-12}
through:  & $\beta_1$ & $\delta$ && $\beta_1$ & $\delta$ && $\beta_1$ & $\delta$ && $\beta_1$ & $\delta$ \\
\hline
July 1, 2014 & 
0.14   &  0.69 && 
-- & -- &&
-- & -- &&
0.067 & 0.37\\

Aug 1, 2014 & 
0.12 & 0.67 &&
0.13 & 0.97 &&
-- & -- &&
0.10 & 0.51 \\

Sep 1, 2014 & 
0.13 & 0.64 &&
0.13 & 0.75 &&
0.11 & 0.79 &&
0.11 & 0.49 \\

Oct 1, 2014 & 
0.13 & 0.58 &&
0.13 & 0.69 &&
0.12 & 0.68 &&
0.13 & 0.37 \\

Nov 1, 2014 & 
0.13 & 0.53 &&
0.12 & 0.64 &&
0.11 & 0.60 &&
0.13 & 0.36 \\

Dec 1, 2014 & 
0.14 & 0.45 &&
0.11 & 0.66 &&
0.12 & 0.51 &&
0.14 & 0.28 \\
\hline
\end{tabular}
\caption{Parameter estimates for each fit in Figure \ref{fig:MultiForecast_nok}, with $k$ fixed equal to 1 in all cases. Each model is fitted to the data up through the dates given on the left.}
\label{tab:MultiForecast_nok}
\end{table*}

\begin{table*}[h!]
\centering
\def\arraystretch{1.4}
\begin{tabular}{| c | c  c     c    c  c     c    c  c     c    c  c |}
\hline
Using data & \multicolumn{2}{c}{All countries}  && \multicolumn{2}{c}{Guinea} && \multicolumn{2}{c}{Liberia} && \multicolumn{2}{c|}{Sierra Leone}\\
\cline{2-3}\cline{5-6} \cline{8-9} \cline{11-12}
through:  & $\beta_1$ & $\delta$ && $\beta_1$ & $\delta$ && $\beta_1$ & $\delta$ && $\beta_1$ & $\delta$ \\
\hline
July 1, 2014 & 
0.14 & 0.68 &&
-- & -- &&
-- & -- &&
0.067 & .37 \\

Aug 1, 2014 & 
0.12 & 0.67 &&
0.13 & 0.97 &&
-- & -- &&
0.10 & 0.51 \\

Sep 1, 2014 & 
0.13 & 0.64 &&
0.13 & 0.75 &&
0.11 & 0.79 &&
0.11 & 0.49 \\

Oct 1, 2014 & 
0.13 & 0.58 &&
0.13 & 0.69 &&
0.12 & 0.68 &&
0.13 & 0.37 \\

Nov 1, 2014 & 
0.13 & 0.53 &&
0.12 & 0.64 &&
0.11 & 0.60 &&
0.13 & 0.36 \\

Dec 1, 2014 & 
0.14 & 0.45 &&
0.11 & 0.66 &&
0.12 & 0.50 &&
0.14 & 0.28 \\
\hline
\end{tabular}
\caption{Parameter estimates for each fit in Figure \ref{fig:MultiForecast_k0pt4}, with $k$ fixed equal to $0.4$ in all cases. Each model is fitted to the data up through the dates given on the left.}
\label{tab:MultiForecast_k0pt4}
\end{table*}

\end{document}